 \documentclass[prc,amsmath,showpacs,floatfix,eqsecnum,nofootinbib]{revtex4}
\usepackage{axodraw}
\usepackage{graphicx}
\usepackage{bm}    
\usepackage{pstcol} 

\begin{document}

\title{Constituent Quark-model for Baryons \\         
 Harmonic confinement and Two-body Meson-exchange Potentials }  
\author{Th.A. Rijken}
\affiliation{Institute for Mathematics, Astrophysics and Particle Physics, \\
         University of Nijmegen,
         Nijmegen, the Netherlands}
\date{version: \today}

\begin{abstract}
Soft Two-body potentials between the
constituent quarks of the nucleon are derived using harmonic
oscillator, i.e. gaussian,  quark wave-functions.
The gaussian wave functions are very suited for applications with the ESC 
soft-core interactions, which employ gaussian form factors.
In these notes using the Fourier transformation to 
momentum space the local and non-local contributions of the potentials 
based on the ESC meson-quark-quark vertices are evaluated.
Using the ESC16 parameters translated to the quark-level lead to parameter-free
two-body and three-body diquark and triquark meson-exchange interactions.
Application to the SU(3) baryon-octet and the $\Delta_{33}$-resonance are performed,
within the CQM using a harmonic confinement potential, leading to a 
satisfactory picture with relativistic constituent quarks.
We present two versions for the $N-\Delta$ splitting: (i) model A  with the 
instanton interaction, and (ii) model B with a large color-magnetic interaction
from an almost point like OGE.
The size of the baryons $\approx$ 1 fm.
\end{abstract}

\maketitle

\twocolumngrid
\section{Introduction by Topics}

\noindent {\bf General:}
The interpretation of the ESC-model in the context of QCD is based
on the constituent quark model (CQM). The latter is connected to the low-energy
vacuuum structure of QCD as an instanton-anti-instanton liquid \cite{DY-PE84}
which leads to constituent quark masses at low momenta. 
Then, application of the CQM in the QPC mechanism \cite{Mic69}, 
in e.g. the SU(6)-version of Ref.~\cite{LeY73}, leads to a successful 
match of the description of the couplings with the fitted results in the ESC-models.
It was shown \cite{TAR-QQ14} that for the CQM meson-exchange between quarks
leads by folding to the correct baryon-baryon potentials up to $1/M_B^2$-terms,
i.e. the right central, spin-spin, tensor, spin-orbit, and
quadratic spin-orbit potentials. Based on this correspondence quark-quark (QQ)
and quark-nucleon potentials were constructed \cite{RY24}, which have been applied to the
study of quark-matter \cite{YYR24}.

\noindent In order to check the validity of this approach to QQ-interactions
it is required to apply such a meson-exchange QQ interaction to the baryons
themselves. In this note we derive the matrix elements for the proton (P)
and neutron (N) of the one-boson-exchange (OBE) QQ-potentials.
The masses for the SU(3) octet baryons P, $\Lambda$, $\Sigma$, $\Xi$, and 
$\Delta_{33}(1236)$ are evaluated within the CQM, including the OBE and OGE
potentials in Born-approximation. It turns out that the contributions from
OBE and OGE are marginal, and there are large cancellations between the 
confinement potential and the (relativistic) kinetic energies of the quarks.\\
\noindent {\bf Constituent Quark model and QCD:} In this paper  we work 
within the framework of the CQM. For the baryons we envisage that the 
three constituent quarks are put into a deep, but finite, potential well, 
which we assume of the form $V_{conf} = -C_0 + C_2\ r^2$.  
This is similar to the quark-bag models \cite{Has78} where the quarks are confined 
to a sphere, and also there is a resemblance with the nuclear shell-model. 
In principle this well 
should be derived from the QCD interactions between the quarks, which proved to
be too difficult thus far. The energy levels correspond to
the baryon masses, where we restrict ourselves to the ground states. The 
residual interactions are one-gluon-exchange (OGE) and meson-exchange (ESC) between the
quarks. Rotational invariance in three-dimensional space leads to O(3) invariance,
and the states are symmetric in SU(3) flavor and SU(2) spin, and antisymmetry 
in color SU(3). So the full symmetry group structure is            
  SU$_c$(3)$\otimes$SU$_{sf}$(6)$\otimes$O(3).\\
There are indications from QCD that the confining potential between two quarks rises
linearly with the distance r, i.e $V_{conf} = -a + b r$. For the ground states the 
harmonic and linear potential give similar results \cite{Gro76}. 
This because (i) at small r the radial wave function $u(r)=\psi(r)/r$ is zero 
at the origin, and (ii) at large r in both cases the wave function is decreasing 
exponentially. Therefore, only the intermediate r-region contributes to the energy.\\
\noindent Finally, we note that the quark systems in the confining well are bound states 
by definition.
For confining potentials the situation is different from that with non-confining
potentials. In the latter case for a bound-state it is necessary that the mass is
less than the total mass of the constituents. This is not so with confinement. 
For example in the case of the $\Delta_{33}(1236)$-resonance the sum of the quark masses
is $\approx M_p$ which is 300 MeV less than the $\Delta_{33}$-mass.
In the space of the three-quark system $\Delta_{33}$ is a bound-state, but in the
space of the three-quarks+pion it is a resonance. 

\noindent {\bf ESC, Constituent Quarks, Instantons, and QPC:}
In the CQM the BBM-coupling constants of the ESC-models can be explained nicely
by the quark-pair-creation (QPC) mechanism.
Table~II in \cite{CW10501} shows the buildup of these couplings by the $^3S_1$ and $^3P_0$
quark-pair creation mechanisms, where the latter is dominant by a factor 4.
The calculation of this table uses the constituent quark model (CQM) in the 
SU(6)-version of \cite{LeY73}. Since this calculation uses implicitly the coupling of
the mesons to quarks, it defines the QQM-vertex. Then, OBE-potentials can be 
derived by folding meson-exchange with the quark wave functions of the baryons.
At the baryon level the vertices have in Pauli-spinor space the structure

\onecolumngrid\vspace*{-0.5ex}{\tiny
\noindent\begin{tabular}[t]{c|} \parbox{0.493\hsize}{~} \\
\hline \end{tabular}}
\begin{eqnarray*}
 \bar{u}(p',s') \Gamma u(p,s) &=& \chi_{s'}^{\prime \dagger}\left\{
 \Gamma_{bb}+\Gamma_{bs}\frac{\bm{\sigma}\cdot{\bf p}}{E+M} 
 -\frac{\bm{\sigma}\cdot{\bf p}'}{E'+M'} \Gamma_{sb} 
 -\frac{\bm{\sigma}\cdot{\bf p}}{E+M} \Gamma_{ss} 
 \frac{\bm{\sigma}\cdot{\bf p}'}{E'+M'} \Gamma_{sb}\right\}\ \chi_s \nonumber\\
 &\equiv& \sum_l c_{BB}^{(l)} O_l({\bf p}',{\bf p}) (\sqrt{M'M})^{\alpha_l}\ \
 (l=bb,bs,sb,ss).
\end{eqnarray*}
{\tiny\hspace*{\fill}\begin{tabular}[t]{|c}\hline\parbox{0.49\hsize}{~}\\
\end{tabular}}\vspace*{-0.5ex}
\twocolumngrid

This expansion is general and does not depend on the internal structure of the
baryon. A similar expansion can be made on the quark-level with quark masses $m_Q$ and
coefficients $c_{QQ}^{(l)}$.
Now it appears that in the CQM, i.e. $m_Q=M_B/3$, the QQM-vertices can be 
chosen such that the ratio's $c_{QQ}^{(l)}/c_{BB}^{(l)}$ are constant for each type of
meson \cite{TAR-QQ14}.  Then, by scaling the couplings these coefficients can be made equal.
(Ipso facto this defines a meson-exchange quark-quark interaction.)
This shows that the use of the QPC-model is consistent with the 1/M-expansion.

The observation above can be related to low-energy QCD. 
The two non-perturbative effects in QCD are confinement and chiral symmetry
breaking. The SU(3)$_L\otimes$SU(3)$_R$ chiral symmetry is spontaneously broken to 
an SU(3)$_v$ symmetry at a scale $\Lambda_{\chi SB} \approx 1$ GeV. 
The confinement scale is $\Lambda_{QCD} \approx 100-300$ MeV, which roughly
corresponds to the baryon radius $\approx$ 1 fm.
Due to the complex structure of the QCD vacuum, which can be understood as a
liquid of BPST instantons and anti-instantons \cite{BPST75,DY-PE84}, 
the valence quarks acquire a dynamical or constituent mass \cite{Wei75,Man84,
SHUR84}. 
With the empirical value of the gluon
condensate \cite{SVZ79} as input the instanton density and radius become \cite{SHUR84}
$n_c= 8\cdot 10^{-4}\ {\rm GeV}^{-4}, \rho_c \approx 0.3\ {\rm fm}$. 
With these parameters the non-perturbative vacuum expectation value for 
the quark fields is
$\langle vac|\bar{\psi}\psi|vac\rangle \approx -10^{-2}\ {\rm GeV}^3$.                    
The calculated effective low-momentum quark and gluon mass in the
instanton vacuum \cite{DY-PE84,Hut95}  are $m_Q(p=0)= 345, m_G(p=0)=420$ MeV.    
Note that this quark mass is remarkably close to the constituent mass $M_N/3$,
which gives support to the relations given above.

In \cite{Gloz96a} the coupling of the pseudoscalar mesons, being the 
Goldstone bosons of spontaneous broken chiral symmetry, to the quarks explained 
many features of the hadronic spectrum. Also the quark-quark instanton-exchange 
 interaction \cite{GtH76} can explain the $\pi-\rho$ mass difference.
In the ESC-models we can apparently extend 
the meson-exchange between quarks by proposing to include,
besides the pseudoscalar, all meson nonets: vector, axial-vector, scalar etc.
{\it Since all these meson nonets can be considered as quark-antiquark bound states,
there is no reason to exclude any of these mesons from the quark-quark interactions.}
{\it Furthermore, our preferred value for the constituent quark mass seems to have
a basis in the instanton liquid structure of the QCD vacuum.}

\noindent {\bf Quark wave-functions $J^P=\frac{1}{2}^+$ Baryons:}
In this note we evaluate expectation value of the two-body QQ-potentials 
for the P and N using the D\&D-model \cite{Dal58,HT64}
for the three-quark wave function. 
We estimate that the contribution to the binding will be $\approx -0.?$ MeV. 

\noindent {\it We note that the formalism described in these notes is easily 
generalized to the case where the three-body wave function is a sum over
gaussians. Therefore, using a realistic gaussian expansion of the wave
functions, as for example practiced in the GEM approach of Hiyama and 
Kamimura \cite{Hiy03}, a truly realistic estimate of the contribution
of the OBE-potential to the nucleon mass is feasible
within the framework of these notes.}\\
In this paper we do not distinguish 
between the $\bm{\rho}$ and $\bm{\lambda}$ modes, which would break the
$S_3$-symmetry and make the implementation of the generalized 
Pauli-principle difficult.\\

\noindent {\bf Content:}                            
The contents of these notes is as follows. 
In section \ref{sec:T} (i) The A=3 wave functions for the proton (P) and
neutron (N) are described in momentum space. 
In section \ref{sec:mspace1} the basic integrals for the evaluation of the
matrix elements of the two-body OBE interactions are derived.
In section \ref{sec:EN} the matrix elements of the two-body OBE forces
worked out explicitly for the nucleons.
These are expressed in terms of the matrix elements of the isospin/spin 
operators and basic integrals. 
The same is done in section~\ref{sec:conf1} for the one-gluon-exchange (OGE)
potential.
In section~\ref{sec:conf2} the Nambu-Jona-Lasinio (NJL) form of the instanton 
interaction and the choice of the confining potential are described 
and applied to the calculation of the baryon masses.
The same is done in section~\ref{sec:conf1} for the one-gluon-exchange (OGE)
potential and the color-magnetic interaction.
In section~\ref{sec:results} the results for the $V_2$-contribution
to the nucleon mass are given and discussed for the parameters of the ESC16 model. 
	
\noindent At the end of these notes several appendices are included for 
spelling out some details of the calculations.
In Appendix~\ref{sec:mspace2} the details of the basic functions are given.   
In Appendix~\ref{app:T.d} the momentum space integrals for the three-body
matrix elements are listed.
In Appendix~\ref{app:GDD} we work out the momentum space 
integrals for the general case where the initial and final
states are sums of gaussians of the Dalitz-Downs type. 
this opens the possibility to apply this work for e.g. GEM wave-functions.
In Appendix~\ref{app:OBE} the OBE quark-quark potentials are given in
momentum space.
Similarly, in Appendix~\ref{app:OBE2} the "additional" OBE quark-quark potentials due to
the extra meson-quark-quark vertices are given in
momentum space.
In Appendix~\ref{app:iso} the matrix elements of the isospin- and 
spin-operators in three-quark space for the nucleon are evaluated.
In Appendix~\ref{app:mom2} the expectation value of the non-relativistic 
kinetic energy is recalculated using the cartesian momenta including explicitly 
the CM-constraint on the momenta of the quarks.
\onecolumngrid
\begin{flushleft}
\rule{16cm}{0.5mm}
\end{flushleft}
\vspace*{25mm}

\begin{figure}[hhhbt]
\begin{center}
 \resizebox{21.25cm}{!}
 {\includegraphics[1.0in,9.0in][9in,10in]{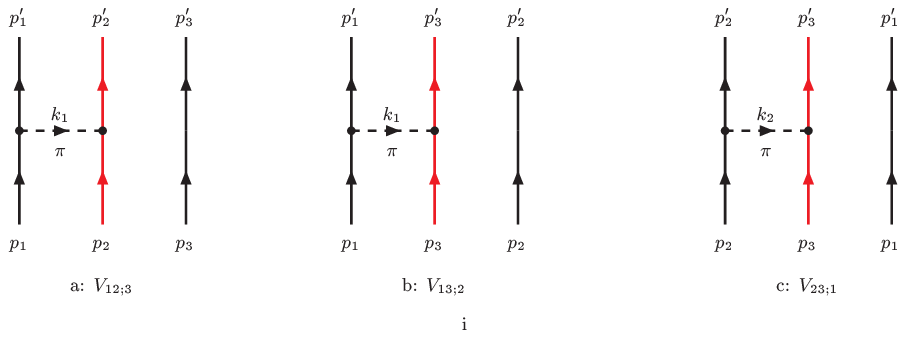}}
\end{center}
\caption{\sl The Born-Feynman diagrams for two-body forces
$V_{12;3}, V_{13;2}, V_{23;1}$} 
\label{fig.twbf2-born} 
\end{figure}
\onecolumngrid
\section{A=3 Dalitz-Downs Model}            
\label{sec:T}  

\subsection{Wave functions for the proton P(uud) and neutron N(udd)}   
\label{app:T.a}  
The 3Q wave function is assumed to be of the following form
\cite{Dal58,HT64}:                  
\begin{equation}
 \psi_{3Q}(r_1,r_2,r_3)= N_3\exp\left[-\frac{1}{2}\lambda\left({\bf x}_{12}^2
 +{\bf x}_{23}^2+{\bf x}_{31}^2\right)\right],
\label{eq:T.1}\end{equation}
where
$
 {\bf x}_{12} ={\bf x}_1-{\bf x}_2,\ {\bf x}_{23} ={\bf x}_2-{\bf x}_3,\ 
 {\bf x}_{31} ={\bf x}_3-{\bf x}_1.  
$
The Jacobian coordinates for the three-particle system are
\begin{subequations}
\begin{eqnarray}
 \bm{\rho} = \frac{1}{\sqrt{2}}\left({\bf x}_1-{\bf x}_2\right) 
 \hspace{1.0cm} &,&\hspace{0.5cm}
  {\bf x}_1 = \frac{1}{\sqrt{6}}\bm{\lambda}+\frac{1}{\sqrt{2}} \bm{\rho} 
  +\frac{1}{\sqrt{3}} {\bf R}, \\
 \bm{\lambda} = \frac{1}{\sqrt{6}}\left({\bf x}_1+{\bf x}_2-2{\bf x}_3\right) 
 &,&\hspace{0.5cm}
  {\bf x}_2 = \frac{1}{\sqrt{6}}\bm{\lambda}-\frac{1}{\sqrt{2}} \bm{\rho} 
  +\frac{1}{\sqrt{3}} {\bf R}, \\
 {\bf R} = \frac{1}{\sqrt{3}}\left({\bf x}_1+{\bf x}_2+{\bf x}_3\right) 
 \hspace{1mm} &,&\hspace{0.5cm}
  {\bf x}_3 = -\sqrt{\frac{2}{3}}\bm{\lambda} +\frac{1}{\sqrt{3}} {\bf R}.    
\label{eq:T.3}\end{eqnarray}
\end{subequations}
The differences expressed in the Jacobi-coordinates are 
\begin{eqnarray*}
&& {\bf x}_1-{\bf x}_2 = \sqrt{2} \bm{\rho},\ \ 
 {\bf x}_1-{\bf x}_3 = \sqrt{\frac{1}{2}} \bm{\rho}+\sqrt{\frac{3}{2}} \bm{\lambda}, \\
&& {\bf x}_2-{\bf x}_3 = -\sqrt{\frac{1}{2}} \bm{\rho}+\sqrt{\frac{3}{2}} \bm{\lambda},    
\end{eqnarray*}
which leads to the expression
$
 {\bf x}_{12}^2+{\bf x}_{13}^2+{\bf x}_{23}^2 = 
 3\left(\bm{\rho}^2+\bm{\lambda}^2\right), 
$
and the three-nucleon wave function (\ref{eq:T.1}) becomes           
\begin{equation}
 \psi_{3Q}(r_1,r_2,r_3)= N_3\exp\left[-\frac{3}{2}\lambda\left(
 \bm{\rho}^2+\bm{\lambda}^2 \right)\right] \equiv \psi_{QN}(\bm{\rho},\bm{\lambda}).
\label{eq:T.6}\end{equation}
The normalization is
\begin{equation}
 1 = \int d^3\rho \int d^3\lambda\ |\psi(\bm{\rho},\bm{\lambda})|^2 =
 N_3^2\left(\frac{\pi}{3\lambda}\right)^3 \rightarrow 
 N_3=\left(\frac{3\lambda}{\pi}\right)^{3/2}.
\label{eq:T.7}\end{equation}

\subsection{Momentum-representation D\&D model}                               
\label{app:T.b}  
\noindent {\bf 1.\ Wave function}: 
The momentum-space the 3Q-wave function is defined by
\begin{eqnarray}
 \widetilde{\psi}_{3Q}({\bf p}_\rho,{\bf p}_\lambda) &=& N_3 
 \int\int d^3\rho d^3\lambda\ e^{i\left({\bf p}_\rho\cdot\bm{\rho}
 + {\bf p}_\lambda\cdot\bm{\lambda}\right)}
 \exp\left[-\frac{3}{2}\lambda \left(\bm{\rho}^2+\bm{\lambda}^2\right)\right]
 \nonumber\\ &=& 
 \widetilde{N}_3 \exp\left[-\frac{1}{6\lambda}\left({\bf p}_\rho^2+
 {\bf p}_\lambda^2\right)\right],\ \ {\rm with}\ 
 \widetilde{N}_3 = \left(\frac{4\pi}{3\lambda}\right)^{3/2},
\label{eq:T.11}\end{eqnarray}
and in configuration-space 
\begin{equation}
 \psi_{3Q}(\bm{\rho},\bm{\lambda}) = \int\int\frac{d^3p_\rho}{(2\pi)^3}
 \frac{d^3p_\lambda}{(2\pi)^3}\ e^{-i\left({\bf p}_\rho\cdot\bm{\rho}
 +{\bf p}_\lambda\cdot\bm{\lambda}\right)}\ 
 \widetilde{\psi}_{3Q}({\bf p}_\rho,{\bf p}_\lambda), 
\label{eq:T.12}\end{equation}
and normalization
\begin{equation}
 \int\int \frac{d^3p_\rho}{(2\pi)^3} \frac{d^3p_\lambda}{(2\pi)^3} 
 |\widetilde{\psi}_{3Q}(\bm{p}_\rho,{\bf p}_\lambda)|^2=1.
\label{eq:T.13}\end{equation}

\noindent Using the momentum-space wave functions of this subsection, given the
momentum-space $\widetilde{V}_3$-potential, the integrals occurring in the
matrix elements can executed analytically.\\

\noindent {\bf 2.\ Momentum-space Matrix elements}:                         
We first translate the momenta that occur in the potentials in the 
$(\bm{\rho},\bm{\lambda})$-language. 
For the initial state the momenta are 
\begin{subequations}
\begin{eqnarray}
&& {\bf p}_1 = \sqrt{\frac{1}{6}}{\bf p}_\lambda+\sqrt{\frac{1}{2}} {\bf p}_\rho
 +\sqrt{\frac{1}{3}} {\bf P}, \\
&& {\bf p}_2 = \sqrt{\frac{1}{6}}{\bf p}_\lambda-\sqrt{\frac{1}{2}} {\bf p}_\rho
 +\sqrt{\frac{1}{3}} {\bf P}, \\
&& {\bf p}_3 = -\sqrt{\frac{2}{3}}{\bf p}_\lambda
 +\sqrt{\frac{1}{3}} {\bf P},
\label{eq:T.16}\end{eqnarray}
\end{subequations}
where  $\sqrt{3} {\bf P} = {\bf P}_i = \sum_{i=1}^3 {\bf p}_i$, and similarly 
for the final state momenta.
In passing we note that with these definitions
\begin{eqnarray*}
 \sum_{i=1}^3 {\bf p}_i\cdot{\bf x}_i &=& 
{\bf p}_\rho\cdot{\bf x}_\rho + {\bf p}_\lambda\cdot{\bf x}_\lambda
 +{\bf P}\cdot{\bf R}.
\end{eqnarray*}
\noindent We work in the overall CM-momentum frame,
i.e. for the total momentum in the initial and final state we have
${\bf P}= {\bf P}_f=0$. Then, the customary momenta 
$({\bf q}_i=({\bf p}_i'+{\bf p}_i)/2$ and ${\bf k}_i={\bf p}_i'-{\bf p}_i$ 
become in the $(\bm{\rho},\bm{\lambda})$-language
\begin{subequations}
\begin{eqnarray}
 {\bf k}_1 = \frac{1}{\sqrt{6}} {\bf k}_\lambda+\frac{1}{\sqrt{2}}{\bf k}_\rho
 &,& 
   {\bf q}_1 = \frac{1}{\sqrt{6}} {\bf q}_\lambda+\frac{1}{\sqrt{2}}{\bf q}_\rho, \\
 {\bf k}_2 = \frac{1}{\sqrt{6}} {\bf k}_\lambda-\frac{1}{\sqrt{2}}{\bf k}_\rho
 &,& 
   {\bf q}_2 = \frac{1}{\sqrt{6}} {\bf q}_\lambda-\frac{1}{\sqrt{2}}{\bf q}_\rho, \\
{\bf k}_3 = -\sqrt{\frac{2}{3}}{\bf k}_\lambda \hspace{1.0cm} 
 &,& 
   {\bf q}_3 = -\sqrt{\frac{2}{3}}{\bf q}_\lambda.
\label{eq:T.17}\end{eqnarray}
\end{subequations}
For the squares occurring in the wave functions and potentials we obtain
\begin{subequations}
\begin{eqnarray}
&& {\bf p}_\rho^{\prime 2}+{\bf p}_\rho^2 = 
2({\bf q}_\rho^2+\frac{1}{4}{\bf k}_\rho^2),\ \ 
   {\bf p}_\lambda^{\prime 2}+{\bf p}_\lambda^2 = 
2({\bf q}_\lambda^2+\frac{1}{4}{\bf k}_\lambda^2), \\ 
&& {\bf k}_1^2 = \frac{1}{6} {\bf k}_\lambda^2+\frac{1}{2}{\bf k}_\rho^2 
 +\frac{1}{\sqrt{3}} {\bf k}_\rho\cdot{\bf k}_\lambda, \\
&& {\bf k}_2^2 = \frac{1}{6} {\bf k}_\lambda^2+\frac{1}{2}{\bf k}_\rho^2 
 -\frac{1}{\sqrt{3}} {\bf k}_\rho\cdot{\bf k}_\lambda, \\
&& {\bf k}_3^2 = \frac{2}{3} {\bf k}_\lambda^2.
\label{eq:T.18}\end{eqnarray}
\end{subequations}
Working in the three-body CM-system, i.e. ${\bf P}=0$, the transformation between 
the different coordinates leads to 
\begin{equation}
 d^3p_\rho d^3 p_\lambda d^3P = \left\| \begin{array}{ccc}    
 \frac{\partial p_\rho}{\partial p_1} & \frac{\partial p_\rho}{\partial p_2} 
 & \frac{\partial p_\rho}{\partial p_3} \\
 \frac{\partial p_\lambda}{\partial p_1} & \frac{\partial p_\lambda}{\partial p_2}
 & \frac{\partial p_\lambda}{\partial p_3} \\
 \frac{\partial P}{\partial p_1} & \frac{\partial P}{\partial p_2}
 & \frac{\partial P}{\partial p_3} \\
\end{array}\right\|\ d^3 p_1 d^3p_2 d^3p_3= d^3p_1 d^3p_2 d^3p_3,\ 
 d^3p_1 d^3 p'_1 = d^3 q_1 d^3k_1.  
\label{eq:T.19}\end{equation}

In the case of a two-body interaction $V_2$ we take ${\bf k}_3=0$ and hence 
 ${\bf k}_2=-{\bf k}_1 \equiv {\bf k}$. In the case of the three-body
interaction $V_3$ one has ${\bf k}_1+{\bf k}_2+{\bf k}_3=0$. 
With the setting of the Jacobi-coordinates in momentum space the
matrix elements of the interactions can be evaluated using the 
momentum space representation of the potentials.

\section{ V$_2$ Three-body matrix elements in Momentum Space}    
\label{sec:mspace1}
The three-body matrix element of the two-body potential $V_2$ is
\begin{eqnarray}
&& \langle {\bf p}'_1, {\bf p}'_2, {\bf p}'_3| V_2|{\bf p}_1, {\bf p}_2, {\bf p}_3\rangle
= \langle {\bf p}'_1, {\bf p}'_2| V_2|{\bf p}_1, {\bf p}_2\rangle\cdot 
 \langle {\bf p}'_3|{\bf p}_3\rangle =\nonumber\\ && 
 (2\pi)^3\ \langle {\bf p}'_1, {\bf p}'_2| V_2|{\bf p}_1, {\bf p}_2\rangle\cdot 
 \delta^3({\bf p}'_3-{\bf p}_3) = \nonumber\\ &&      
 (2\pi)^3\ \delta^3({\bf p}_1'+{\bf p}'_2+{\bf p}'_3-{\bf p}_1-{\bf p}_2-{\bf p}_3)        
  \cdot(2\pi)^3 \langle {\bf p}'_1, {\bf p}'_2| v_2|{\bf p}_1, {\bf p}_2\rangle\cdot 
 \delta^3({\bf p}'_3-{\bf p}_3).                      
\label{eq:T.1a}\end{eqnarray}
The factor $(2\pi)^3$ is due to the normalization of the one-particle momentum states,
$({\bf p}'|{\bf p}) = (2\pi)^3 \delta^3({\bf p}'-{\bf p})$.\\
The $V_2$ interaction in momentum space for the central-, spin-spin-,
tensor-, spin-orbit-, and quadratic-spin-orbit has factors:
 $ 1, {\bf  k}^2, {\bf q}^2, {\bf k}\times{\bf q}$. We consider the $V_{12;3}$
potential. Then, for $V_2$ we have ${\bf k}_2=-{\bf k}_1$, and for the non-local
potentials ${\bf q}^2 \rightarrow ({\bf q}_1^2+{\bf q}_2^2)/2$. 
The evaluation
of the three-body matrix elements using harmonic oscillator wave
functions the overlap integrals $I_3(i,j)$ are given in this section.\\
\noindent In Appendix~\ref{app:T.d} we list a complete set of Gaussian
integrals that enables to do all momentum space integrals relevant 
for this paper. Among them integrals quadratic in the components 
of the vectors ${\bf k}_1$ and ${\bf k}_2$. 
\noindent We define
\begin{subequations} \label{eq:T.15}
\begin{eqnarray}
&& H_{[k,l]} \equiv  \langle \psi_3|
 \left({\bf k}^2\right)^k 
 \left({\bf q}^2\right)^l\ G_0({\bf k}^2;m^2,\Lambda^2)
| \psi_3\rangle,\ \ {\rm with} \\
&& G_0({\bf k}^2;m^2,\Lambda^2) = e^{-{\bf k}^2/\Lambda^2}\
 \bigl[ {\bf k}^2+m^2 \bigr]^{-1}.
\end{eqnarray}
\end{subequations}
We also define the "diffractive" matrix element by 
\begin{subequations} \label{eq:T.20}
\begin{eqnarray}
&& D_{[k,l]} \equiv  \langle \psi_3|
 \left({\bf k}^2\right)^k 
 \left({\bf q}^2\right)^l\ G_D({\bf k}^2;\Lambda^2)
| \psi_3\rangle,\ \ {\rm with} \\
&& G_D({\bf k}^2;\Lambda^2) = e^{-{\bf k}^2/\Lambda^2}.
\end{eqnarray}
\end{subequations}

\noindent {\bf a.\ Evaluation $V_2$ expectation values:}                      
For diagram (a) in Fig.~\ref{fig.twbf2-born} we 
evaluate in momentum space the basic integral
\begin{eqnarray}
 H_{[0,0]} &=& (2\pi)^3\widetilde{N}_3^2\
 \int \frac{d^3p_\rho' d^3p_\lambda'}{(2\pi)^6} 
 \int \frac{d^3p_\rho d^3p_\lambda}{(2\pi)^6}\left\{
 \exp\left[-\frac{1}{6\lambda}({\bf p}_\rho^{\prime 2}+{\bf p}_\lambda^{\prime 2})
\right]
 \right.\cdot\nonumber\\ && \times\left.
 \exp\left[-\frac{1}{6\lambda}({\bf p}_\rho^{2}+{\bf p}_\lambda^{2})
\right]\ \frac{e^{-{\bf k}^2/\Lambda^2}}{{\bf k}^2+m^2}
\right\}\nonumber\\ &=& (2\pi)^{-9 }
 \widetilde{N}_3^2\ \int_0^\infty d\alpha\ e^{-\alpha m^2}\cdot
 \int d^3p_\rho' d^3p_\lambda' \int d^3p_\rho d^3p_\lambda 
 \cdot\nonumber\\ && \times\left\{
 \exp\left[-\frac{1}{6\lambda}\left(
 {\bf p}_\rho^{\prime 2}+{\bf p}_\lambda^{\prime 2}
 +{\bf p}_\rho^{2}+{\bf p}_\lambda^{2}\right)\right]\ 
 e^{-\gamma {\bf k}^2} \right\},
\label{eq:T.21}\end{eqnarray}
where $\gamma = \alpha+\Lambda^{-2}$.\\[1mm]

\noindent {\bf b.\ Cartesian momenta:} 
Since the potentials $V_2$ are expressed in the cartesian momenta 
${\bf k}_i, (i=1,2,3)$ it is convenient
to express the integral in (\ref{eq:T.19}) in terms of these variables. (This is also 
the case for the non-local momenta ${\bf q}_i, (i=1,2,3)$ when the contribution
of these terms is non-vanishing, of course.) 
In cartesian coordinates the exponential factor from the wave functions has
\begin{eqnarray}
 {\bf p}_\rho^{\prime 2}+{\bf p}_\lambda^{\prime 2} 
 +{\bf p}_\rho^2+{\bf p}_\lambda^2 = 4\biggl[
 ({\bf q}_1^2+{\bf q}_1\cdot{\bf q}_2+{\bf q}_2^2) +
 \frac{1}{4}({\bf k}_1^2+{\bf k}_1\cdot{\bf k}_2+{\bf k}_2^2)\biggr].
\label{eq:T.30}\end{eqnarray}
In cartesian momenta we get 
\begin{eqnarray}
&& H_{[0,0]} = (2\pi)^{-9 }\ \widetilde{N}_3^2\ 
\int_0^\infty d\alpha\ e^{-\alpha m^2}\cdot
\int d^3q_1 d^3k_1 \int d^3q_2 d^3k_2 
 \cdot\nonumber\\ && \times
 \exp\bigg\{-\frac{1}{6\lambda}\biggl[
 4({\bf q}_1^2+{\bf q}_1\cdot{\bf q}_2+{\bf q}_2^2) 
 +({\bf k}_1^2+{\bf k}_1\cdot{\bf k}_2+{\bf k}_2^2)\biggr]\bigg\}\cdot
 e^{-\gamma {\bf k}^2}. 
\label{eq:T.31}\end{eqnarray}
In Appendix~\ref{sec:mspace2} the details of the three-body matrix elements
of $V_2$ are given, and below we summarize the results.\\

\noindent {\bf c.\ Resume:} We rewrite the basic matrix element integral is (\ref{eq:T.31}) 
as follows:  
\begin{eqnarray}
 H_{[0,0]} &=& {N}_{[0,0]} \int_0^\infty d\alpha\ e^{-\alpha m^2}\cdot
 \int d^3q_1 d^3k_1 \int d^3q_2 d^3k_2 
 \cdot\nonumber\\ && \times
 \exp\bigg\{-\frac{1}{6\lambda}\biggl[
 4({\bf q}_1^2+{\bf q}_1\cdot{\bf q}_2+{\bf q}_2^2) 
 +({\bf k}_1^2+{\bf k}_1\cdot{\bf k}_2+{\bf k}_2^2)\biggr]\bigg\}\cdot
 e^{-\gamma {\bf k}^2} \nonumber\\ &\equiv& 
\int_0^\infty d\alpha\ e^{-\alpha m^2}\ F_{[0,0]}(\alpha,\beta,\gamma), 
\label{eq:T.81}\end{eqnarray}
where $N_{[0,0]}= (2\pi)^{-9 }\widetilde{N}_3^2$, and 
$\beta= 1/6\lambda, \gamma=\alpha+1/\Lambda^2$. Then,
\begin{eqnarray}
&& F_{[0,0]}(\alpha,\beta,\gamma) = N_{[0,0]} \int d^3q_1 d^3k_1 \int d^3q_2 d^3k_2\ 
 e^{-\gamma {\bf k}^2} 
 \cdot\nonumber\\ && \times
 \exp\bigg\{-\frac{1}{6\lambda}\biggl[
 4({\bf q}_1^2+{\bf q}_1\cdot{\bf q}_2+{\bf q}_2^2) 
 +({\bf k}_1^2+{\bf k}_1\cdot{\bf k}_2+{\bf k}_2^2)\biggr]\bigg\} = N_{[0,0]}\cdot
\nonumber\\ && \times
  \left(\frac{\pi^2}{12\beta^2}\right)^{3/2}\ 
  \left(\frac{\pi}{\beta+\gamma}\right)^{3/2} =
 (2\pi)^{-6}\left(\frac{16\pi}{3}\right)^{3/2}\ \Lambda^3 \left(1+\frac{\Lambda^2R_N^2}{18}
 + \frac{\Lambda^2}{m^2}\ \bar{\alpha}\right)^{-3/2}.
\label{eq:T.82}\end{eqnarray}
 with $\bar{\alpha}= \alpha m^2$. For $H_{[m,n]}$ we have
\begin{subequations} \begin{eqnarray}
 F_{[2,0]} &=& \frac{3}{2}(\beta+\gamma)^{-1}\ F_{[0,0]} = \frac{3}{2} \Lambda^2\
 \left(1+\frac{\Lambda^2R_N^2}{18}+\frac{\Lambda^2}{m^2}\ \bar{\alpha}\right)^{-1}\
 F_{[0,0]}, \\
 F_{[4,0]} &=& \frac{15}{4}(\beta+\gamma)^{-2}\ F_{[0,0]} = \frac{15}{4} \Lambda^4\
 \left(1+\frac{\Lambda^2R_N^2}{18}+\frac{\Lambda^2}{m^2}\ \bar{\alpha}\right)^{-2}\
 F_{[0,0]}, \\
 F_{[0,2]} &=& \frac{1}{3\beta}\ F_{[0,0]}= 6 \Lambda^2\ (\Lambda R_N)^{-2}\
 F_{[0,0]}, \\
 F_{[2,2]} &=& \frac{1}{2\beta}(\beta+\gamma)^{-1}\ F_{[0,0]} = 9\Lambda^4\
 (\Lambda R_N)^{-2}\
 \left(1+\frac{\Lambda^2R_N^2}{18}+\frac{\Lambda^2}{m^2}\ \bar{\alpha}\right)^{-1}\
\label{eq:T.83}\end{eqnarray} \end{subequations}
The tensor operator matrix element  has a factor $-k_i k_j$, which gives
\begin{eqnarray}
 F^T_{i,j}([0,0]) &=& -[2(\beta+\gamma)]^{-1}\ F_{[0,0]}\ \delta_{ij} =
 -\frac{1}{2} \Lambda^2 \left(1+\frac{\Lambda^2 R_N^2}{18}+
  \frac{\Lambda^2}{m^2}\ \bar{\alpha}\right)^{-1}\ F_{[0,0]}\ \delta_{ij}.
\label{eq:T.84}\end{eqnarray} 
The quadratic spin-orbit operator matrix element  has a factor $-k_i k_j q_m q_n$, which gives
\begin{eqnarray}
 F^Q_{i,m;j,n}([0,0]) &=& [6\beta(\beta+\gamma)]^{-1}\ F_{[0,0]}\ \delta_{i,j} \delta_{m,n} =
  3\Lambda^4 (\Lambda R_N)^{-2} \left(1+\frac{\Lambda^2 R_N^2}{18}+
  \frac{\Lambda^2}{m^2}\ \bar{\alpha}\right)^{-1}\cdot\nonumber\\ && \times  F_{[0,0]}\ 
\delta_{i,j} \delta_{m,n}.
\label{eq:T.85}\end{eqnarray} 
\noindent {\it For the $G_{[n,m]}$ functions the correspondent $F_{[n,m]}$ are the same as
those above, but with $\bar{\alpha}=0$.}\\

\noindent {\bf d.\ Explicit expressions:}
From Appendix~\ref{sec:mspace2} we obtain for $H_{[0,0]}$, the expression 
\begin{equation}
 H_{[0,0]} = {\cal N}_{[0,0]} \int_0^\infty d\alpha\ e^{-\alpha m^2}\
 \left(\frac{\pi}{\alpha+A}\right)^{3/2}\ ,\ 
 A= \left(1+\frac{\Lambda^2R_N^2}{18}\right)/\Lambda^2,
\label{eq:T.86}\end{equation} 
where ${\cal N}_{[0,0]}= (3\pi^2\lambda^2)^{3/2} N_{[0,0]}$.
The $\alpha$-integral, called $J_1$ (\ref{eq:T.35b}), is worked out in 
Appendix~\ref{sec:mspace2} with the result
\begin{equation}
 H_{[0,0]} = (2\pi\sqrt{\pi})\ {\cal N}_{[0,0]}\ m\ \biggl[
\frac{1}{\sqrt{\pi A m^2}}-e^{A m^2}\ {\it Erfc}\left(\sqrt{A m^2}\right)\biggr].
\label{eq:T.87}\end{equation} 
Also, $G_{[0,0]}= F_{[0,0]}(\alpha=0,\beta,\gamma)= \pi\
{\cal N}_{[0,0]}\ A^{-3/2}$.
For $H_{[2,0]}$ the integral expression is
\begin{equation}
 H_{[2,0]} = (3/2\pi)\ {\cal N}_{[0,0]} \int_0^\infty d\alpha\ e^{-\alpha m^2}\
 \left(\frac{\pi}{\alpha+A}\right)^{5/2},    
\label{eq:T.88}\end{equation} 
which, using the $J_2$-integral (\ref{eq:T.54}), 
\begin{equation}
 H_{[2,0]} = (2\pi\sqrt{\pi})\ {\cal N}_{[0,0]}\ m^3\ \biggl[
 e^{A m^2}\ {\it Erfc}\left(\sqrt{A m^2}\right) +
\frac{1}{2\sqrt{\pi}} (Am^2)^{-3/2}\left(1-2 A m^2\right)\biggr],
\label{eq:T.89}\end{equation} 
with the relation $H_{[2,0]}=G_{[0,0]}-m^2 H_{[0,0]}$ (check!).

\noindent For the presentation of the QQ-potential contributions to 
the nucleon mass it is useful to introduce the dimensionless $B_{[k,l]}$ as follows
\begin{eqnarray}
&& H_{[0,0]} = m\ B_{[0,0]},\ H_{[2,0]} = m^3\ B_{[2,0]}, 
   H_{[0,2]} = m^3\ B_{[0,2]},\ H_{[2,2]} = m^5\ B_{[2,2]}.  
\label{eq:T.64}\end{eqnarray}
Similarly, for the Pomeron we define 
\begin{eqnarray}
&& G_{[0,0]} = \frac{\Lambda^3}{{\cal M}^2}\ D_{[0,0]},\ 
 G_{[2,0]} = \frac{\Lambda^5}{{\cal M}^2}m^3\ D_{[2,0]}, 
   G_{[0,2]} = \frac{\Lambda^5}{{\cal M}^2}\ D_{[0,2]}.  
\label{eq:T.65}\end{eqnarray}
\noindent {\bf Remark:} The tensor-integral gives a $\delta_{ij}$ factor.
Contraction with 
$\sigma_{1,i}\sigma_{2,j}-(1/3)(\bm{\sigma}_1\cdot\bm{\sigma}_2)\ \delta_{i,j}$
gives zero. Therefore, for s-wave quarks the tensor-potential gives no
contribution, which is logical.

\begin{flushleft}
\rule{16cm}{0.5mm}
\end{flushleft}
\section{Kinetic Energy Three-quark System}
\label{sec:KINEN}    
\noindent {\bf 1.\ Quark-contribution}:
For equal quark masses $m_i=m_Q\ (i=1,2,3)$ the non-relativistic kinetic energy 
operator is \cite{kin-energy}
\begin{equation}
  T_{op}= \left[{\bf p}_1^2+{\bf p}_2^2+{\bf p}_3^2\right]/(2m_Q)
   = \frac{1}{2m_Q}\left[{\bf p}_\lambda^2+{\bf p}_\rho^2\right].
\label{eq:KIN1}\end{equation}
Then, 
\begin{eqnarray}
 T &=& \langle \Psi_3| T_{op}|\Psi_3\rangle = 
 \Pi_{i=1}^3 \biggl[\int \frac{d^3p_i}{(2\pi)^3}\biggr]\ \Psi^*({\bf p}_i)\
  \left[{\bf p}_1^2+{\bf p}_2^2+{\bf p}_3^2\right]/(2m_Q)
  \Psi({\bf p}_i) \nonumber\\
 &=&  \widetilde{N}_3^2\
 \int\int \frac{d^3p_\lambda d^3p_\rho}{(2\pi)^6}\ 
 \exp\biggl[-\frac{1}{3\lambda}\left({\bf p}_\lambda^2+{\bf p}_\rho^2\right)
 \biggr]\ \left({\bf p}_\lambda^2+{\bf p}_\rho^2\right)/(2m_Q)
  = \widetilde{N}_3^2
 \cdot\nonumber\\ &\times&
 (2\pi)^{-6} (3\pi\lambda)^{3/2}\ \frac{3}{2\pi} (3\pi\lambda)^{5/2}/m_Q
  = 9\lambda/(2m_Q) = (27/2) (m_Q R_N)^{-2}\ m_Q.
\label{eq:KIN2}\end{eqnarray}
Here is used $\widetilde{N}_3 = (4\pi/3\lambda)^{3/2}$ and 
$\lambda= 3R_N^{-2}$. With $R_N=1$ fm and $m_Q$=312.75 MeV one gets 
$\langle T \rangle \approx (9/2) m_Q$ which implies per quark a 
kinetic energy $\approx 470$ MeV. Clearly the quarks move relativistically, and
the non-relativistic formula is wrong.  \\

\noindent {\bf 2.\ de Broglie estimation}: 
An alternative derivation is as follows: Using the de Broglie
relation between momentum and wave-length $p=h/\lambda$, one has
for each quark 
\begin{equation}
 pc \approx  2\pi\ \frac{\hbar c}{2R_N} \rightarrow
 \frac{{\bf p}^2}{2m_Q} \approx 
 \frac{\pi^2}{2}\frac{(\hbar c)^2}{m_Q R_N^2}= 
 \frac{\pi^2}{2}\frac{(\hbar c)^2}{(m_Q R_N)^2}\ m_Q.
\label{eq:KIN3}\end{equation}
With $\hbar c = 197.325$ MeVfm we obtain for $R_N=1$ fm
the kinetic energy per quark $1.6 m_Q = 500$ MeV, which agrees roughly with the more exact 
result in (\ref{eq:KIN2}).\\
\noindent {\bf 3.\ Relativistic Energy Expectation-value}:   
First we derive a gaussian-type of presentation for the relativistic energy. Using     
integral representations, see \cite{TAR91}, we derive for the relativistic
energy a gaussian-type of expression
\begin{eqnarray}
E({\bf p}) &=& \sqrt{{\bf p}^2+m^2} =\frac{{\bf p}^2+m^2}{\sqrt{{\bf p}^2+m^2}} =
 \frac{2}{\pi}\int_0^\infty d\lambda\ \frac{ {\bf p}^2+m^2}{{\bf p}^2+m^2+\lambda^2}
\nonumber\\ &=& ({\bf p}^2+m^2)\cdot \frac{2}{\pi}\int_0^\infty d\lambda\ \int_0^\infty d\alpha\
\exp\bigl[-\alpha( {\bf p}^2+m^2+\lambda^2)\bigr]
\nonumber\\ &=& 
({\bf p}^2+m^2)\cdot\frac{1}{\sqrt{\pi}}\int_0^\infty \frac{d\alpha}{\sqrt{\alpha}}\
 e^{-\alpha m^2}\ e^{-\alpha {\bf p}^2}.
\label{eq:KIN11}\end{eqnarray}
Then, the expression for the relativistic kinetic energy of the three-quark system becomes
\begin{eqnarray}
 E_T &=& \bigl\langle\sum_{i=1}^3 \sqrt{{\bf p}_i^2+m^2}\bigr\rangle = \frac{1}{\sqrt{\pi}} \int_0^\infty 
 \frac{d\alpha}{\sqrt{\alpha}} e^{-\alpha m^2}\ 
\bigl\langle\sum_{i=1}^3 ({\bf p}_i^2+m^2)\ e^{-\alpha {\bf p}_i^2}\bigr\rangle.
\label{eq:KIN12}\end{eqnarray}
The evaluation of the expectation value in (\ref{eq:KIN12}) involves only gaussian integrals and is
straightforward. We remind the formulas, with ${\bf P}=0$,
\begin{eqnarray*}
 {\bf p}_1^2 &=& \frac{1}{6} {\bf p}_\lambda^2+\frac{1}{2} {\bf p}_\rho^2
               +\frac{1}{\sqrt{3}} {\bf p}_\lambda\cdot{\bf p}_\rho, \\
 {\bf p}_2^2 &=& \frac{1}{6} {\bf p}_\lambda^2+\frac{1}{2} {\bf p}_\rho^2
               -\frac{1}{\sqrt{3}} {\bf p}_\lambda\cdot{\bf p}_\rho, \\
 {\bf p}_3^2 &=& ({\bf p}_1+{\bf p}_2)^2 = \frac{2}{3} {\bf p}_\lambda^2.
\end{eqnarray*}
\noindent {\bf (a)} For quark 1 the expectation of the kinetic energy is given by
\begin{eqnarray}
 \bigl\langle E_T \bigr\rangle_1 &=& \biggl\langle \Psi\bigg| \frac{1}{\sqrt{\pi}}
\int_0^\infty \frac{d\alpha}{\sqrt{\alpha}}\ e^{-\alpha m^2}\
 ({\bf p}_1^2+m^2)\ e^{-\alpha {\bf p}_1^2}\big|\Psi\biggr\rangle 
 = \frac{1}{\sqrt{\pi}} \int_0^\infty \frac{d\alpha}{\sqrt{\alpha}}\ e^{-\alpha m^2} 
 \cdot\nonumber\\ && \times
\bigg\{\ \widetilde{N}_3^2 \int\frac{d^3p_\lambda d^3p_\rho}{(2\pi)^6}\ 
 \exp\left[-\frac{1}{3\lambda}({\bf p}_\lambda^2+{\bf p}_\rho^2)\right]\  
 ({\bf p}_1^2+m^2)\ e^{-\alpha {\bf p}_1^2} \biggr\}              
\label{eq:KIN13}\end{eqnarray}
The momentum integral $I \equiv \bigl\{ \ldots \bigr\}$ is
\begin{eqnarray}
I &=& \biggl(-\frac{d}{d\alpha}+m^2\biggr)\
 \widetilde{N}_3^2 \int\frac{d^3p_\lambda d^3p_\rho}{(2\pi)^6}\ 
 \exp\biggl[-\biggl( \frac{1}{3\lambda}\left[ {\bf p}_\lambda^2+{\bf p}_\rho^2 \right]\  
 \nonumber\\ &&
 +\alpha \left[\frac{1}{6}{\bf p}_\lambda^2+\frac{1}{2}{\bf p}_\rho^2 +
\frac{1}{\sqrt{3}} {\bf p}_\lambda\cdot{\bf p}_\rho\right]\biggr)\biggr]
 \equiv \left(-\frac{d}{d\alpha}+m^2\right)\ J
\label{eq:KIN14}\end{eqnarray}
where 
\begin{subequations} \label{eq:KIN15}
\begin{eqnarray}
J &=& \widetilde{N}_3^2 \int\frac{d^3p_\lambda d^3p_\rho}{(2\pi)^6}\ 
 \exp\bigg[-\bigl\{ a {\bf p}_\lambda^2+c {\bf p}_\lambda\cdot{\bf p}_\rho + b {\bf p}_\rho^2
 \bigr\}\bigg],\ \rm{where} \\
 a&=& \frac{1}{3\lambda}+\frac{\alpha}{6}\ ,\
 b =  \frac{1}{3\lambda}+\frac{\alpha}{2}\ ,\ c= \frac{\alpha}{\sqrt{3}}.
\end{eqnarray}
\end{subequations}
From the integrals in Eqn.~(\ref{eq:T.d.1}) we have
\begin{eqnarray}
 J(a,b,c) &=& \widetilde{N}_3^2 (2\pi)^{-6} \left(\frac{4\pi^2}{4ab-c^2}\right)^{3/2}=
 (1+2\lambda \alpha)^{-3/2},\ -\frac{d}{d\alpha} J(a,b,c)=
 3\lambda \left(1+2\lambda\alpha\right)^{-5/2}.
\label{eq:KIN16}\end{eqnarray}
The integral in (\ref{eq:KIN14}) becomes
\begin{eqnarray}
I(\alpha,\lambda) &=& \bigl(1+2\alpha\lambda\bigr)^{-3/2}
\biggl[m^2+3\lambda\bigl(1+2\alpha\lambda\biggr)^{-1}\biggr] \nonumber\\ &=&
m^2 \left(1+6\frac{\alpha m^2}{m^2R_N^2}\right)^{-3/2}\biggl[1+9(mR_N)^{-2}\
\left(1+6\frac{\alpha m^2}{m^2R_N^2}\right)^{-1}\biggr].
\label{eq:KIN17}\end{eqnarray}
Because of the symmetry, the total kinetic energy is three times that for quark 1, so
\begin{eqnarray}
\bigg\langle E_T \biggr\rangle &=& \frac{3m^2}{\sqrt{\pi}} \int_0^{\infty}
\frac{d\alpha}{\sqrt{\alpha}}\ e^{-\alpha m^2} \biggl[
\left(1+6\frac{\alpha m^2}{m^2R_N^2}\right)^{-3/2}
+9(mR_N)^{-2}\ \left(1+6\frac{\alpha m^2}{m^2R_N^2}\right)^{-5/2}\biggr] 
\nonumber\\ &=& \frac{6m}{\sqrt{\pi}}\int_0^\infty dy\ e^{-y^2} \biggl[
\biggl(1+\frac{6y^2}{m^2R_N^2}\biggr)^{-3/2}
+9 (mR_N)^{-2} \biggl(1+\frac{6y^2}{m^2R_N^2}\biggr)^{-5/2}\biggr].
\label{eq:KIN18}\end{eqnarray}
We remark that 
\begin{eqnarray*}
 \lim_{R_N \rightarrow \infty} E_T = \frac{3m}{\sqrt{\pi}} \int_0^\infty 
\frac{d\alpha}{\sqrt{\alpha}}\ e^{-\alpha m^2} = 
 \frac{6m}{\sqrt{\pi}} \int_0^\infty dy\ e^{-y^2}= 3m.
\end{eqnarray*}
In a concise form we write
\begin{subequations}\label{eq:KIN20}
\begin{eqnarray}
 E_T(mR_N) &=& \frac{1}{\sqrt{6}} (m R_N)^3\
 \biggl[ K_3+\frac{3}{2} K_5 \biggr]\ m,\ T_{rel}=E_T-3m,\\
 K_n(m R_N) &=& \frac{1}{\sqrt{\pi}}
 \int_0^\infty dy\ e^{-y^2}\ \left(y^2+d^2\right)^{-n/2}\ {\rm with}\ 
 d=m R_N/\sqrt{6}.
\end{eqnarray}
\end{subequations}
In Table~\ref{tab:FK3} the numerical results are shown for the kinetic energies
 (K.E.'s) as a function of the radius $R_N$.
\begin{table}
 \caption{Kinetic energy $E_T$ as a function of $R_N$. 
 Listed are the integrals $K_{3,5}$, the non-relativistic K.E. $T_{NR}$ and 
the relativistic K.E. $T_R$ per quark.}
\begin{center}
\begin{ruledtabular}
\begin{tabular}{c|c|cc|cc|c} 
  $R_N$[fm] & $\bigl\langle p \bigr\rangle$ & $K_3$  & $K_5$ & $T_{NR}(1Q)$ & $T_R(1Q)$ & $T_R(3Q)$ \\
 \hline
 0.50  &  519.4 & 8.37 &57.66 & 2241.0 & 820.8 & 2462.5 \\
 0.60  &  432.8 & 5.62 &27.45 & 1556.2 & 653.6 & 1960.9 \\
 0.70  &  371.0 & 3.99 &14.60 & 1143.4 & 536.1 & 1608.3 \\
 0.80  &  324.6 & 2.94 &8.40  &  875.4 & 448.4 & 1345.2 \\
 0.90  &  288.6 & 2.24 & 5.14 &  691.7 & 380,8 & 1142.5 \\
 1.00  &  259.7 & 1.75 & 3.30 &  560.2 & 327.6 &  982.9 \\
 1.20  &  216.4 & 1.13 & 1.52 &  389.1 & 250.2 &  750.7 \\
 1.40  &  185.5 & 0.77 & 0.78 &  285.8 & 197.6 &  592.8 \\
 1.60  &  162.3 & 0.55 & 0.44 &  218.8 & 160.0 &  480.0 \\
 1.80  &  144.3 & 0.41 & 0.26 &  172.9 & 132.2 &  396.6 \\
 2.00  &  130.0 & 0.31 &0.163 &  140.1 & 111.0 &  333.0 \\
\end{tabular}
\end{ruledtabular}
\end{center}
\label{tab:FK3} 
\end{table}


\noindent {\bf 4.\ Average quark momentum:} The expectation value for ${\bf p}_1^2$ is
given by
\begin{eqnarray}
\bigl\langle {\bf p}_1^2 \bigr\rangle &=&
 \widetilde{N}_3^2 \int\frac{d^3p_\lambda d^3p_\rho}{(2\pi)^6}\ {\bf p}_1^2\
 \exp\left[-\frac{1}{3\lambda}({\bf p}_\lambda^2+{\bf p}_\rho^2)\right] 
\nonumber\\ &=&
 \widetilde{N}_3^2 \int\frac{d^3p_1 d^3p_2}{(2\pi)^6}\ {\bf p}_1^2\
 \exp\bigg[-\left(a {\bf p}_1^2+b {\bf p}_2^2+c {\bf p}_1\cdot{\bf p}_2\right)\biggr] 
\nonumber\\ &=&
  \widetilde{N}_3^2\ (2\pi)^{-6}\ \frac{3b}{2\pi^2}\ 
 \left(\frac{4\pi^2}{4ab-c^2}\right)^{5/2}= \sqrt{3}\ R_N^{-2}.
\label{eq:KIN23}\end{eqnarray}
So, the average quark momentum is $\langle p \rangle = 3^{1/4}\ R_N^{-1}$. 
The average K.E. $\langle T(1Q)\rangle = (\langle p \rangle)^2/2m_Q$ 
matches with $T_{NR}(1Q)$ in Table~\ref{tab:FK3}. The average relativistic
energy is $E_{av}= \sqrt{p_{av}^2+m_Q^2}$, Defining the average quark mass by
$m_{av}=(E_{av}+m_Q)/2$ gives for $R_N=1$ fm a value $m_{av}=469=1.5 m_Q$ MeV.
Since the quarks are relativistic it is better in the QQ-potentials to
make the replacements $1/(4m_q^2) \rightarrow 1/(4m_{av}^2)$, which gives for the
the tensor, spin-orbit a reduction by a factor $\approx 6$, and for the 
quadratic spin-orbit a reduction by $\approx 39$.  
This makes these potential more reasonable, without having to do a 
fully relativistic calculation. In Appendix~\ref{app:relfacts} a more exact,
but rather complicated, way of including relativistic effects is described. \\

\noindent {\bf 5.\ CM subtraction}:   
Considering the 3-quark system residing in a central harmonic confining potential 
we subtract the zero-mode energy from the kinetic energy (?!). 
With
\begin{equation}
 V_{conf}= C_2\ r_N^2 = \frac{1}{2} m_N \omega_{CM}^2\ {\bf r}_N^2
\label{eq:KIN31}\end{equation}
one has 
\begin{equation}
 E_{CM}= \frac{3}{2} \hbar \ \omega_{CM},\ {\rm with}\ 
\omega_{CM}=\sqrt{\frac{2C_2}{3m_Q}}.      
\label{eq:KIN32}\end{equation}
Using $C_2= 315$ MeV fm$^{-2}$ and $m_Q= M_p/3$ one obtains $E_{CM} \approx 231$ MeV.

\begin{flushleft}
\rule{16cm}{0.5mm}
\end{flushleft}
\vspace*{25mm}

\begin{figure}[hhhbt]
\begin{center}
 \resizebox{21.25cm}{!}
 {\includegraphics[1.0in,9.0in][9in,10in]{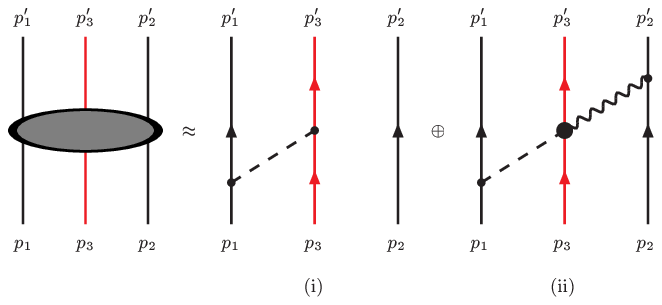}}
\end{center}
\caption{\sl Three-particle amplitude (a) and the Born-Feynman 
 graphs type (i) and (ii)} 
\label{fig.tbf0} \end{figure}
\begin{flushleft}
\rule{16cm}{0.5mm}
\end{flushleft}
\section{Nucleon Mass from Two-body forces}
\label{sec:EN}    
In this note we calculate the contribution to the nucleon-mass from
the two-body QQ-potentials, see graph (i) in Fig.~\ref{fig.tbf0}.
The contributions from the three-body QQQ-potentials, see graph (ii) 
in Fig.~\ref{fig.tbf0}, will be derived in another note \cite{THAR.3bf}.\\
Here, explicit formulas are given for the contributions to the nucleon-energy,
i.e. nucleon mass, from the two-body QQ-potentials. 
The formulas below are based on the potentials in section~\ref{app:OBE}.
Below, the contributions from the local, non-local, and "additional" potentials are
listed separately. ("Additional" = contributions to potentials due to the extra
meson-quark-quark vertices, which have been introduced in order to match with
the potentials at the baryon-level.)

\noindent {\it 
Below we compute the contributions from the 
potentials for the graphs (a)-(c) of Fig.~\ref{fig.twbf2-born}
to the expectation values 
 $E_N = \langle \Psi_{N}|V_2|\Psi_{N}\rangle$
for the different OBE-potentials. The $V_{13;2}$ and $V_{23;1}$ 
give identical results in the case of the nucleon. Therefore, 
we multiply the results for $V_{12;3}$ by a factor 3
to obtain the total answer.}\\
We remark that terms proportional to ${\bf q}_i$ and/or ${\bf k}_i$
vanish due to the integrations in the matrix elements 
$\langle \Psi_{3Q}| V_2 |\Psi_{3Q}\rangle$, which implies no contributions 
from the spin-orbit potentials. This is logical because of the absence of
P-waves etc. in the quark wave functions.

\begin{flushleft}
\rule{16cm}{0.5mm}
\end{flushleft}
{\bf 1. Nucleon:}
The isospin-spin operators that occur in $E_N$ are $\bm{\tau}_1\cdot\bm{\tau}_2$, 
$\bm{\sigma}_1\cdot\bm{\sigma}_2$, and the product
 $\bm{\tau}_1\cdot\bm{\tau}_2\ \bm{\sigma}_1\cdot\bm{\sigma}_2$. 
The symmetrized spin-isospin part of the nucleon state is
\begin{eqnarray}
 \Psi_N &=& \frac{1}{\sqrt{2}}\left(\vphantom{\frac{A}{A}} 
 \phi_{M,S} \chi_{M,S}+\phi_{M,A} \chi_{M,A}\right).
\label{eq:iso.1c}\end{eqnarray}
In Appendix~\ref{app:iso} 
the nucleon matrix elements of the spin-isospin operators are derived, with the result, 
see (\ref{eq:spin.16}):  
\begin{subequations} \label{eq:T.c}
\begin{eqnarray}
&& \left(\Psi_N| \bm{\tau}_1\cdot\bm{\tau}_2 |\Psi_N\right) =  
   \left(\Psi_N| \bm{\tau}_1\cdot\bm{\tau}_3 |\Psi_N\right) =  
   \left(\Psi_N| \bm{\tau}_2\cdot\bm{\tau}_3 |\Psi_N\right) = -1, \\
&& \left(\Psi_N| \bm{\sigma}_1\cdot\bm{\sigma}_2 |\Psi_N\right) =  
   \left(\Psi_N| \bm{\sigma}_1\cdot\bm{\sigma}_3 |\Psi_N\right) =  
   \left(\Psi_N| \bm{\sigma}_2\cdot\bm{\sigma}_3 |\Psi_N\right) = -1, \\
&& \left(\Psi_N| \bm{\tau}_1\cdot\bm{\tau}_2\ \bm{\sigma}_1\cdot\bm{\sigma}_2|\Psi_N\right) =  
   \left(\Psi_N| \bm{\tau}_1\cdot\bm{\tau}_3\ \bm{\sigma}_1\cdot\bm{\sigma}_3|\Psi_N\right) =  
\nonumber\\ &&
   \left(\Psi_N| \bm{\tau}_2\cdot\bm{\tau}_3\ \bm{\sigma}_2\cdot\bm{\sigma}_3|\Psi_N\right) = +5.  
\end{eqnarray}
\end{subequations}
The antisymmetry of the full nucleon state is provided by the color part of the
wave function being the singlet $SU(3)_c$-irrep.

\noindent {\it Here, including a factor 3 takes into account of the similar
contributions from $V_{13;2}$ and $V_{23;1}$.}\\

\noindent {\bf 2.} $\bm{\Lambda}$: For the $\Lambda$ the spin-isospin matrix elements 
 are, see (\ref{eq:spin.16}), 
\begin{subequations} \label{eq:T.d}
\begin{eqnarray}
&& \left(\Psi_\Lambda| \bm{\tau}_1\cdot\bm{\tau}_2 |\Psi_\Lambda\right) =
   \left(\Psi_\Lambda| \bm{\tau}_1\cdot\bm{\tau}_3 |\Psi_\Lambda\right) =  
   \left(\Psi_\Lambda| \bm{\tau}_2\cdot\bm{\tau}_3 |\Psi_\Lambda\right) = -1, \\
&& \left(\Psi_\Lambda| \bm{\sigma}_1\cdot\bm{\sigma}_2 |\Psi_\Lambda\right) =
   \left(\Psi_\Lambda| \bm{\sigma}_1\cdot\bm{\sigma}_3 |\Psi_\Lambda\right) =  
   \left(\Psi_\Lambda| \bm{\sigma}_2\cdot\bm{\sigma}_3 |\Psi_\Lambda\right) = -1, \\
&& \left(\Psi_\Lambda| \bm{\tau}_1\cdot\bm{\tau}_2\
    \bm{\sigma}_1\cdot\bm{\sigma}_2|\Psi_\Lambda\right) =
   \left(\Psi_\Lambda| \bm{\tau}_1\cdot\bm{\tau}_3\  
    \bm{\sigma}_1\cdot\bm{\sigma}_3|\Psi_\Lambda\right) =
\nonumber\\ &&
   \left(\Psi_\Lambda| \bm{\tau}_2\cdot\bm{\tau}_3\ 
    \bm{\sigma}_2\cdot\bm{\sigma}_3|\Psi_\Lambda\right) = +2.  
\end{eqnarray}
\end{subequations}
The total three-body matrix element has three terms
 $\langle V_2 \rangle = \langle V_{12;3}\rangle + \langle V_{13;2}\rangle
 + \langle V_{23;1}\rangle$, 
where $V_{12;3} = V_{UD}(12), V_{13;2} = V_{US}(13), V_{23;1}= V_{DS}(23)$.\\

\noindent {\bf 3.} $\bm{\Sigma^+}$: The wave functions for $\Psi_{\Sigma^+}$ is
\begin{eqnarray*}
&& \phi_{M,S}= \frac{1}{\sqrt{6}}\biggl[(us+su)u-2uus\biggr],\ 
   \phi_{M,A}= \frac{1}{\sqrt{2}}(us-su)u,  
\end{eqnarray*}
and for for the spin wave functions $\chi_{M,S}$ and $\chi_{M,A}$ similarly as
for the proton P. This gives, see (\ref{eq:spin.17}),
\begin{subequations} \label{eq:T.e}
\begin{eqnarray}
&& \left(\Psi_\Sigma| \bm{\tau}_1\cdot\bm{\tau}_2 |\Psi_\Sigma\right) = -\frac{1}{6}\ ,\
   \left(\Psi_\Sigma| \bm{\tau}_1\cdot\bm{\tau}_3 |\Psi_\Sigma\right) = 
   \left(\Psi_\Sigma| \bm{\tau}_2\cdot\bm{\tau}_3 |\Psi_\Sigma\right) = +\frac{2}{3}, \\
&& \left(\Psi_\Sigma| \bm{\sigma}_1\cdot\bm{\sigma}_2 |\Psi_\Sigma\right) =  
   \left(\Psi_\Sigma| \bm{\sigma}_1\cdot\bm{\sigma}_3 |\Psi_\Sigma\right) =  
   \left(\Psi_\Sigma| \bm{\sigma}_2\cdot\bm{\sigma}_3 |\Psi_\Sigma\right) = -1, \\
&& \left(\Psi_\Sigma| \bm{\tau}_1\cdot\bm{\tau}_2\ 
    \bm{\sigma}_1\cdot\bm{\sigma}_2|\Psi_\Sigma\right) = -\frac{1}{6}\ ,\
   \left(\Psi_\Sigma| \bm{\tau}_1\cdot\bm{\tau}_3\ 
    \bm{\sigma}_1\cdot\bm{\sigma}_3|\Psi_\Sigma\right) = +\frac{1}{3}\ ,\
\nonumber\\ &&
   \left(\Psi_\Sigma| \bm{\tau}_2\cdot\bm{\tau}_3\ 
    \bm{\sigma}_2\cdot\bm{\sigma}_3|\Psi_\Sigma\right) = +\frac{1}{3}.
\end{eqnarray}
\end{subequations}
The total three-body matrix element has three terms
$\langle V_2 \rangle = \langle V_{12;3}\rangle + \langle V_{13;2}\rangle
 + \langle V_{23;1}\rangle$, 
where $V_{12;3} = V_UD(12), V_{13;2} = V_US(13), V_{23;1}= V_{DS}(23)$.

\noindent {\bf 4.} $\bm{\Xi^0}$: 
In this case the matrix elements are, see (\ref{eq:spin.18}),
\begin{subequations} \label{eq:T.f}
\begin{eqnarray}
&& \left(\Psi_\Xi| \bm{\tau}_1\cdot\bm{\tau}_2 |\Psi_\Xi\right) = 
   \left(\Psi_\Xi| \bm{\tau}_1\cdot\bm{\tau}_3 |\Psi_\Xi\right) = 
   \left(\Psi_\Xi| \bm{\tau}_2\cdot\bm{\tau}_3 |\Psi_\Xi\right) = 0, \\
&& \left(\Psi_\Xi| \bm{\sigma}_1\cdot\bm{\sigma}_2 |\Psi_\Xi\right) =  
   \left(\Psi_\Xi| \bm{\sigma}_1\cdot\bm{\sigma}_3 |\Psi_\Xi\right) =  
   \left(\Psi_\Xi| \bm{\sigma}_2\cdot\bm{\sigma}_3 |\Psi_\Xi\right) = -1, \\
&& \left(\Psi_\Xi| \bm{\tau}_1\cdot\bm{\tau}_2\ 
    \bm{\sigma}_1\cdot\bm{\sigma}_2|\Psi_\Xi\right) = 
   \left(\Psi_\Xi| \bm{\tau}_1\cdot\bm{\tau}_3\ 
    \bm{\sigma}_1\cdot\bm{\sigma}_3|\Psi_\Xi\right) = 
\nonumber\\ &&
   \left(\Psi_\Xi| \bm{\tau}_2\cdot\bm{\tau}_3\ 
    \bm{\sigma}_2\cdot\bm{\sigma}_3|\Psi_\Xi\right) = 0.            
\end{eqnarray}
\end{subequations}

\noindent {\bf 5.} $\bm{\Delta}_{33}^{++}$: 
In this case the matrix elements are, see (\ref{eq:spin.22}),
\begin{subequations} \label{eq:T.g}
\begin{eqnarray}
&& \left(\Psi_\Delta| \bm{\tau}_1\cdot\bm{\tau}_2 |\Psi_\Delta\right) = 
   \left(\Psi_\Delta| \bm{\tau}_1\cdot\bm{\tau}_3 |\Psi_\Delta\right) = 
   \left(\Psi_\Delta| \bm{\tau}_2\cdot\bm{\tau}_3 |\Psi_\Delta\right) = +1, \\
&& \left(\Psi_\Delta| \bm{\sigma}_1\cdot\bm{\sigma}_2 |\Psi_\Delta\right) =  
   \left(\Psi_\Delta| \bm{\sigma}_1\cdot\bm{\sigma}_3 |\Psi_\Delta\right) =  
   \left(\Psi_\Delta| \bm{\sigma}_2\cdot\bm{\sigma}_3 |\Psi_\Delta\right) = +1, \\
&& \left(\Psi_\Delta| \bm{\tau}_1\cdot\bm{\tau}_2\ 
    \bm{\sigma}_1\cdot\bm{\sigma}_2|\Psi_\Delta\right) = 
   \left(\Psi_\Delta| \bm{\tau}_1\cdot\bm{\tau}_3\ 
    \bm{\sigma}_1\cdot\bm{\sigma}_3|\Psi_\Delta\right) = 
\nonumber\\ &&
   \left(\Psi_\Delta| \bm{\tau}_2\cdot\bm{\tau}_3\ 
    \bm{\sigma}_2\cdot\bm{\sigma}_3|\Psi_\Delta\right) = 1.            
\end{eqnarray}
\end{subequations}
\begin{flushleft}
\rule{16cm}{0.5mm}
\end{flushleft}

\subsection{Mass from Local Two-body forces}
\label{app:EN.a}  
Contributions to $E_N$ from local QQ-potentials, given in subsection \ref{app:OBE.b}.\\

\noindent {\bf (a)} Pseudoscalar-meson exchange $J^{PC}=0^{-+}$:
\begin{eqnarray}
 E^{(P)}_{12;3} &=& -g^p_{13}g^p_{24}\ \left(\frac{m_P^3}{12M_yM_n}\right)\ 
 \biggl[B_{[2,0]}+3B_{[0,0]}(T)\biggr] (\bm{\sigma}_1\cdot\bm{\sigma}_2).
\label{app.ennps}\end{eqnarray}
\noindent {\bf (b)} Vector-meson exchange $J^{PC}=1^{--}$:
\begin{eqnarray}
&& E^{(V)}_{12;3} =  g^v_{13} g^v_{24}\ m_V\cdot\nonumber\\ && \times
\biggl( B_{[0,0]}-\frac{m_V^2}{4M_yM_n}\biggl[ 2 + 
           \left(\kappa^v_{24}\frac{M_y}{\cal M} 
      +\kappa^v_{13}\frac{M_n}{\cal M}\right)\biggr]\ B_{[2,0]}
  +\kappa^v_{13}\kappa^v_{24} \frac{m_V^4}{16{\cal M}^{2}M_yM_n}\ B_{[4,0]}                    
\nonumber\\ && 
    - \frac{m_V^2}{6M_yM_n} \biggl\{ (1+\kappa^v_{13}\frac{M_y}{{\cal M}})
           (1+\kappa^v_{24}\frac{M_n}{{\cal M}})\ B_{[2,0]}
          -\kappa^v_{13}\kappa^v_{24} \frac{m_V^2}{8{\cal M}^{2}} B_{[4,0]}\biggr\}
 (\bm{\sigma}_1\cdot\bm{\sigma}_2)
\nonumber\\ && +\frac{m_V^2}{4M_yM_n}
 \biggl\{ (1+\kappa^v_{13}\frac{M_y}{{\cal M}}) (1+\kappa^v_{24}\frac{M_n}{{\cal M}}) 
 B_{[0,0]}(T)
 -\kappa^v_{13}\kappa^v_{24}\frac{m_V^2}{8{\cal M}^{2}} B_{[2,0]}(T)
\biggr\})           
 (\bm{\sigma}_1\cdot\bm{\sigma}_2)
\nonumber\\ && 
 -\frac{m_V^4}{16M_y^{2}M_n^{2}}
\biggl\{ 1+4(\kappa^v_{24}+\kappa^v_{13}) \frac{\sqrt{M_yM_n}}{{\cal M}}  
   +8\kappa^v_{13}\kappa^v_{24}\frac{M_yM_n}{{\cal M}^{2}}\biggr\} B_{[0,0]}(Q)
 (\bm{\sigma}_1\cdot\bm{\sigma}_2).
\label{app.ennvc1}\end{eqnarray}
\noindent {\bf (c)} Scalar-meson exchange $J^{PC}=0^{++}$:
\begin{eqnarray}
 E^{(S)}_{12;3} &=& -g^s_{13}g^s_{24} m_S \biggl(
      \left\{ B_{[0,0]} + \frac{m_S^2}{4M_yM_n} B_{[2,0]}\right\}
      -\frac{m_S^4}{16M_y^2M_n^2} B_{[0,0]} (\bm{\sigma}_1\cdot\bm{\sigma}_2) 
      \biggr).
\label{app.ennsc1}\end{eqnarray}
\noindent {\bf (d)} Axial-vector-meson exchange $J^{PC}=1^{++}$:
\begin{eqnarray}
&& E^{(A)}_{12;3} = -g^a_{13}g^a_{24}\ m_A\cdot\nonumber\\
 && \times  \biggl( \bigg\{
    B_{[0,0]}-\frac{m_A^2}{6M_yM_n}\left[ 4 +\left(\kappa_{24}^a \frac{M_n}{{\cal M}}
    +\kappa_{13}^a\frac{M_y}{{\cal M}}\right)\right]\ B_{[2,0]} 
    +\kappa_{13}^a \kappa_{24}^a \frac{m_A^4}{12{\cal M}^2M_yM_n} B_{[4,0]}\biggr\}
      \nonumber\\ && +\biggl\{ 1-2 \left(\kappa_{24}^a\frac{M_n}{{\cal M}}
         +\kappa_{13}^a \frac{M_y}{{\cal M}}\right) B_{[0,0]}(T)
         +\kappa_{13}^a \kappa_{24}^a \frac{m_A^4}{4{\cal M}^2M_yM_n} B_{[2,0]}(T)
         \biggr\}  \nonumber\\ &&     
     +\left[\frac{2m_A^2}{M_yM_n}\right]\ B'_5\biggr)\ 
         (\bm{\sigma}_1\cdot\bm{\sigma}_2).
\label{app.ennax1}\end{eqnarray}
\noindent {\bf (e)} Axial-vector-meson exchange $J^{PC}=1^{+-}$:
\begin{eqnarray}
 E^{(B)}_{12;3} &=& + 
       f^B_{13}f^B_{24}\ \frac{(M_n+M_y)^2}{m_B^2} \frac{m_B^3}{12M_yM_n} \biggl(
       \biggl[ B_{[0,0]}- \frac{m_B^2}{4M_yM_n}B_{[2,0]})\biggr]
       \nonumber\\ && + 3    
       \biggl[ B_{[0,0]}(T)-\frac{m_B^2}{4M_yM_n} B_{[2,0]}(T)\biggr] 
       \biggr) (\bm{\sigma}_1\cdot\bm{\sigma}_2).
\label{app.ennax2}\end{eqnarray}
\noindent {\bf (f)} Diffractive exchange $J^{PC}=0^{++}$:
\begin{eqnarray}
 E^{(D)}_{12;3} &=& +g^d_{13}g^d_{24} \left(\frac{m_P^3}{\Lambda^2}\right) \biggl(
      \left\{ D_{[0,0]} + \frac{m_P^2}{4M_yM_n} D_{[2,0]}\right\}
      -\frac{m_P^4}{16M_y^2M_n^2} D_{[0,0]} (\bm{\sigma}_1\cdot\bm{\sigma}_2) 
      \biggr).
\label{app.diffr3}\end{eqnarray}
\noindent {\bf (g)} Gluon exchange $J^{PC}=1^{--}$:
\begin{eqnarray}
&& E^{(G)}_{12;3} =  g_{QCD}^2\ m_G\cdot\nonumber\\ && \times
\biggl( B_{[0,0]}-\frac{m_G^2}{4M_yM_n}\biggl[ 2 + \kappa_G
           \left(\frac{M_y}{\cal M} +\frac{M_n}{\cal M}\right)\biggr]\ B_{[2,0]}
  +\kappa_G^2 \frac{m_G^4}{16{\cal M}^{2}M_yM_n}\ B_{[4,0]}                    
\nonumber\\ && 
    - \frac{m_G^2}{6M_yM_n} \biggl\{ (1+\kappa_G\frac{M_y}{{\cal M}})
           (1+\kappa_G\frac{M_n}{{\cal M}})\ B_{[2,0]}
          -\kappa_G^2 \frac{m_G^2}{8{\cal M}^{2}} B_{[4,0]}\biggr\}
 (\bm{\sigma}_1\cdot\bm{\sigma}_2)
\nonumber\\ && +\frac{m_G^2}{4M_yM_n}
 \biggl\{ (1+\kappa_G\frac{M_y}{{\cal M}}) (1+\kappa_G\frac{M_n}{{\cal M}}) 
 B_{[0,0]}(T) -\kappa_G^2\frac{m_G^2}{8{\cal M}^{2}} B_{[2,0]}(T) \biggr\})           
 (\bm{\sigma}_1\cdot\bm{\sigma}_2) \nonumber\\ && 
 -\frac{m_G^4}{16M_y^{2}M_n^{2}} \biggl\{ 1+8\kappa_G \frac{\sqrt{M_yM_n}}{{\cal M}}  
   +8\kappa_G^2\frac{M_yM_n}{{\cal M}^{2}}\biggr\} B_{[0,0]}(Q)
 (\bm{\sigma}_1\cdot\bm{\sigma}_2).
\label{app.enngl1}\end{eqnarray}

\subsection{Mass from Non-local Two-body forces}
\label{app:EN.b}  
Contributions to $E_N$ from non-local QQ-potentials, given in subsection 
\ref{app:OBE.c}.\\

\noindent {\bf (a)} Pseudoscalar-meson exchange $J^{PC}=0^{-+}$:
\begin{eqnarray}
&& E^{(P)}_{12;3} = E^{(P)}_{12;3}+g^p_{13}g^p_{24}\  
  \left(\frac{m_P^3}{24M_yM_n}\right) 
 \cdot\nonumber\\ && \times
 \biggl( B_{[4,2]} +\frac{1}{2} B_{[6,0]}+ 
 3 \biggl[B_{[2,2]}(T) + B_{[4,0]}(T) \biggr] \biggr)\
 (\bm{\sigma}_1\cdot\bm{\sigma}_2).
\label{app.ennps2}\end{eqnarray}
\noindent {\bf (b)} Vector-meson exchange $J^{PC}=1^{--}$:
\begin{eqnarray}
 && E^{(V)}_{12;3} = E^{(V)}_{12;3} +g^v_{13}g^v_{24}\frac{3m_V^3}{2M_yM_n}
  \biggl[ B_{[0,2]}+\frac{1}{4} B_{[2,0]}\biggr].
\label{app.obevc3}\end{eqnarray}
\noindent {\bf (c)} Scalar-meson exchange $J^{PC}=0^{++}$:
\begin{eqnarray}
 && E^{(S)}_{12;3} = E^{(S)}_{12;3}+ g^s_{13} g^s_{24}\
 \frac{m_S^3}{2M_yM_n} \biggl(B_{[0,2]}+\frac{1}{4} B_{[2,0]}\biggr).
\label{app.obesc3}\end{eqnarray}
\noindent {\bf (d)} Axial-vector-meson exchange $J^{PC}=1^{++}$:
\begin{eqnarray}
 && E^{(A)}_{12;3} = E^{(A)}_{12;3} - g^a_{13} g^a_{24}\ 
 \frac{3m_A^3}{2M_yM_n}\biggl( B_{[0,2]}+\frac{1}{4}B_{[2,0]}\biggr)\
  (\bm{\sigma}_1\cdot\bm{\sigma}_2).
\label{app.obeax4}\end{eqnarray}
\noindent {\bf (e)} Axial-vector-meson exchange $J^{PC}=1^{+-}$:
\begin{eqnarray}
 && E^{(B)}_{12;3} = V^{(B)}_{12;3} +
       f^B_{13}f^B_{24}\frac{(M_n+M_y)^2}{m_B^2} \left(\frac{m_B^5}{8M_y^2M_n^2}\right)
\cdot\nonumber\\ && \times 
\biggl\{\biggl[B_{[2,2]}+\frac{1}{4}B_{[4,0]}\biggr] +3
        \biggl[B_{[2,2]}(T)+\frac{1}{4}B_{[4,0]}(T)\biggr] \biggr)\
  (\bm{\sigma}_1\cdot\bm{\sigma}_2).
\label{app.obeax5}\end{eqnarray}
\noindent {\bf (c)} Diffractive  exchange $J^{PC}=0^{++}$:
\begin{eqnarray}
 && E^{(D)}_{12;3} = E^{(D)}_{12;3}- g^d_{13} g^d_{24}\ \left(\frac{m_P^6}{\Lambda^5}\right)
 \frac{m_P^2}{2M_yM_n} \biggl(D_{[0,2]}+\frac{1}{4} D_{[2,0]}\biggr).
\label{app.diffr4}\end{eqnarray}
\noindent {\bf (d)} Gluon exchange $J^{PC}=1^{--}$:
\begin{eqnarray}
 && E^{(G)}_{12;3} = E^{(V)}_{12;3} +g_{QCD}^2\
 \frac{3m_G^3}{2M_yM_n}
  \biggl[ B_{[0,2]}+\frac{1}{4} B_{[2,0]}\biggr].
\label{app.obegl2}\end{eqnarray}

\subsection{Mass from Additional Two-body forces}
\label{app:EN.c}  
\noindent {\bf a} Pseudoscalar-meson exchange $J^{PC}=1^{--}$: no extra contributions.\\

\noindent {\bf b} Vector-meson exchange $J^{PC}=1^{--}$:
     \begin{eqnarray}  
&&       \Delta E^{(V)}_{12;3} =  - \left(\frac{m_V^3}{4{\cal M} m_Q}\right)
   \biggl\{\bigl[g^v_{13}f^v_{24}+f^v_{13}g^v_{24}\bigr]\ B_{[2,0]}
    +\bigg\{ \left(g^v_{13}+f^v_{13}\frac{M_y}{{\cal M}}\right)
           f^v_{24}\left(1+\frac{M_y}{m_Q}\right)
 \nonumber\\ && 
           +f^v_{13}\left(g^v_{24}+f^v_{24}\frac{M_n}{\cal M}\right)\
           \left(1+\frac{M_n}{m_Q}\right) \biggr\}\
            \left(\frac{m_V^2}{4M_yM_n}\right) 
            \biggl[\frac{2}{3}B_{[4,0]}-B_{[2,0]}(T)\biggr]                  
             (\bm{\sigma}_1\cdot\bm{\sigma}_2) 
 \nonumber\\ && 
     +\biggl\{ \left(1+4\frac{\sqrt{M_yM_n}}{m_Q}\right)(g^v_{13}f^v_{24}+f^v_{13}g^v_{24})
     +8f^v_{13}f^v_{24}\frac{\sqrt{M_yM_n}}{\cal M} \biggr\}
     \left(\frac{m_V^4}{16M_y^2M_n^2}\right)\cdot\nonumber\\ && \times  B_{[0,0]}(Q)\
             (\bm{\sigma}_1\cdot\bm{\sigma}_2).
\label{app.obevc4}\end{eqnarray}
\noindent {\bf c} Scalar-meson exchange $J^{PC}=0^{++}$: 
      \begin{eqnarray} 
       \Delta E^{(S)}_{12;3}&=& - g^s_{13}g^s_{24}
       \left(\frac{m_S^3}{2M_yM_n}\right)
       \biggl( B_{[2,0]} -\frac{m_S^4}{16M_y^2M_n^2}\ B_{[2,0]}(Q)
       (\bm{\sigma}_1\cdot\bm{\sigma}_2) \biggr).
\label{app.obesc4}\end{eqnarray}
\noindent {\bf d} Axial-vector-meson exchange $J^{PC}=0^{++}$: no additional contributions.

\begin{flushleft}
\rule{16cm}{0.5mm}
\end{flushleft}

\section{ Instantons, Confining Potentials}           
\label{sec:conf2}       
The SU(3) generalization of the 't Hooft interaction for the (u,d,s) quarks 
in the NJL-form, see Appendix~\ref{app:instnlag}, reads
\begin{equation}
{\cal L}_{uds} = G_I\left[\vphantom{\frac{A}{A}} 
 (\bar{\psi}\lambda_0\psi)^2+(\bar{\psi}i\gamma_5\bm{\lambda}\psi)^2
 -(\bar{\psi}\bm{\lambda}\psi)^2-(\bar{\psi}i\gamma_5\lambda_0\psi)^2\right],
\label{njl.1}\end{equation}
with $G_I=\lambda_{ud}/4$, and 
where $\psi =(u,d,s)$ i.e. the flavor $\{3\}$-irrep spinor field, 
$\lambda_a, a=1,8$ are the Gell-Mann matrices, and $\lambda_0=(2/\sqrt{3}) {\bf 1}$,  
see Appendix~\ref{app:instnlag}.\\
\noindent For the U,D quarks, and written in the quark fields, it reads
\begin{eqnarray}
{\cal L}_{ud} &=& G_I \sum_{i>j=1}^2\biggl[\biggl\{ (\bar{q}_i q_i) (\bar{q}_j q_j)
-(\bar{q}_i \bm{\tau}_i q_i)\cdot (\bar{q}_j \bm{\tau}_j q_j)
\biggr\} \nonumber\\ && +\biggl\{
 (\bar{q}_i \gamma_5 q_i) (\bar{q}_j \gamma_5 q_j) 
-(\bar{q}_i \gamma_5 \bm{\tau}_i q_i)\cdot (\bar{q}_j\gamma_5 \bm{\tau}_j q_j)
\biggr\}\biggr].
\label{njl.2}\end{eqnarray}
The quark-quark momentum-space instanton potential $V_{I,12}({\bf p}',{\bf p})$ 
is obtained from the constituent quark Dirac spinors as follows
\begin{eqnarray*}
 (\bar{q}q)^2 &\rightarrow& 1-\frac{1}{4 m_Q^2}\biggl(2{\bf p}'\cdot{\bf p}+
 i\left(\bm{\sigma}_1+\bm{\sigma}_2\right)\cdot{\bf p}'\times{\bf p}\biggr)
 \biggr), \nonumber\\ 
 (\bar{q}\gamma_5 q)^2 &\rightarrow& -\frac{1}{4m_Q^2}\ 
 \bm{\sigma}_1\cdot({\bf p}'-{\bf p})\ \bm{\sigma}_2\cdot({\bf p}'-{\bf p})
\end{eqnarray*}
Noting that ${\cal H}_I=-{\cal L}_I$, and using the momenta 
${\bf k}={\bf p}'-{\bf p}$ and ${\bf q}= ({\bf p}'+{\bf p})/2$
the instanton exchange potential between $q_1$ and $q_2$ becomes \cite{factor-2}
\begin{eqnarray}
&& V_{I,12}({\bf p}',{\bf p}) = -2G_I \biggl(1-\bm{\tau}_1\cdot\bm{\tau}_2\biggr)\
 \biggl[\biggl\{\biggl(1+\frac{{\bf k}^2}{8m_Q^2}-\frac{{\bf q}^2}{2m_Q^2}\biggr)
\nonumber\\ &&
  -\frac{i}{4m_Q^2} \left(\bm{\sigma}_1+\bm{\sigma}_2\right)\cdot
 {\bf p}'\times{\bf p}  
 +\frac{1}{16m_Q^4} [\bm{\sigma}_1\cdot{\bf p}'\times{\bf p}]\
 [\bm{\sigma}_2\cdot{\bf p}'\times{\bf p}]\biggr\}
\nonumber\\ &&
  -\biggl\{ \frac{{\bf k}^2}{12m_Q^2}\ \bm{\sigma}_1\cdot\bm{\sigma}_2  
 +\frac{1}{4m_Q^2}\left( \bm{\sigma}_1\cdot{\bf k}\ \bm{\sigma}_2\cdot{\bf k}
 -\frac{1}{3} \bm{\sigma}_1\cdot\bm{\sigma}_2\ {\bf k}^2 \right) \biggr\}\biggr].
\label{njl.3}\end{eqnarray}
Now, the factor $(1-\bm{\tau}_1\cdot\bm{\tau}_2)$ is +2, and 0 for respectively the
proton P and the $\Delta_{33}$. 
\begin{equation}
(1-\bm{\tau}_1\cdot\bm{\tau}_2) = \biggr\{\begin{array}{cc} +2 & P(938) \\
 0 & \Delta_{33}(1236) \end{array},\ 
(1-\bm{\tau}_1\cdot\bm{\tau}_2)(\bm{\sigma}_1\cdot\bm{\sigma}_2)  = 
\biggr\{\begin{array}{cc} -6 & P(938) \\ 0 & \Delta_{33}(1236) \end{array}.  
\label{njl.4}\end{equation}
\noindent {\it For SU(3) the coefficients in (\ref{njl.4}) become 
$(4/3-\bm{\lambda}_1\cdot\bm{\lambda}_2)$ and 
$(4/3-\bm{\lambda}_1\cdot\bm{\lambda}_2)(\bm{\sigma}_1\cdot\bm{\sigma}_2)$ which 
corresponds to the matrix elements given in (\ref{njl.4}), see Appendix~\ref{app:uni}.      
For the baryon-octet the contribution of the instantons is universal, giving a down-shift
and an up-shift of the mass for the baryon octet and decuplet respectively, 
producing a mass splitting between the octet and decuplet.}\\

\noindent In configuration space, with the addition of the gaussian
cut-off, for the proton and the 33-resonance the effective local QQ-potential,
see {\it e.g.} \cite{NRS78} for the momentum- and configuration space formulas, is
\begin{eqnarray}
 V_{I,loc}(r) &=& -2(1-\bm{\tau}_1\cdot\bm{\tau}_2)\ G_I\
 \left(\frac{\Lambda_I}{2\sqrt{\pi}}\right)^3 \left[ 1
 +\frac{\Lambda_I^2}{2m_Q^2}\left(3-\frac{1}{2}\Lambda_I^2 r^2\right)
\right.\cdot\nonumber\\ && \times \left.
 \left(1-\frac{1}{3}\bm{\sigma}_1\cdot\bm{\sigma}_2\right)\right]\
 \exp\left[-\frac{1}{4}\Lambda_I^2r^2\right] \rightarrow
\nonumber\\ &&
 -4 G_I \left(\frac{\Lambda_I}{2\sqrt{\pi}}\right)^3 \left[ 1
 +\frac{2\Lambda_I^2}{3m_Q^2}\left(3-\frac{1}{2}\Lambda_I^2 r^2\right)\right]
 \exp\left[-\frac{1}{4}\Lambda_I^2r^2\right].             
\label{njl.6}\end{eqnarray}
The last expression in (\ref{njl.6}) is for each pair in the nucleon, where 
$\bm{\tau}_1\cdot\bm{\tau}_2= \bm{\sigma}_1\cdot\bm{\sigma}_2=-1$.\\
From the $\pi-\rho$ splitting $G_I = \lambda_{ud}/4= (3.5-5.0)$ GeV$^{-2}$, and for 
$\Lambda_I = {\cal M} = 1$ GeV
the potential is attractive $V_{I,loc}(0) \approx -2.4 M_p$. This leads to the
$N-\Delta$ splitting caused by the instantons.
In these notes we call the model with this instanton-splitting model-A.
Writing $G_I = C_I/{\cal M}^2$, the contribution to the nucleon and 
the 33-resonance mass is 
\begin{eqnarray}
 E^{(I)}_{12;3} &=& -2C_I\ (1-\bm{\tau}_1\cdot\bm{\tau}_2)\ 
 \left(\frac{\Lambda_I^3}{{\cal M}^2}\right) \biggl(
      \left\{ D_{[0,0]} + \frac{\Lambda_I^2}{4M_yM_n} D_{[2,0]}
      -\frac{\Lambda_I^2}{2M_yM_n}\ D_{[0,2]} \right\}
 \nonumber\\ && +\frac{\Lambda_I^2}{12 M_yM_n}\bigg\{ D_{[2,0]} + 3 D_{[0,0]}(T)
	+\frac{3\Lambda_I^2}{4 M_y M_n} D_{[0,0]}(Q) \biggr\}
(\bm{\sigma}_1\cdot\bm{\sigma}_2) \biggr).
\label{njl.8}\end{eqnarray}
 The confining potential is taken of the same form as in Eq.~(\ref{sec:conf.3b})
i.e. 
\begin{equation}
 V_{conf} = -C'_0 + \biggl[C'_2\ r^2\ e^{-m_C^2 r^2}\biggr].  
\label{njl.9}\end{equation}
The contribution to $E_{conf}$ is 
\begin{eqnarray}
 E_{conf}(12;3) &=& -C'_0 +\left(\frac{\pi}{m_C^2}\right)^{3/2} \biggl[ 
       +\frac{C'_2}{m_C^2}\bigg\{\frac{3}{2}\ G_{conf}^{(0)} 
  - \frac{1}{4m_C^2}G_{conf}^{(2)}\bigg\} \biggr],       
\label{njl.10}\end{eqnarray}
where $G_{conf}^{(0)}= (2m_C)^3 D_{[0,0]}(m_C^2)$ and 
 $G_{conf}^{(2)}= (2m_C)^5 D_{[2,0]}(m_C^2)$.     

\begin{flushleft}
\rule{16cm}{0.5mm}
\end{flushleft}

\section{ Gluons, Confining Potentials}           
\label{sec:conf1}       
The QCD one gluon-exchange (OGE) has the form
$V_{OGE} = g_{QCD}^2\ (\bm{\lambda}^c_1\cdot\bm{\lambda}^c_2)\ V_V(m_G,r,\Lambda_G)$, 
where $V_V$ is the OBE vector exchange potential, and $\lambda^c_i\ (i=1,8)$ are the Gell-Mann
matrices. Here, $m_G \approx 420$ MeV,
which is the mass of the gluon propagator in the
"liquid instanton model" \cite{Hut95}.

\noindent Apart from the OGE potential the potential for the three-quark
system consists of a single-quark potential $V_{conf}$ and a two-quark
potential $V_{mm}$, where the latter is the color-magnetic moment
 interaction. We distinguish between the OGE and the phenomenological $V_{mm}$. \\

\noindent {\bf a)\ OGE:} the contribution to the nucleon mass is given by the 
same formula as those from vector-meson exchange making the substitution:
$m_V \rightarrow m_G$, and $g^v_{13}g^v_{24} \rightarrow g_{QCD}^2$, and
$\kappa^v_{13},\kappa^v_{24} \rightarrow \kappa_G$. For the "current quarks" 
$\kappa_G=0$ since this quark has at low energies no internal structure.
However, "constituent quarks" presumably have internal gluonic structure
because of the dressing, and hence in principle $\kappa_G \neq 0$. 
Also, the quark-gluon coupling for constituent quarks can be expected to have
a form factor with a cut-off $\Lambda_{QCD} \approx 1$ GeV.
{\it Although the mass splitting between the nucleon and the 33-resonance, as well as the mass splitting
between the $\pi$ and the $\rho$,  could be attributed totally to OGE, see {\it e.g.} 
Ref.~\cite{RGG75,Close79}, important contributions from instantons are also possibly present. 
Utilizing the sensitivity w.r.t. to the cut-off room for the latter contributions can be 
made.}
The gluon-quark coupling is described by the Lagrangian
\begin{equation}
{\cal H}_I = g\bar{\psi}(\lambda_a/2)\biggl[ \gamma^\mu A^\mu_a + \frac{\kappa}{4{\cal M}}
 \sigma_{\mu\nu} \left(\partial^\mu A^\nu_a-\partial^\nu A^\mu_a\right)\biggr] \psi.
\label{sec:conf.3a}\end{equation}
\noindent In configuration space the OGE potential, see {\it e.g.} \cite{NRS78} Eqn.~(32),
for the (12)-pair reads
\begin{eqnarray}
&& V_{12}(OGE) = \frac{g_{QCD}^2}{4\pi} m \biggl[ \biggl(\phi_C^0+\frac{m_G^2}{2M_yM_n}
 \phi_C^1-\frac{3}{4M_yM_n}\left(\bm{\nabla}^2\phi_C^0+\phi_C^0\bm{\nabla}^2\right)\biggr)
+\frac{m_G^2}{6M_yM_n}\ \phi_C^1\ 
\cdot\nonumber\\ && \hspace{0mm} \times
(\bm{\sigma}_1\cdot\bm{\sigma}_2)
-\frac{m_G^2}{4M_yM_n}\ \phi_T^0                                    
-\frac{3m_G^2}{2M_yM_n}\ \phi_{SO}^0\ {\bf L}\cdot{\bf S}
+\frac{m_G^4}{16 M_y^2M_n^2}\ \frac{3}{(m_Gr)^2}\ \phi_T^0 Q_{12}\biggr] 
({\bf F}^c_1\cdot{\bf F}^c_2).
\label{sec:conf.3c}\end{eqnarray}
Here $ M_n=m_Q, M_y=m_Q'$, and ${\bf F}=\bm{\lambda}/2$.
For the quark pairs (13) and (23) similar expressions apply.
For the octet baryons and the $\Delta_{33}$ 
$(\bm{\sigma}_i\cdot\bm{\sigma}_j)$ is -1 and +1 respectively. 
Similarly, for ${\bf F}^c_i\cdot{\bf F}^c_j=(\bm{\lambda}^c_i\cdot\bm{\lambda}^c_j)/4$ 
one has -2/3 for both the octet baryons and the $\Delta_{33}$-resonance 
(see Table~\ref{tab:ff1} below). 
This because, in contrast to flavor and spin 
in the baryons, the color and spin are not intertwined.\\
The point-like limits are given by 
$\lim_{\Lambda \rightarrow \infty} \phi_C^0 = \exp(-m_Gr)/(m_Gr)$ etc.\\

\noindent {\bf b)\ $\bm{V_{conf}, V_{mm}}$}: 
We choose a color-singlet central confining potential and 
a color-octet ("magnetic") spin-spin potential.
We restrict the contribution to the region of the   
nucleon, i.e. for $r < R$, with R= quark radius of the nucleon.
An attractive procedure is the multiply the confining potential by a 
Wood-Saxon type of function. However, this makes the integrals for the
three-body matrix element very complicated. Therefore, we choose to work here with
a gaussian cut-off: 
\begin{eqnarray}
 V_{conf} +V_{mm} &=& -C_0 + \biggl[C_2\ r^2 
 -\frac{1}{4}C_1\ (\bm{\sigma}_1\cdot\bm{\sigma}_2)
 (\bm{\lambda}^c_1\cdot\bm{\lambda}^c_2)\biggr]\ e^{-m_C^2 r^2}.  
\label{sec:conf.3b}\end{eqnarray}
Here, we choose $ m_C \approx 0.74$ fm$^{-1}$ which means 
that $V_{conf}$ is reduced by a factor 2 at $r= 1$ fm.
Then, in momentum space
\begin{eqnarray}
 \biggl[\widetilde{V}_{conf} + \widetilde{V}_{mm}\biggr]({\bf k}^2) &=& 
-C_0 +\left(\frac{\pi}{m_C^2}\right)^{3/2}\biggl[
 \frac{C_2}{m_C^2}\bigg\{\frac{3}{2}- \frac{{\bf k}^2}{4m_C^2}\biggr\}
\nonumber\\ && 
-\frac{1}{4}C_1\ (\bm{\sigma}_1\cdot\bm{\sigma}_2)
 (\bm{\lambda}^c_1\cdot\bm{\lambda}^c_2)\biggr]\
 \exp\biggl[-\frac{{\bf k}^2}{4m_C^2}\biggr].
\label{sec:conf.4}\end{eqnarray}
We note that (\ref{sec:conf.3b}) is a cut-off modified potential in Ref.~\cite{Rib80}.\\
The parameters in \cite{Rib80} are $C_0= +230$ MeV, 
$C_2= +93.75 R_0^{-2}= +314.47$ MeVfm$^{-2}$, with
$R_0=0.546$ fm, and $C_1=+293.7$ MeV.\\
Assuming that the confinement potential $V_{conf}$ is a scalar-exchange the
contribution to the nucleon mass is  
\begin{eqnarray}
 E_{conf}(12;3) &=& -C_0 +\left(\frac{\pi}{m_C^2}\right)^{3/2} \biggl[ 
       +\frac{C_2}{m_C^2}\bigg\{\frac{3}{2}\ G_{conf}^{(0)} 
  - \frac{1}{4m_C^2}G_{conf}^{(2)}\bigg\} \nonumber\\ && \hspace{2cm}
 -\frac{1}{4}C_1\ G_{conf}^{(0)} (\bm{\sigma}_1\cdot\bm{\sigma}_2)
(\bm{\lambda}^c_1\cdot\bm{\lambda}^c_2)\biggr],       
\label{sec:conf.6}\end{eqnarray}
where $G_{conf}^{(0)}= (2m_C)^3 D_{[0,0]}(m_C^2)$ and 
 $G_{conf}^{(2)}= (2m_C)^5 D_{[2,0]}(m_C^2)$.\\
   
\noindent To make the color spin-spin more like a $\delta^3({\bf r})$ function it is useful 
to take $m_C \rightarrow m_{C_0}$ and $m_C \rightarrow m_{C_1}$ for the central and
spin-spin potential respectively.
For example $m_{C_0} \approx 10$ MeV and $m_{C_1} \approx 200$ MeV. The formulas
above can readily be adapted to accommodate this.\\
\noindent {\it In models this phenomenological spin-spin interaction is often used to generate 
the $N-\Delta$ and $\pi-\rho$ mass splittings. If one includes the OGE potential 
this interaction is unnecessary, hence $C_1=0$.}

\noindent {\bf c)\ Color-Spin factor}: In Table~\ref{tab:ff1} the color factor is given. 
For the other pairs, because of the complete antisymetrization,  one has
\begin{equation}
 (\bm{\lambda}_1\cdot\bm{\lambda}_2) = (\bm{\lambda}_1\cdot\bm{\lambda}_3) =             
 (\bm{\lambda}_2\cdot\bm{\lambda}_3).              
\label{sec:conf.5}\end{equation}
Since the diquarks are in the $\{\bar{3}\}$ color-irrep one has
\begin{eqnarray}
&& \bm{\lambda}^c_1\cdot\bm{\lambda}^c_2 = \frac{1}{2}\left(\bm{\lambda}^c_1+\bm{\lambda}^c_2\right)^2
-\frac{1}{2}\left((\bm{\lambda}^c_1)^2+(\bm{\lambda}^c_2)^2\right)
\label{sec:conf.7}\end{eqnarray}
We have ${\bf F}_i=\bm{\lambda}_i/2$, and 
\begin{eqnarray*}
&& \langle {\bf F}^2\rangle = \langle {\bf I}^2\rangle
 +2 \langle I_z\rangle +\frac{3}{4}Y^2
\end{eqnarray*}
which for the quarks ($I_c=1/2, I_{c,z}=+1/2, Y_c=1/3$) 
gives $\langle {\bf F}^2\rangle = 0,\ 4/3,\ 4/3,\ 10/3,\ 3,\ 6$ for the 
SU(3)-irreps $\{1\}, \{3\}, \{3^*\}, \{6\}, \{8\}, \{10\}$ respectively. 
Then, the color factor for the $\{\bar{3}\}_c$-irreps becomes
$ \bm{\lambda}_1\cdot\bm{\lambda}_2 = -8/3$, which applies to the nucleon as well as to the 
$\Delta_{33}$.

For the spin operators one has summing over three quarks
\begin{eqnarray}
&& \bm{\sigma}_1\cdot\bm{\sigma}_2 +\bm{\sigma}_1\cdot\bm{\sigma}_3
+\bm{\sigma}_2\cdot\bm{\sigma}_3 = \frac{1}{2}\left(\bm{\sigma}_1+\bm{\sigma}_2
+\bm{\sigma}_3\right)^2-\frac{9}{2} = 
\nonumber\\ && 2S(S+1)-\frac{9}{2} = \biggl\{\begin{array}{cc}
 \Delta_{33}: S=3/2 \rightarrow +3 \\
  N_{11}    : S=1/2 \rightarrow -3 \end{array}
\label{sec:conf.8}\end{eqnarray}
Because of the SU(4)-symmetry w.r.t. spin-flavor one has  
$\bm{\sigma}_1\cdot\bm{\sigma}_2 =\bm{\sigma}_1\cdot\bm{\sigma}_3
=\bm{\sigma}_2\cdot\bm{\sigma}_3 = \pm$ for respectively the $\Delta_{33}$ and the
nucleon. Therefore, the $N-\Delta$ mass-splitting from OGE is due to the spin-spin force.\\

\noindent {\bf d)\ Remark}: In \cite{RGGD12,Rib80} the confining potential is taken 
to be a scalar color-octet exchange potential. In \cite{Gloz97} the confining potential 
is color-singlet scalar exchange of the form $ V_{conf} = C_0 + C_1\ r^2$,
where $C_0$ is adjusted to give the 939 MeV for the nucleon mass, and
depends on the other parts of the total Q-Q potential. For the GBE-model
\cite{Gloz96a,Gloz96b} in \cite{Gloz97} table III the fitted GBE parameters are
$C_0$= -416 MeV, $C_1$= 2.33 MEVfm$^{-2}$.\\
\noindent Since the GBE-model approach is also that of Manohar-Georgi, we
choose in this work the confining potential in (\ref{sec:conf.3b}).

\begin{table}
 \caption{Color and Spin matrix elements, ${\bf F}=\bm{\lambda}^c/2$.}
\begin{center}
\begin{ruledtabular}
\begin{tabular}{|ccc|cc} & & & & \\
  S & I & C & $\langle \bm{\lambda}^c_1\cdot\bm{\lambda}^c_2\rangle$ &
 $\langle \bm{\sigma}_1\cdot\bm{\sigma}_2\rangle$ \\
 & & & & \\ \hline
 0 & 0 & $\{3^*\}$ &-8/3 & -1 \\
 0 & 1 & $\{6\}$   &+8/3 & -1 \\
 1 & 0 & $\{6\}$   &+8/3 & +1 \\
 1 & 1 & $\{3^*\}$ &-8/3 & +1 \\
\end{tabular}
\end{ruledtabular}
\end{center}
\label{tab:ff1}
\end{table}
\noindent {\bf e)\ N-$\Delta$-splitting I:} In \cite{RGG75} the mass splitting 
between the nucleon and the $\Delta_{33}$-resonance is given by the
expectation of the spin-spin force
\begin{eqnarray}
\Delta_M &=& -\frac{\pi}{2}\ \delta^3({\bf r})\
\biggl\langle \frac{4 \bm{\sigma}_i\cdot\bm{\sigma}_j}{3m_i m_j}\biggr\rangle.
\label{eq:DNsplit.1}\end{eqnarray}
Here (ij) is the quark pair. Because of the symmetry of the quark wave
functions we evaluate this for the pair (12) and multiply the results by 3.
The calculation of d in eq. 4b of Ref.~\cite{RGG75} is as follows
\begin{eqnarray}
d &=& \frac{\pi}{2}\biggl\langle \Psi_0|\delta({\bf r}_{12})|\Psi_0\biggr\rangle
 = \lim_{\Lambda \rightarrow \infty} \frac{\pi}{2}
\biggl\langle \Psi_0\bigg| \frac{\Lambda^3}{8\pi\sqrt{\pi}}
 \exp\left[-\frac{1}{4}\Lambda^2{\bf r}_{12}^2\right] \bigg| \Psi_0\biggr\rangle
\nonumber\\ &\Rightarrow& \frac{\Lambda^3}{16\sqrt{\pi}}\ N_3^2
\int d^3\rho\int d^3\lambda\ e^{-3\lambda(\bm{\rho}^2=\bm{\lambda}^2)}\
e^{-\Lambda^2\bm{\rho}^2/2} \nonumber\\ &=& \frac{\Lambda^3}{16\sqrt{\pi}}
 \left(1+\frac{\Lambda^2}{6\lambda}\right)^{-3/2}  
 \longrightarrow  \sqrt{\frac{2}{\pi}}
\frac{27}{16}\ R_N^{-3}\ (\Lambda \rightarrow \infty).
\label{eq:DNsplit.2}\end{eqnarray}
This gives for mass shift of the spin-spin force
\begin{eqnarray}
\Delta M_{12} &=& -\frac{2}{3}\alpha_s\cdot -d\ 
\frac{4\langle\bm{\sigma}_1\cdot\bm{\sigma}_2 \rangle}{3m_im_j} = 
+\frac{8}{9} \alpha_s \langle\bm{\sigma}_1\cdot\bm{\sigma}_2 \rangle\
(d/m_Q^2).
\label{eq:DNsplit.3}\end{eqnarray}
Using that 
$\langle\bm{\sigma}_1\cdot\bm{\sigma}_2 \rangle$ is +1 and -1 for the
$\Delta_{33}$ and nucleon respectively, and multiplying by the number 
of pairs (3), one gets
\begin{eqnarray}
 \Delta_M(I) &=& M_\Delta-M_N = \frac{16}{9}\ \alpha_s\ (d/m_Q^2)=
 3\sqrt{\frac{8}{\pi}}\ \alpha_s\ \left(m_Q R_N\right)^{-2}\ R_N^{-1}.
\label{eq:DNsplit.4}\end{eqnarray}
For $R_N= 1$ fm, $m_Q= M_p/3 = 312$ MeV, one obtains 
$\Delta_M = 603.0\ \alpha_s$ MeV. With $\alpha_s= 0.48$ the mass shift is 
289 MeV.\\
 
\noindent {\bf f)\ N-$\Delta$-splitting II:} Using the formulas of these notes,
we get in the massless and point-coupling limits
\begin{eqnarray}
\lim_{\Lambda\rightarrow \infty}\lim_{m \rightarrow 0} H_{[0,0]} &=& 
 6\sqrt{2\pi}\ {\cal N}_{[0,0]}\ R_N^{-1},\ 
\lim_{\Lambda\rightarrow \infty}\lim_{m \rightarrow 0} H_{[2,0]} = 
 (18\pi)^{3/2}\ {\cal N}_{[0,0]}\ R_N^{-3}. 
\label{eq:DNsplit.6}\end{eqnarray}
Then, in the same limits the OGE gives
\begin{eqnarray}
 E^{(G)}_{12;3} &\Rightarrow& -(2\pi)^{-3}\ 27(2\pi\sqrt{2\pi})\ R_N^{-3}\ 
 \frac{g_{QCD}^2}{2m_Q^2}\ \left[1+\frac{1}{3} \langle\bm{\sigma}_1\cdot\bm{\sigma}_2\rangle
\right].
\label{eq:DNsplit.7}\end{eqnarray}
This leads to the spin-splitting, via adding the color factor -2/3 and using $g_{QCD}^2=4\pi \alpha_s$,
\begin{eqnarray}
 \Delta_M(II) &=& M_\Delta-M_N = \frac{3\alpha_s}{\pi^2}\ (2\pi\sqrt{2\pi})\ 
\left(m_Q R_N\right)^{-2} R_N^{-1}.
\label{eq:DNsplit.8}\end{eqnarray}
This leads to the ratio
\begin{equation}
 \Delta_{M}(I)/\Delta_{M}(II) = 1.    
\label{eq:DNsplit.9}\end{equation}
\noindent {\bf Corollary: this checks our formula with the literature \cite{RGG75}!}

\begin{flushleft}
\rule{16cm}{0.5mm}
\end{flushleft}

 \begin{table}[hbt]
\begin{center}
\caption{ESC08c (rationalized) 
coupling constants, $F/(F+D)$-ratio's, mixing angles etc.
 The values with $\star )$ have
 been determined in the fit to the $YN$-data. The other parameters
 are theoretical input or determined by the fitted parameters and
 the constraint from the $NN$-analysis. }
\label{tab.su3par} 
\begin{tabular}{ccccll} \hline\hline & & & & &  \\
  mesons    &   & $\{1\}$  & $\{8\}$  &  $F/(F+D)$   &    angles   \\
            &   &          &          &           &             \\
\hline
            &   &          &          &           &             \\
  ps-scalar  & f & 0.3389   &  0.2684 & $\alpha_{P}=0.3650$
             & $\theta_{P} = -11.40^{0~\ast)}$ \\
            &   &          &          &           &             \\
  vector    & g & 3.1983   &  0.5793  & $\alpha^{e}_{V}=1.0^{\ast)}$ &
 $\theta_{V} =\ \ 39.10^{0~\ast)}$   \\
            &   &          &          &           &             \\
            & f &--2.2644  &  3.7791  & $\alpha^{m}_{V}=0.4655^{\ast)}$ &  \\
            &   &          &          &           &             \\
  axial(A)  & g &--0.8826  &--0.8172  & $\alpha_{A}=0.3830$ &
 $\theta_{A} =-50.00^{0~\ast)}$ \\
            &   &          &          &           &             \\
            & f &--6.2681  &--1.6521  & $\alpha^{p}_{A}=0.3830^{\ast)}$ \\
            &   &          &          &           &             \\
  axial(B)  & f &--0.9635  &--2.2598  & $\alpha_{B}=0.4000^{\ast)}$ &
 $\theta_{B} = 35.26^{0~\ast)}$ \\
            &   &          &          &           &             \\
  scalar    & g &  3.2369  &  0.5393  & $\alpha_{S}=1.0000$ &
 $\theta_{S} =\ \ 44.00^{0~\ast)}$ \\
            &   &          &          &           &             \\
  diffractive   & $g_P$ &  2.7191  & $g_O$= 4.1637 & $f_O$= -3.8859  &
 $a_{PB}=\ \ 0.39^{\ast)}$\\
            &   &          &          &           &             \\
\hline
\end{tabular}
\end{center}
 \end{table}

\section{ Results and Discussion}                                             
\label{sec:results}
\subsection{ Coupling Constants, $F/(F+D)$ Ratios, and Mixing Angles}             
\label{sec:15b} 
In Table~\ref{table5} we give the ESC16 meson masses, and the fitted 
couplings and cut-off parameters \cite{ESC16a,ESC16b}.
Note that the axial-vector couplings for the
B-mesons are scaled with $m_{B_1}$.
The mixing for the pseudo-scalar, vector, and scalar mesons, as well as the 
handling of the diffractive potentials, has been described elsewhere, see
e.g. Refs.~\cite{MRS89,RSY99}. The mixing scheme of the axial-vector mesons is completely
similar as for the vector etc. mesons, except for the mixing angle.        
As mentioned above, we searched for solutions where 
all OBE-couplings are compatible with the QPC-predictions. This time the QPC-model
contains a mixture of the $^3P_0$ and $^3S_1$ mechanism, whereas in 
Ref.~\cite{Rij04a} only the $^3P_0$-mechanism was considered.
For the pair-couplings all $F/(F+D)$-ratios were fixed to the predictions of
the QPC-model. 

\begin{table}
\caption{Meson couplings and parameters employed in the ESC16-potentials.
         Coupling constants are at ${\bf k}^{2}=0$.
         An asterisk denotes that the coupling constant is constrained via SU(3).
         The masses and $\Lambda$'s are given in MeV.}
\label{table4}
\begin{center}
\begin{ruledtabular}
\begin{tabular}{crccr} \hline\hline
meson & mass & $g/\sqrt{4\pi}$ & $f/\sqrt{4\pi}$ & \multicolumn{1}{c}{$\Lambda$}\\
\hline
 $\pi$         &  138.04 &           &   0.2684   &  1030.96\     \\
 $\eta$        &  547.45 &           & \hspace{1mm}0.1368$^\ast$   & ,, \hspace{5mm} \\
 $\eta'$       &  957.75 &           &   0.3181   &  ,, \hspace{5mm} \\ \hline
 $\rho$        &  768.10 &  0.5793   &   3.7791   &    680.79\    \\
 $\phi$        & 1019.41 &--1.2384$^\ast$ & \hspace{2mm}2.8878$^\ast$ & 
 ,, \hspace{5mm}  \\
 $\omega$      &  781.95 &  3.1149   & --0.5710 \hspace{0.5mm}   &    734.21\\ \hline
 $a_1 $        & 1270.00 &--0.8172   & --1.6521   &   1034.13\   \\
 $f_1 $        & 1420.00 &  0.5147 &   4.4754  &    
  ,,  \hspace{5mm} \\
 $f_1'$        & 1285.00 &--0.7596   & --4.4179 &      ,, \hspace{5mm}  \\ \hline
 $b_1 $        & 1235.00 &           & --2.2598   &   1030.96    \\
 $h_1 $        & 1380.00 &           & \hspace{2mm}--0.0830$^\ast$   &     
 ,,  \hspace{5mm} \\
 $h_1'$        & 1170.00 &           & --1.2386   &      ,, \hspace{5mm}  \\ \hline
 $a_{0}$       &  962.00 &  0.5393   &            &    830.42\    \\
 $f_{0}$       &  993.00 &--1.5766$^\ast$   &            &  
 ,, \hspace{5mm}  \\
 $\varepsilon$ &  620.00 &  2.9773   &            &   1220.28 \\ \hline
 Pomeron       &  212.06 &  2.7191   &            &              \\
 Odderon       &  268.81 &  4.1637   & --3.8859   &              \\
\hline
\end{tabular}
\end{ruledtabular}
\end{center}
\label{table5}
\end{table}
One notices that all the BBM $\alpha$'s have values rather close to that 
which are expected from the QPC-model. In the ESC08c solution $\alpha_A \approx 0.31$,
which is not too far from $\alpha_A \sim 0.4$. 
As in previous works, e.g. Ref.~\cite{NRS78}, $\alpha_V^e=1$ is 
kept fixed.
Above, we remarked that the axial-nonet parameters may be sensitive to whether
or not the heavy pseudoscalar nonet with the $\pi$(1300) are included.

In Table~\ref{table4} we show the OBE-coupling constants and the 
gaussian cut-off's $\Lambda$. The used  $\alpha =: F/(F+D)$-ratio's 
for the OBE-couplings are:
pseudo-scalar mesons $\alpha_{pv}=0.365$, 
vector mesons $\alpha_V^e=1.0, \alpha_V^m=0.472$, 
and scalar-mesons $\alpha_S=1.00$, which is calculated using the physical 
$S^* =: f_0(993)$ coupling etc.
\subsection{ Model A: Instanton interactions}                                     
\label{sec:15d} 
In model A the mass splitting between the nucleon and the 33-resonance
is produced by the four-quark instanton Lagrangian. In Table~\ref{table:I.1}
the baryon masses are shown with $V_{OBE}=0$. The mass of the $\Xi(1321)$ is
about 100 MeV too large, which could be repaired by taking the quark radius 
$R=0.95$ fm reducing the kinetic energy contribution.

\noindent In Table~\ref{table:I.2} shows that the contribution of $V_{OBE}$
is small. The contributions of the ESC-potential are small by themselves 
and moreover there are big cancellations. In Table~\ref{table:I.2} $C_I$ and
$\Lambda_I$ are different from Table~\ref{table:I.1}, while the $V_I$ is
about the same. This shows that there is a strong correlation between these
parameters. Checked should be the consistency of $(C_I,\Lambda_I)$ with
those for the $\pi-\rho$ splitting.

\noindent {\it Looking at the contributions from $V_{OBE}$ displayed in Table~\ref{table:I.2}
it is clear that also with model A  a good match with the baryon
masses is quite possible.}
\begin{table}
\caption{Contributions Baryon masses from 
the confinement central potential and the instanton interaction (V$_{conf}$),
the kinetic energy (E$_{kin}$), and constituent quark masses. 
Quark-radii are $R= 0.95, 0.95, 0.875, 0.875, 0.850$ for 
P, $\Delta_{33}$, $\Lambda$, $\Sigma^+$, and $\Xi$ respectively.
The quark masses are $m_N=312.75$ and $m_S=500$ in MeV. The "confinement parameters are
$C'_0=760, C'_2= 93.75$ MeV. With $G_I$=2.8 GeV$^{-2}$ and $\Lambda_I=1$ GeV.
 The instanton quark-quark
interaction gives -324.4 MeV for P,$\Lambda,\Sigma,\Xi$, and 0 MeV for $\Delta_{33}$.
The CM-energy subtraction is 231 MeV.
}
\label{table:I.1}
\begin{center}
\begin{ruledtabular}
\begin{tabular}{c|ccc|ccc|c} \hline\hline
 baryon & $ V_{OBE}$ & $V_{conf}$ & $V_{OGE}$ & $V_{tot}$ & $E_{kin}$ & 
 $\sum_{i=1}^3 m_i$ & Mass \\
\hline
 $P(939)$ & ---  & -528  & ---  & -852 & +827 & 938.26 & 914 \\    
 $\Delta_{33}(1236)$ & ---  & -528  & --- & -525 & +827 & 938.26 &1238 \\    
 $\Lambda(1115)$ & ---  & -528  & --- & -852 & +878 &1125.50 & 1151 \\   
 $\Sigma(1189) $ & ---  & -528  & --- & -852 & +878 &1125.50 & 1151 \\   
 $\Xi(1321)    $ & ---  & -528  & --- & -852 & +843 &1312.75 & 1304 \\   
\hline
\end{tabular}
\end{ruledtabular}
\end{center}
\end{table}
\begin{table}
\caption{Contributions Baryon masses from the ESC QQ-potential (V$_{OBE}$),
the confinement central potential and the instanton interaction (V$_{conf}$),
the one-gluon-exchange interactions (OGE), the kinetic energy (E$_{kin}$), and 
constituent quark masses. 
In OBE the quark-meson Gaussian cut-off mass is $\Lambda_{QQM}=500$ MeV.
Quark-radii are $R= 0.80, 0.90, 0.775, 0.850, 0.850$ fm for
P, $\Delta_{33}$, $\Lambda$, $\Sigma^+$, and $\Xi$ respectively.
The quark masses are $m_N=312.75$ and $m_S=500$ in MeV. The "confinement parameters are
$C'_0=760, C'_2= 93.75$ MeV. With $G_I$=2.8 GeV$^{-2}$ and $\Lambda_I=1$ GeV,
 the instanton quark-quark
interaction gives -318.0 MeV for P,$\Lambda,\Sigma,\Xi$, and 0 MeV for $\Delta_{33}$.
The CM-energy subtraction is 231 MeV.
}
\label{table:I.2}
\begin{center}
\begin{ruledtabular}
\begin{tabular}{c|ccc|ccc|c} \hline\hline
 baryon & $ V_{OBE}$ & $V_{conf}$ & $V_{INST}$ & $V_{tot}$ & $E_{kin}$ & 
 $\sum_{i=1}^3 m_i$ & Mass \\
\hline
 $P(939)$            & -288  & -525 & -318 &-1131 &+1110 & 938.26 & 918 \\    
 $\Delta_{33}(1236)$ & -42.4 & -525 &  0.0 & -568 & +912 & 938.26 &1282 \\    
 $\Lambda(1115)$     & -248  & -525 & -318 &-1090 &+1090 &1125.50 & 1221 \\   
 $\Sigma(1189) $     & -21.1 & -525 & -318 & -864 & +925 &1125.50 & 1186 \\   
 $\Xi(1321)    $     & -21.1 & -525 & -318 & -864 & +843 &1312.75 & 1300 \\   
\hline
\end{tabular}
\end{ruledtabular}
\end{center}
\end{table}

\subsection{ Model B: Color Magnetic interactions}                                
\label{sec:15e} 
In Table~\ref{table.C1} the baryon masses are tabulated coming from the 
OGE-potentials, the confinement potential,
the quark kinetic energies, the CM-energy subtraction, and the quark masses.
In Table~\ref{table.C2} the baryon masses are tabulated coming from the 
ESC16 OBE QQ-potentials, OGE-potentials, the confinement potential,
the quark kinetic energies, the CM-energy subtraction, and the quark masses.
\begin{table}
\caption{Contributions Baryon masses from 
the confinement central potential $V_{conf}$, 
the "magnetic" spin-spin interaction V$_{mm}=0$,
the one-gluon-exchange interactions (OGE), the kinetic energy (E$_{kin}$), and 
constituent quark masses. Quark-radii are $R= 0.95, 0.95, 0.90, 0.90, 0.90$ fm for 
P, $\Delta_{33}$, $\Lambda$, $\Sigma^+$, and $\Xi$ respectively.
The quark masses are $m_N=312.75$ and $m_S=500$ in MeV. The "confinement parameters are
$C_0=395, C_1= 0, C_2= 93.75$ MeV. The CM-energy subtraction is 231 MeV.
}
\label{table.C1}
\begin{center}
\begin{ruledtabular}
\begin{tabular}{c|ccc|ccc|c} \hline\hline
 baryon & $ V_{OBE}$ & $V_{conf}$ & OGE & $V_{tot}$ & $E_{kin}$ & 
 $\sum_{i=1}^3 m_i$ & Mass \\
\hline
 $P(939)$            & ---  & -394  & -411 & -805 & +827 & 938.26 & 961 \\    
 $\Delta_{33}(1236)$ & ---  & -394  & -135 & -529 & +827 & 938.26 &1237 \\    
 $\Lambda(1115)$     & ---  & -394  & -411 & -805 & +833 &1125.50 & 1154 \\   
 $\Sigma(1189) $     & ---  & -394  & -411 & -805 & +833 &1125.50 & 1154 \\   
 $\Xi(1321)    $     & ---  & -394  & -411 & -805 & +755 &1312.75 & 1263 \\   
\hline
\end{tabular}
\end{ruledtabular}
\end{center}
\end{table}
\begin{table}
\caption{Contributions Baryon masses from the ESC QQ-potential (V$_{OBE}$),
the confinement central potential $V_{conf}$, the "magnetic" spin-spin interaction 
$V_{mm}=0$,
the one-gluon-exchange interactions (OGE), the kinetic energy (E$_{kin}$), and 
constituent quark masses. 
In OBE the quark-meson Gaussian cut-off mass is $\Lambda_{QQM}=500$ MeV.
Quark-radii are $R= 0.80, 0.90, 0.90, 0.935, 0.935$ for 
P, $\Delta_{33}$, $\Lambda$, $\Sigma^+$, and $\Xi$ respectively.
The quark masses are $m_N=312.75$ and $m_S=500$ in MeV. 
The gluon mass $m_G=420$ MeV, $\Lambda_{QCD}= 1000$ MeV, $g^2_{QCD}/4\pi= 0.48$.
The "confinement parameters are $C_0=395, C_1= 0, C_2= 93.75$ MeV. 
The CM-energy subtraction is 231 MeV.
}
\label{table.C2}
\begin{center}
\begin{ruledtabular}
\begin{tabular}{c|ccc|ccc|c} \hline\hline
 baryon & $ V_{OBE}$ & $V_{conf}$ & $V_{OGE}$ & $V_{tot}$ & $E_{kin}$ & 
 $\sum_{i=1}^3 m_i$ & Mass \\
\hline
 $P(939)$            & -288  & -394  & -432  &-1120 &+1110 & 938.26 &  937 \\    
 $\Delta_{33}(1236)$ & -42.4 & -394  & -140  & -577 & +912 & 938.26 & 1273 \\    
 $\Lambda(1115)$     & -50.2 & -394  & -432  & -877 & +833 &1125.50 & 1082 \\   
 $\Sigma(1189) $     & +177  & -394  & -432  & -650 & +776 &1125.50 & 1251 \\   
 $\Xi(1321)    $     & +177  & -394  & -432  & -604 & +700 &1312.75 & 1485 \\   
\hline
\end{tabular}
\end{ruledtabular}
\end{center}
\end{table}

\noindent CHECK: From Table~\ref{table.C2} it is seen that 
$R_\delta > R_P > R_\Lambda > R_\Sigma > R_\Xi$. The strong magnetic
repulsion in the $\Delta_{33}$-resonance makes the 'bag" larger. Furthermore,
the S-quark is slower than the U-,D-quark, which makes the order
of the radii not illogical. Of course, the differences between the 
$\{8\}$-baryons are small and there could be other reasons.


\subsection{ Summary and Conclusions}                                             
\label{sec:15g} 
\noindent {\bf In summary:} {\it The picture of this quark model is that
of the sixties. This is a picture of quarks moving in a deep 
potential well. Here we have constituent quarks moving relativistically
in a deep harmonic potential well. The depth of the well is the same
as for charmonium suggesting universality, which is pleasing in view
of the flavor-blindness of the gluons.}\\
\noindent We stress that we have evaluated the baryon masses in Born-approximation (B.A.).
Therefore, to properly evaluate model A, model B, or a mix of these, the three-body 
Lippmann-Schwinger or Schr\"{o}dinger equation should be solved.

\noindent {\bf Conclusion:} 
The contributions from OBE are not large if the meson-quark form factor 
cut-off $\Lambda_{QQM}\approx 500$ MeV. For example for $\Lambda_{QQM} =1$ GeV the
OBE is very large. This because the interaction is essentially short range ($r \approx 0.5$ fm),
and therefore very cut-off dependent.\\
For $\alpha_s=0.48$ and $\Lambda_{QCD}= 1$ GeV the $N-\Delta$ mass splitting is 
reproduced (model B). 
The same is true by using the instanton interaction, without OGE (model A). 

\noindent This opens the possibility to fit simultaneously the $N-\Delta$ and
$\pi-\rho$ splitting, using both mechanisms for these splittings. 
This because the OGE is rather dependent on the gluon-quark-quark cut-off.
Decreasing $\Lambda_{QCD}$ diminishes the $N-\Delta$ splitting, making room
for the presence of instanton interactions.
So, there is a possibility to fit both the $N-\Delta$ and
$\pi-\rho$ splitting, using both mechanisms for these splittings, consistent
with (perturbative) QCD and instanton physics.

\noindent There are very large cancellations between the confinement potential 
and the (relativistic) kinetic energies of the quarks. The inclusion of the ESC meson-exchange 
potential between the quarks is perfectly compatible with the picture of the
baryons in the CQM. An important condition is that the ESC QQ-potential is rather soft.
This also legitimates the application of the quark-quark ESC-potential to quark matter.\\

\noindent {\bf Thinking that there will be truth in both models A and B,
 a mix of these models is most likely the correct picture!}
For example taking $(C_I,\Lambda_I)$ the same as for the $\pi-\rho$ mass
splitting, the rest of the N-$\Delta$ splitting can be attributed 
to the color magnetic moment spin-spin interaction.

\section*{Acknowledgments }
Discussions with Y. Yamamoto are gratefully acknowledged. 
His stimulating work on the quark-matter created the strong motivation 
necessary for the start of this enterprise.

\appendix

\begin{flushleft}
\rule{16cm}{0.5mm}
\end{flushleft}

\section{Details $V_2$ Three-body momentum-space Integrals}       
\label{sec:mspace2}
\noindent {\bf $H_{[0,0]}$ in cartesian momenta:} Since the potentials $V_2$ 
are expressed in the cartesian momenta ${\bf k}_i, (i=1,2,3)$ it is convenient
to express the integral in (\ref{eq:T.19}) in terms of these variables. (This is also 
the case for the non-local momenta ${\bf q}_i, (i=1,2,3)$ when the contribution
of these terms is non-vanishing, of course.) 
In cartesian coordinates the exponential factor from the wave functions has
\begin{eqnarray}
 {\bf p}_\rho^{\prime 2}+{\bf p}_\lambda^{\prime 2} 
 +{\bf p}_\rho^2+{\bf p}_\lambda^2 = 4\biggl[
 ({\bf q}_1^2+{\bf q}_1\cdot{\bf q}_2+{\bf q}_2^2) +
 \frac{1}{4}({\bf k}_1^2+{\bf k}_1\cdot{\bf k}_2+{\bf k}_2^2)\biggr].
\label{eq:T.30b}\end{eqnarray}
For the following it is useful to introduce the short-hand 
\begin{equation}
  {\cal N}_{[0,0]} \equiv (2\pi)^{-9 }(3\pi^2\lambda^2)^{3/2}
  \widetilde{N}_3^2 = (2\pi)^{-3}.
\label{eq:T.32a}\end{equation}
Then, we get 
\begin{eqnarray}
&& H_{[0,0]} = (2\pi)^{-9 }
 \widetilde{N}_3^2\ \int_0^\infty d\alpha\ e^{-\alpha m^2}\cdot
 \int d^3q_1 d^3k_1 \int d^3q_2 d^3k_2 
 \cdot\nonumber\\ && \times
 \exp\bigg\{-\frac{1}{6\lambda}\biggl[
 4({\bf q}_1^2+{\bf q}_1\cdot{\bf q}_2+{\bf q}_2^2) 
 +({\bf k}_1^2+{\bf k}_1\cdot{\bf k}_2+{\bf k}_2^2)\biggr]\bigg\}\cdot
 e^{-\gamma {\bf k}^2} = {\cal N}_{[0,0]}
\cdot\nonumber\\ && \times 
 \int_0^\infty d\alpha\ e^{-\alpha m^2}\cdot
 \int d^3k_1 d^3k_2 
 \exp\bigg\{-\frac{1}{6\lambda}\biggl[
 {\bf k}_1^2+{\bf k}_1\cdot{\bf k}_2+{\bf k}_2^2\biggr]\bigg\}\cdot
 e^{-\gamma {\bf k}^2},                  
\label{eq:T.31b}\end{eqnarray}
where in the last step the ${\bf q}$-integrations are performed.
Using ${\bf k}_2=-{\bf k}_1$ brings (\ref{eq:T.31}) into the form
\begin{eqnarray}
 H_{[0,0]} &=& {\cal N}_{[0,0]}
 \int_0^\infty d\alpha\ e^{-\alpha m^2}
 \int d^3k \exp\left[-\frac{1}{6\lambda} {\bf k}^{2}\right]\
 \exp\left[-\gamma {\bf k}^2 \right].
\label{eq:T.32}\end{eqnarray}
Doing the ${\bf k}$-integration we obtain
\begin{eqnarray}
 H_{[0,0]} &=& {\cal N}_{[0,0]}
 \int_0^\infty d\alpha\ e^{-\alpha m^2}\cdot
 \int d^3k\ \exp\left[-\left\{
  \left(\frac{1}{6\lambda}+\gamma\right) {\bf k}^2
 \right\}\right] \nonumber\\ 
  &=&  {\cal N}_{[0,0]} \int_0^\infty d\alpha\ e^{-\alpha m^2}\cdot
 \left(\frac{\pi}{(\alpha+A)}\right)^{3/2},\ \ A = \frac{1}{6\lambda}+\frac{1}{\Lambda^2}.
\label{eq:T.33}\end{eqnarray}
The integral in (\ref{eq:T.33}) can be worked out explicitly. Defining 
$x=\alpha+A$ the integral reads
\begin{eqnarray}
 J_1(m,A) &=& \int_0^\infty d\alpha\ e^{-\alpha m^2}\ 
 \left(\frac{\pi}{(\alpha+A)}\right)^{3/2} = -2\pi\frac{d}{dA}\
 \int_0^\infty d\alpha\ e^{-\alpha m^2}\ 
 \left(\frac{\pi}{(\alpha+A)}\right)^{1/2} \nonumber\\       
 &=& -2\pi \frac{d}{dA}\left[ e^{A m^2} \int_A^\infty
 \frac{dx}{\sqrt{x}}\ e^{-x m^2}\right] = -4\pi\frac{d}{dA}\left[
 e^{A m^2} \int_{\sqrt{A}}^\infty dy\ e^{-m^2y^2}\right]
\nonumber\\ &=& -\left(2\pi\sqrt{\pi}/m\right)\ \frac{d}{dA}\left[ e^{A m^2}\
{\it Erfc}\bigl(\sqrt{A m^2}\bigr) \right] 
\nonumber\\ &=& 
 -(2\pi\sqrt{\pi}) m\ \biggl[ e^{A m^2} {\it Erfc}\bigl(\sqrt{A m^2}\bigr)
 -\frac{1}{\sqrt{\pi A m^2}}\biggr].
\label{eq:T.35b}\end{eqnarray}
With $\lambda= 3 R_N^{-2}$ one has 
$A=\big(1+\Lambda^2 R_N^2/18\bigr)/\Lambda^2$ and 
\begin{eqnarray}
  A m^2 &=& \frac{m^2}{\Lambda^2}\left(1+\frac{1}{18}\Lambda^2 R_N^2\right).
\label{eq:T.36}\end{eqnarray}
Finally, the expression for $H_{[0,0]}$ becomes
\begin{eqnarray}
 H_{[0,0]} &=&  {\cal N}_{[0,0]}\ J_1 = {\cal N}_{[0,0]}\cdot
 2\pi\sqrt{\pi} m\ \biggl[ \frac{1}{\sqrt{\pi A m^2}}
 -e^{A m^2} {\it Erfc}\bigl(\sqrt{A m^2}\bigr) \biggr].
\label{eq:T.37}\end{eqnarray}

\noindent {\bf b.}\ Factor ${\bf k}^2$ in $V_2$: Writing 
${\bf k}^2= ({\bf k}^2+m^2)-m^2$ a new integral occurs which 
is purely gaussian 
\begin{eqnarray}
 G_{[0,0]} &\equiv& \langle \psi_3|I_3 | \psi_3\rangle = \widetilde{N}_3^2\
 \int \frac{d^3p_\rho' d^3p_\lambda'}{(2\pi)^6} 
 \int \frac{d^3p_\rho d^3p_\lambda}{(2\pi)^6}\left\{
 \exp\left[-\frac{1}{6\lambda}({\bf p}_\rho^{\prime 2}+{\bf p}_\lambda^{\prime 2})
\right]
 \right.\cdot\nonumber\\ && \times\left.
 \exp\left[-\frac{1}{6\lambda}({\bf p}_\rho^{2}+{\bf p}_\lambda^{2})
\right]\ e^{-{\bf k}^2/\Lambda^2} \right\}
 = (2\pi)^{-9 } \widetilde{N}_3^2\ 
 \int d^3p_\rho' d^3p_\lambda' \int d^3p_\rho d^3p_\lambda 
 \cdot\nonumber\\ && \times\left\{
 \exp\left[-\frac{1}{6\lambda}\left(
 {\bf p}_\rho^{\prime 2}+{\bf p}_\lambda^{\prime 2}
 +{\bf p}_\rho^{2}+{\bf p}_\lambda^{2}\right)\right]\ 
 e^{-\gamma {\bf k}^2} \right\},\ \ {\rm where}\  \gamma = \Lambda^{-2}.
\label{eq:T.41}\end{eqnarray}
Following the same steps as above from (\ref{eq:T.20}), but now without
the $\alpha$-integral etc., one gets
\begin{eqnarray}
 G_{[0,0]} &=& (2\pi)^{-9 }(3\pi\lambda)^3 \widetilde{N}_3^2\
 \int d^3k_\rho \int d^3k_\lambda\ 
 \exp\left[-\frac{1}{12\lambda}\left( 
 {\bf k}_\rho^{2}+{\bf k}_\lambda^{2}\right)\right] 
 \cdot\nonumber\\ && \times
 \exp\left[-\gamma\left\{\frac{1}{2}{\bf k}_\rho^2 +\frac{1}{6}{\bf k}_\lambda^2 
 +\frac{1}{\sqrt{3}} {\bf k}_\rho\cdot{\bf k}_\lambda\right\}\right],
\label{eq:T.42}\end{eqnarray}
which reads in cartesian coordinates, see (\ref{eq:T.32}),
\begin{eqnarray}
 G_{[0,0]} &=& {\cal N}_{[0,0]} 
 \int d^3k \exp\left[-\frac{1}{6\lambda} {\bf k}^{2}\right]\
 \exp\left[-\gamma {\bf k}^2 \right] \nonumber\\
 &=& {\cal N}_{[0,0]} \left(\frac{6\lambda\pi}{1+6\gamma\lambda}\right)^{3/2}= 
  {\cal N}_{[0,0]} \Lambda^3\left(\frac{\pi}{1+\frac{1}{18} \Lambda^2 R_N^2}\right)^{3/2}.
\label{eq:T.43}\end{eqnarray}
The integral for the matrix element with an extra ${\bf k}^2$ is denoted as
$H_{[2,0]}$, which is 
\begin{eqnarray}
 H_{[2,0]}&=& G_{[0,0]}-m^2 H_{[0,0]}.  
\label{eq:T.44}\end{eqnarray}
The integral with a factor ${\bf k}^4$ in the integrand, i.e. $H_{[4,0]}$
is easily found as follows. We write 
${\bf k}^4/({\bf k}^2+m^2)= ({\bf k}^2-m^2+m^4/({\bf k}^2+m^2)$. The term with ${\bf k}^2$
leads to $G_{[2,0]}=-(d/d\gamma)G_{[0,0]}$ which is, see (\ref{eq:T.43}), 
\begin{equation}
 G_{[2,0]} = {\cal N}_{[0,0]} \frac{3}{2\pi}\left(\frac{6\lambda\pi}{1+6\gamma\lambda}\right)^{5/2}=
  {\cal N}_{[0,0]} \frac{3}{2\pi}\Lambda^5\left(\frac{\pi}{1+\frac{1}{18}\Lambda^2 R_N^2}\right)^{5/2}.
\label{eq:T.45}\end{equation}
Then we find
\begin{equation}
H_{[4,0]} = G_{[2,0]}-m^2 G_{[0,0]}+m^4 H_{[0,0]}.
\label{eq:T.46}\end{equation}

\noindent {\bf c.}\ Factor ${\bf q}^2=({\bf q}_1^2+{\bf q}_2^2)/2$ in $V_2$: 
The ${\bf q}$-integrals, see Eqn.~(\ref{eq:T.20}) gives the factor 
\begin{eqnarray}
 I_q &=& \int d^3q_1 \int d^3q_2\ \frac{1}{2}
 \left({\bf q}_1^2 + {\bf q}_2^2\right)
 \exp\left[-\frac{4}{6\lambda}\left( {\bf q}_1^2+{\bf q}_1\cdot{\bf q}_2+{\bf q}_2^2
 \right)\right] 
\label{eq:T.47}\end{eqnarray}
Using 
\begin{eqnarray*}
&& J= \int d^3q\ \int d^3q_2\ \exp\biggl[-\left(a {\bf q}_1^2+c {\bf q}_1\cdot{\bf q}_2
 +b {\bf q}_2^2\right) = \left(\frac{4\pi^2}{4ab-c^2}\right)^{3/2},
\end{eqnarray*}
one gets with a factor $\alpha {\bf q}_1^2+\beta {\bf q}_2^2+\gamma {\bf q}_1\cdot{\bf q}_2$
in the integrand
\begin{eqnarray*}
 J &\rightarrow& \frac{3}{8\pi^2}\bigl[4\alpha b + 4 \beta a-2\gamma c \bigr]\
\left(\frac{4\pi^2}{4ab-c^2}\right)^{5/2}
\end{eqnarray*}
Application to the integral (\ref{eq:T.47}) with $a=b=c=4/6\lambda$ and 
$\alpha=1/2, \beta=1/6, \gamma=0$ one gets 
 $I_q = 2\lambda\ (3\pi^2\lambda^2)^{3/2}$.
Therefore, after doing the ${\bf q}$-integrals we have 
\begin{eqnarray}
 H_{[0,2]} &=& 2\lambda {\cal N}_{[0,0]} \int_0^\infty\!\! d\alpha\ e^{-\alpha m^2}\!\!
 \int d^3k_1\!\!d^3k_2\ 
 \exp\left[-\frac{1}{6\lambda}\left( 
 {\bf k}_1^2+{\bf k}_1\cdot{\bf k}_2+{\bf k}_2^2\right)\right]\cdot e^{-\gamma {\bf k}^2}.
\label{eq:T.48}\end{eqnarray}
Then, comparing with the expression (\ref{eq:T.31}) for $H_{[0,0]}$ one gets  
\begin{eqnarray}
 H_{[0,2]} &=& 2\lambda\ {\cal N}_{[0,0]}\cdot 
 2\pi\sqrt{\pi}\ m\biggl[ \frac{1}{\sqrt{\pi A m^2}}
 -e^{A m^2} {\it Erfc}\bigl(\sqrt{A m^2}\bigr) \biggr]  
  = 2\lambda\ {\cal N}_{[0,0]}\ J_1.   
\label{eq:T.49}\end{eqnarray}

With a factor ${\bf q}^2{\bf k}^2$ in the integral, using again 
${\bf k}^2= ({\bf k}^2+m^2)-m^2$, we need $G_{[0,2]}$. Doing the ${\bf q}$-integral
we get 
\begin{eqnarray}
 G_{[0,2]} &=& 2\lambda\ {\cal N}_{[0,0]}
 \int d^3k \exp\left[-\frac{1}{6\lambda} {\bf k}^{2}\right]\
 \exp\left[-\gamma {\bf k}^2 \right] 
 = 2\lambda\ {\cal N}_{[0,0]} \left(\frac{6\lambda\pi}{1+6\gamma\lambda}\right)^{3/2}. 
\label{eq:T.49a}\end{eqnarray}
Then, it can be verified easily that
\begin{equation}
 H_{[2,2]} = G_{[0,2]}-m^2 H_{[0,2]}.
\label{eq:T.49b}\end{equation}

\noindent {\bf d.}\ Factor $q_i k_j$ in the integrand, which occurs for the 
spin-orbit, gives zero in the overlap integral.

\noindent {\bf e.}\ For the tensor the overlap integral is    
\begin{eqnarray}
 I_{ij} &=& \widetilde{N}_3^2\
 \int \frac{d^3p_\rho' d^3p_\lambda'}{(2\pi)^6} 
 \int \frac{d^3p_\rho d^3p_\lambda}{(2\pi)^6}\left\{
 \exp\left[-\frac{1}{6\lambda}({\bf p}_\rho^{\prime 2}+{\bf p}_\lambda^{\prime 2})
\right]
 \right.\cdot\nonumber\\ && \times\left.
 \exp\left[-\frac{1}{6\lambda}({\bf p}_\rho^{2}+{\bf p}_\lambda^{2})
\right]\ \frac{e^{-{\bf k}^2/\Lambda^2}}{{\bf k}^2+m^2}\right\}\cdot
  k_{1,i} k_{2,j}.                                          
\label{eq:T.51}\end{eqnarray}
In terms of Cartesian coordinates (\ref{eq:T.51}) reads
\begin{eqnarray}
 I_{ij} &=& \widetilde{N}_3^2\
 \int \frac{d^3q_1 d^3k_1}{(2\pi)^6} 
 \int \frac{d^3q_2 d^3k_2}{(2\pi)^6}\left\{
 \exp\left[-\frac{4}{6\lambda}({\bf q}_1^2+{\bf q}_1\cdot{\bf q}_2+{\bf q}_2^2)
\right]
 \right.\cdot\nonumber\\ && \times\left.
 \exp\left[-\frac{1}{6\lambda}({\bf k}_1^2+{\bf k}_1\cdot{\bf k}_2+{\bf k}_2^2)
\right]\ \frac{e^{-{\bf k}^2/\Lambda^2}}{{\bf k}^2+m^2}\right\} 
\cdot k_{1,i} k_{2,j}.
\label{eq:T.52}\end{eqnarray}
Performing the ${\bf q}$-integrations in (\ref{eq:T.52}) giving the expression
 ({\it why not factor $(3\pi\lambda)^{3/2}$ ?})
\begin{eqnarray*}
 I_{ij} &=&  (2\pi)^{-9 } (3\pi^2 \lambda^2)^{3/2}\widetilde{N}_3^2\
 \int d^3k_1 d^3k_2\ k_{1,i}k_{2,j}
 \cdot\nonumber\\ && \times\left\{
 \exp\left[-\frac{1}{6\lambda}({\bf k}_1^2+{\bf k}_1\cdot{\bf k}_2+{\bf k}_2^2)
\right]\ \frac{e^{-{\bf k}^2/\Lambda^2}}{{\bf k}^2+m^2}\right\}.
\end{eqnarray*}
Using ${\bf k}_2=-{\bf k}_1$, i.e. insert a factor $\delta({\bf k}_1+{\bf k}_2)$,
 the above expression reduces to
\begin{eqnarray}
&& I_{ij} = -{\cal N}_{[0,0]}
 \int d^3k \left\{ \exp\left[-\frac{1}{6\lambda}{\bf k}^2
\right]\ \frac{e^{-{\bf k}^2/\Lambda^2}}{{\bf k}^2+m^2}\right\}\cdot k_{i}k_{j}= 
\nonumber\\ && -{\cal N}_{[0,0]}
 \int_0^\infty d\alpha\ e^{-\alpha m^2} \int d^3
 e^{-(\alpha+A){\bf k}^2}\cdot k_{i}k_{j} =
\nonumber\\ &&  -{\cal N}_{[0,0]}\cdot
 \frac{1}{2\pi}\int_0^\infty d\alpha\ e^{-\alpha m^2}
 \left(\frac{\pi}{\alpha+A}\right)^{5/2}\ \delta_{ij}.
\label{eq:T.53}\end{eqnarray}
with $A=1/6\lambda+\gamma$, and where (\ref{eq:T.d.1}) is used in the last step.
This result shows that the tensor two-body interaction $V_2$ leads to
spin-spin term in the three-body matrix element.
The remaining $\alpha$-integral is related to $J_1$ in (\ref{eq:T.35b}) 
\begin{eqnarray}
 J_2(m,A) &\equiv& \frac{1}{2\pi}\int_0^\infty d\alpha\ e^{-\alpha m^2}
 \left(\frac{\pi}{\alpha+A}\right)^{5/2} = -\frac{1}{3}\frac{d}{dA} J_1(m,A)
 \nonumber\\ &=& 
  -\frac{1}{3} m^2\ \left[J_1(m,A) - \pi m\ (Am^2)^{-3/2}\right].
\label{eq:T.54}\end{eqnarray}
So, 
\begin{equation}
 I_{ij} = -{\cal N}_{[0,0]} J_2(m,A)\ \delta_{ij} \equiv H_{[1,1]}\ \delta_{i,j}.
\label{eq:T.55}\end{equation}
With this result the three-body integral of the tensor operator $P_3$ is
\begin{equation}
 H_3(m,\Lambda) = \biggl[H_{[1,1]}-\frac{1}{3} H_{[0,0]}\biggr]\ 
 \bm{\sigma}_1\cdot\bm{\sigma}_2
\label{eq:T.55a}\end{equation}

\noindent {\bf f.}\ Factor $q_{1,i}q_{2,j}$ in the integrand, which occurs in the
$P_5'$ Pauli-invariant, in cartesian coordinates the overlap integral is
\begin{eqnarray}
 I_{ij} &=& \widetilde{N}_3^2\ 
 \int \frac{d^3q_1 d^3k_1}{(2\pi)^6} 
 \int \frac{d^3q_2 d^3k_2}{(2\pi)^6}\left\{
 \exp\left[-\frac{4}{6\lambda}({\bf q}_1^2+{\bf q}_1\cdot{\bf q}_2+{\bf q}_2^2)
\right]
 \right.\cdot\nonumber\\ && \times\left.
 \exp\left[-\frac{1}{6\lambda}({\bf k}_1^2+{\bf k}_1\cdot{\bf k}_2+{\bf k}_2^2)
\right]\ \frac{e^{-{\bf k}^2/\Lambda^2}}{{\bf k}^2+m^2}\right\} 
\cdot q_{1,i} q_{2,j}.
\label{eq:T.71}\end{eqnarray}
The ${\bf q}$-integrations give a factor $-(\lambda/2)(3\pi^2\lambda^2)^{3/2}\delta_{ij}$,
and hence
\begin{eqnarray*}
 I_{ij} &=& -(\lambda/2) {\cal N}_{[0,0]}
 \int d^3k_1 d^3k_2\biggl\{
 \exp\left[-\frac{1}{6\lambda}({\bf k}_1^2+{\bf k}_1\cdot{\bf k}_2+{\bf k}_2^2)
\right]\ \frac{e^{-{\bf k}^2/\Lambda^2}}{{\bf k}^2+m^2}\biggr\}\ \delta_{ij}.
\label{eq:T.72}\end{eqnarray*}
Comparing the remaining ${\bf k}$-integrals with those for $H_{[0,0]}$ we find that
\begin{equation}
 I_{ij} = -(\lambda/2)\ H_{[0,0]}\ \delta_{ij}.
\label{eq:T.73}\end{equation}
With this result the three-body integral of the non-local tensor operator 
$P_5'$ is 
\begin{equation}
 H_5'(m,\Lambda) = 
 -9\lambda\ {\cal N}_{[0,0]}(\lambda)\ J_1(m,A)\ (\bm{\sigma}_1\cdot\bm{\sigma}_2)
\label{eq:T.73a}\end{equation}

\noindent {\bf g.}\ For the quadratic spin-orbit the overlap integral is    
\begin{eqnarray}
&& I_3(Q_{12})_{ij} = \widetilde{N}_3^2\
 \int \frac{d^3p_\rho' d^3p_\lambda'}{(2\pi)^6} 
 \int \frac{d^3p_\rho d^3p_\lambda}{(2\pi)^6}\left\{
 \exp\left[-\frac{1}{6\lambda}({\bf p}_\rho^{\prime 2}+{\bf p}_\lambda^{\prime 2})
\right]
 \right.\cdot\nonumber\\ && \times\left.
 \exp\left[-\frac{1}{6\lambda}({\bf p}_\rho^{2}+{\bf p}_\lambda^{2})
\right]\ \frac{e^{-{\bf k}^2/\Lambda^2}}{{\bf k}^2+m^2}\right\}\cdot
 ({\bf q}_1\times{\bf k}_1)_i ({\bf q}_2\times{\bf k}_2)_j
\label{eq:T.61}\end{eqnarray}
In terms of cartesian coordinates (\ref{eq:T.61}) reads
\begin{eqnarray}
&& I_3(Q_{12})_{ij} = \widetilde{N}_3^2\ 
 \int \frac{d^3q_1 d^3k_1}{(2\pi)^6} 
 \int \frac{d^3q_2 d^3k_2}{(2\pi)^6}\left\{
 \exp\left[-\frac{4}{6\lambda}({\bf q}_1^2+{\bf q}_1\cdot{\bf q}_2+{\bf q}_2^2)
\right]
 \right.\cdot\nonumber\\ && \times\left.
 \exp\left[-\frac{1}{6\lambda}({\bf k}_1^2+{\bf k}_1\cdot{\bf k}_2+{\bf k}_2^2)
\right]\ \frac{e^{-{\bf k}^2/\Lambda^2}}{{\bf k}^2+m^2}\right\}\ 
 ({\bf q}_1\times{\bf k}_1)_i ({\bf q}_2\times{\bf k}_2)_j
\label{eq:T.62}\end{eqnarray}
Working out the cross products we have
\begin{eqnarray*}
&& ({\bf q}_1\times{\bf k}_1)_i ({\bf q}_2\times{\bf k}_2)_j =
\varepsilon_{imn}\varepsilon_{jrs} k_{1,m}k_{2,r} q_{1,n} q_{2,s}
 \end{eqnarray*}
Then, for the overlap integral we use (\ref{eq:T.d.1}) 
\begin{eqnarray*}
 J_{ij}^{[12]}(a,b,c) &=& \int d^3k_1 d^3k_2\ (k_{1,i} k_{2,j})\ e^{-a{\bf k}_1^2
 -c {\bf k}_1\cdot{\bf k}_2-b {\bf k}_2^2} 
 \nonumber\\ &=&
   -\frac{c}{4\pi^2} \left(\frac{4\pi^2}{4ab-c^2}\right)^{5/2}\ 
 \delta_{ij} \equiv J_{[1,2]}(a,b,c) \delta_{ij}, 
 \end{eqnarray*}
for the ${\bf q}$-integrations in (\ref{eq:T.62}) giving the expression
\begin{eqnarray*}
 I_3(Q_{12})_{ij} &=& -(\lambda/2) 
 (2\pi)^{-3} (3\pi^2\lambda^2)^{3/2}\ \widetilde{N}_3^2\ 
 \int \frac{d^3k_1 d^3k_2}{(2\pi)^6}
 \varepsilon_{imn}\varepsilon_{jrn}\ k_{1,m}k_{2,r}
 \cdot\nonumber\\ && \times\left\{
 \exp\left[-\frac{1}{6\lambda}({\bf k}_1^2+{\bf k}_1\cdot{\bf k}_2+{\bf k}_2^2)
\right]\ \frac{e^{-{\bf k}^2/\Lambda^2}}{{\bf k}^2+m^2}\right\}.
\end{eqnarray*}
Using ${\bf k}_2=-{\bf k}_1$, i.e. insert a factor $\delta({\bf k}_1+{\bf k}_2)$,
 the above expression reduces to
\begin{eqnarray}
 I_3(Q_{12})_{ij} &=& +(\lambda/2) {\cal N}_{[0,0]}
 \int d^3k \left\{ \exp\left[-\frac{1}{6\lambda}{\bf k}^2
\right]\ \frac{e^{-{\bf k}^2/\Lambda^2}}{{\bf k}^2+m^2}\right\}\ 
 \varepsilon_{imn}\varepsilon_{jrn}\ k_{m}k_{r} \nonumber\\ &=& 
  +(\lambda/2) {\cal N}_{[0,0]}
 \int_0^\infty d\alpha\ e^{-\alpha m^2} \int \frac{d^3k}{(2\pi)^3}
 e^{-(\alpha+A){\bf k}^2}\ \varepsilon_{imn}\varepsilon_{jrn}\ k_{m}k_{r} 
 \nonumber\\ &=&
  +(\lambda/2) {\cal N}_{[0,0]}\cdot
 \frac{1}{2\pi}\int_0^\infty d\alpha\ e^{-\alpha m^2}
 \left(\frac{\pi}{\alpha+A}\right)^{5/2}\cdot 2\delta_{ij}
 \nonumber\\ &=& +\lambda\ {\cal N}_{[0,0]}\ J_2(m,A)\ \delta_{ij}.
\label{eq:T.63}\end{eqnarray}
with $A=1/6\lambda+\gamma$, and where (\ref{eq:T.d.1}) is used in the last step.
This result shows that the quadratic spin-orbit two-body interaction $V_2$ leads to
spin-spin term in the three-body matrix element.
The remaining $\alpha$-integral has been evaluated above, see (\ref{eq:T.54}).

\noindent With this result the three-body integral of the quadratic spin-orbit 
operator $P_5$ is
\begin{equation}
 H_5(m,\Lambda) = H_{Q_{12}}(m,A)\ (\bm{\sigma}_1\cdot\bm{\sigma}_2),
\label{eq:T.63a}\end{equation}
with the definition $I_3(Q_{12})_{ij}= H_{Q_{12}}(m,A)\ \delta_{ij}$.\\

\begin{flushleft}
\rule{16cm}{0.5mm}
\end{flushleft}
\section{Momentum integrals matrix elements}                      
\label{app:T.d}  
Integrals of matrix elements proportional to ${\bf k}_i$ and
${\bf q}_i$ give zero for s-wave nucleons.  
Terms quadratic and tetratic give non-zero results:\\
\noindent 1.\ The integrals with integrands proportional to two momenta
\begin{subequations}
\begin{eqnarray}
 I_{ij}(a) &=& \int d^3k\ (k_i k_j)\ e^{-a{\bf k}^2} = \frac{1}{2a}
 \left(\frac{\pi}{a}\right)^{3/2}\ \delta_{ij} \equiv I_1(a) \delta_{ij}, \\
 J_0(a,b,c) &=& \int d^3k_1 d^3k_2\ e^{-a{\bf k}_1^2
 -c {\bf k}_1\cdot{\bf k}_2-b {\bf k}_2^2} =
 \left(\frac{4\pi^2}{4ab-c^2}\right)^{3/2}, \\
 J_1(a,b,c) &=& \int d^3k_1 d^3k_2\ k_{1,i}\ e^{-a{\bf k}_1^2
 -c {\bf k}_1\cdot{\bf k}_2-b {\bf k}_2^2} 
 \nonumber\\ &=& -\lim_{{\bf d} \rightarrow 0}\ \nabla_{d,i}
 \int d^3k_1 d^3k_2\ e^{-a{\bf k}_1^2
 -c {\bf k}_1\cdot{\bf k}_2-b {\bf k}_2^2}\ e^{-{\bf d}\cdot{\bf k}_1}
 \nonumber\\ &=& -\lim_{{\bf d} \rightarrow 0}\ \nabla_{d,i}
\left(\frac{\pi}{a}\right)^{3/2}
 \int d^3k_2\ e^{-b {\bf k}_2^2}\ \exp\left[\frac{(c{\bf k}_2+{\bf d})^2}{4a}
 \right] \nonumber\\ &=& \frac{c}{2a}\left(\frac{\pi}{a}\right)^{3/2}
 \int d^3k_2\ k_{2,i} \exp\left[-\left(b-\frac{c^2}{4a}\right) {\bf k}_2^2\right]
 \rightarrow 0, \\
 J_{ij}^{[12]}(a,b,c) &=& \int d^3k_1 d^3k_2\ (k_{1,i} k_{2,j})\ e^{-a{\bf k}_1^2
 -c {\bf k}_1\cdot{\bf k}_2-b {\bf k}_2^2} 
 \nonumber\\ &=&
   -\frac{c}{4\pi^2} \left(\frac{4\pi^2}{4ab-c^2}\right)^{5/2}\ 
 \delta_{ij} \equiv J_{[1,2]}(a,b,c) \delta_{ij}, \\
  J_{ij}^{[11]}(a,b,c) &=& \int d^3k_1 d^3k_2\ (k_{1,i} k_{1,j})\ e^{-a{\bf k}_1^2
  -c {\bf k}_1\cdot{\bf k}_2-b {\bf k}_2^2} 
\nonumber\\ &=&
 +\frac{b}{2\pi^2} \left(\frac{4\pi^2}{4ab-c^2}\right)^{5/2}\ 
\delta_{ij} \equiv J_{[1,1]}(a,b,c) \delta_{ij}, \\
 J_{ij}^{[22]}(a,b,c) &=& \int d^3k_1 d^3k_2\ (k_{2,i} k_{2,j})\ e^{-a{\bf k}_1^2
 -c {\bf k}_1\cdot{\bf k}_2-b {\bf k}_2^2} \nonumber\\ &=&
 +\frac{a}{2\pi^2} \left(\frac{4\pi^2}{4ab-c^2}\right)^{5/2}\ 
\delta_{ij} \equiv J_{[2,2]}(a,b,c) \delta_{ij}.    
\label{eq:T.d.1}\end{eqnarray}
\end{subequations}
\noindent 2.\ For the integrals in the main text we use the same  
notation but it is understood that there are integrals over the
$(\alpha,\beta)$-parameters, i.e. 
\begin{subequations}
\label{eq:T.d.4}
\begin{eqnarray}
 J_{[i,j]} & \rightarrow & \int_0^\infty\! d\alpha \int_0^\infty\! d\beta\
 e^{-\alpha m_1^2} e^{-\beta m_2^2}\ J_{[i,j]}(a,b,c), \\
 K_{[i,j]}^{(1,2)} & \rightarrow & \int_0^\infty\! 
 d\alpha \int_0^\infty\! d\beta\ 
 e^{-\alpha m_1^2} e^{-\beta m_2^2}\ K_{[i,j]}^{(1,2)}(a,b,c), \\
 H_{[i,j]} & \rightarrow & \frac{4}{\pi} \int_0^\infty\! 
 \frac{d\alpha}{\sqrt{\alpha}} 
 \int_0^\infty\! \frac{d\beta}{\sqrt{\beta}}\
 e^{-\alpha m_1^2} e^{-\beta m_2^2}\ H_{[i,j]}(a,b,c).   
\end{eqnarray}
\end{subequations}

\begin{flushleft}
\rule{16cm}{0.5mm}
\end{flushleft}
\section{Generalized D\&D-model}
\label{app:GDD}
In this appendix we consider distinctive gaussian wave functions for the
initial and final state. This enables one to treat the case where the
the wave functions are a sum of gaussians with parameters 
$\lambda_i, i=1..N$. This is akin to description of wave functions in the
GEM-approach \cite{Hiy03}.
Then, for $\Psi_{3N} = \sum_i \psi_{3N}(\lambda_i)$ 
the matrix elements are
\begin{eqnarray*}
&& \langle \Psi_{3N}|V_3| \Psi_{3N}\rangle = \sum_{i,j=1}^N
   \langle \psi_{3N}(\lambda_i)|V_3| \psi_{3N}(\lambda_j). 
\end{eqnarray*}

\noindent Here, we consider the The momentum space wave functions are
\begin{subequations}
\begin{eqnarray}
 \widetilde{\psi}_{3N,i}({\bf p}_\rho,{\bf p}_\lambda) &=&  
 \widetilde{N}_3 \exp\left[-\frac{1}{6\lambda}\left({\bf p}_\rho^2+
 {\bf p}_\lambda^2\right)\right],\ \ {\rm with}\ 
 \widetilde{N}_3 = \left(\frac{4\pi}{3\lambda}\right)^{3/2}, \\
 \widetilde{\psi}_{3N,f}({\bf p}'_\rho,{\bf p}'_\lambda) &=&  
 \widetilde{N}'_3 \exp\left[-\frac{1}{6\lambda'}\left({\bf p}_\rho^{\prime 2}+
 {\bf p}_\lambda^{\prime 2}\right)\right],\ \ {\rm with}\ 
 \widetilde{N}'_3 = \left(\frac{4\pi}{3\lambda'}\right)^{3/2}.   
\label{app:GDD.1}\end{eqnarray}
\end{subequations}

\noindent The generalized basic integral is
\begin{eqnarray}
 G_3 &=& \widetilde{N}'_3\ \widetilde{N}_3\
 \int d^3p_\rho' d^3p_\lambda' \int d^3p_\rho d^3p_\lambda\left\{
 \exp\left[-\frac{1}{6\lambda'}({\bf p}_\rho^{\prime 2}+{\bf p}_\lambda^{\prime 2})
\right]
 \exp\left[-\frac{1}{6\lambda}({\bf p}_\rho^{2}+{\bf p}_\lambda^{2})
\right]
 \right.\cdot\nonumber\\ && \times\left.
\frac{e^{-{\bf k}_1^2/\Lambda_1^2}}{{\bf k}_1^2+m_1^2}
\frac{e^{-{\bf k}_2^2/\Lambda_2^2}}{{\bf k}_2^2+m_2^2}\ 
e^{-{\bf k}_3^2/\Lambda_3^2}
\right\}\nonumber\\ &=& 
 \widetilde{N}_3^2\ \int_0^\infty d\alpha\ \int_0^\infty d\beta\
 e^{-\alpha m_1^2} e^{-\beta m_2^2}\cdot
 \int d^3p_\rho' d^3p_\lambda' \int d^3p_\rho d^3p_\lambda\cdot\nonumber\\ 
 \cdot\nonumber\\ && \times\left\{
 \exp\left[-\frac{1}{6\lambda'}\left(
 {\bf p}_\rho^{\prime 2}+{\bf p}_\lambda^{\prime 2}\right)-\frac{1}{6\lambda}
  \left({\bf p}_\rho^{2}+{\bf p}_\lambda^{2}\right)\right]\ 
 e^{-\gamma_1 {\bf k}_1^2} e^{-\gamma_2 {\bf k}_2^2} e^{-\gamma_3 {\bf k}_3^2}\right\},
\label{app:GDD.2}\end{eqnarray}
where $\gamma_1 = \alpha+\Lambda_1^{-2}$, $\gamma_2 = \beta+\Lambda_2^{-2}$. 
and $\gamma_3 = \Lambda_2^{-2}$.\\

\noindent Changing the $({\bf k},{\bf q})$-integration variables and 
expressing everything in the $(\rho,\lambda)$-variables we write for 
(\ref{app:GDD.2})
\begin{eqnarray*}
 G_3 &=& \widetilde{N}'_3\ \widetilde{N}_3\
 \int_0^\infty d\alpha\ \int_0^\infty d\beta\
 e^{-\alpha m_1^2} e^{-\beta m_2^2}\cdot
 \int d^3q_\rho d^3k_\rho \int d^3q_\lambda d^3k_\lambda\cdot\nonumber\\ 
 \cdot\nonumber\\ && \times
 \exp\left[-\frac{1}{6\lambda'}\left( {\bf q}_\rho^2+{\bf k}_\rho^2/4
 +{\bf q}_\rho\cdot{\bf k}_\rho\right)
 -\frac{1}{6\lambda}\left( {\bf q}_\rho^2+{\bf k}_\rho^2/4
 -{\bf q}_\rho\cdot{\bf k}_\rho\right)\right] 
 \cdot\nonumber\\ && \times
 \exp\left[-\frac{1}{6\lambda'}\left( {\bf q}_\lambda^2+{\bf k}_\lambda^2/4
 +{\bf q}_\lambda\cdot{\bf k}_\lambda\right)
 -\frac{1}{6\lambda}\left( {\bf q}_\lambda^2+{\bf k}_\lambda^2/4
 -{\bf q}_\lambda\cdot{\bf k}_\lambda\right)\right] 
\cdot\nonumber\\ && \times 
 \exp\left[-\left\{\frac{1}{2}(\gamma_1+\gamma_2) {\bf k}_\rho^2
 +\frac{1}{6}(\gamma_1+\gamma_2+4\gamma_3) {\bf k}_\lambda^2 
 +\frac{1}{\sqrt{3}}(\gamma_1-\gamma_2) 
 {\bf k}_\rho\cdot{\bf k}_\lambda\right\}\right].
\label{app:GDD.3}\end{eqnarray*}
{\bf Note}: We remark that in this generalized D\&D-model the terms
proportional to the ${\bf q}_i$ vectors no longer vanish doing the 
momentum space integrations.\\

\noindent Using the notations $\mu=1/\lambda$ and $\mu'=1/\lambda'$ we have
\begin{eqnarray}
 G_3 &=& \widetilde{N}'_3\ \widetilde{N}_3\
 \int_0^\infty d\alpha\ \int_0^\infty d\beta\
 e^{-\alpha m_1^2} e^{-\beta m_2^2}\cdot
 \int d^3q_\rho d^3k_\rho \int d^3q_\lambda d^3k_\lambda\cdot\nonumber\\ 
 \cdot\nonumber\\ && \times
 \exp\left[-\frac{1}{6}\left\{(\mu'+\mu)({\bf q}_\rho^2+{\bf k}_\rho^2/4)
 +(\mu'-\mu){\bf q}_\rho\cdot{\bf k}_\rho\right\} \right] 
 \cdot\nonumber\\ && \times
 \exp\left[-\frac{1}{6}\left\{(\mu'+\mu)({\bf q}_\lambda^2+{\bf k}_\lambda^2/4)
 +(\mu'-\mu){\bf q}_\lambda\cdot{\bf k}_\lambda\right\} \right] 
\cdot\nonumber\\ && \times 
 \exp\left[-\left\{\frac{1}{2}(\gamma_1+\gamma_2) {\bf k}_\rho^2
 +\frac{1}{6}(\gamma_1+\gamma_2+4\gamma_3) {\bf k}_\lambda^2 
 +\frac{1}{\sqrt{3}}(\gamma_1-\gamma_2) 
 {\bf k}_\rho\cdot{\bf k}_\lambda\right\}\right].
\label{app:GDD.4}\end{eqnarray}
\noindent 1.\ The basic integral is 
\begin{eqnarray}
 H_0 &=& \widetilde{N}'_3\ \widetilde{N}_3\
 \int d^3q_\rho d^3k_\rho \int d^3q_\lambda d^3k_\lambda 
 \cdot\nonumber\\ && \times
 \exp\left[-\frac{1}{6}\left\{(\mu'+\mu)({\bf q}_\rho^2+{\bf k}_\rho^2/4)
 +(\mu'-\mu){\bf q}_\rho\cdot{\bf k}_\rho\right\} \right] 
 \cdot\nonumber\\ && \times
 \exp\left[-\frac{1}{6}\left\{(\mu'+\mu)({\bf q}_\lambda^2+{\bf k}_\lambda^2/4)
 +(\mu'-\mu){\bf q}_\lambda\cdot{\bf k}_\lambda\right\} \right] 
\cdot\nonumber\\ && \times 
 \exp\left[-\left\{\frac{1}{2}(\gamma_1+\gamma_2) {\bf k}_\rho^2
 +\frac{1}{6}(\gamma_1+\gamma_2+4\gamma_3) {\bf k}_\lambda^2 
 +\frac{1}{\sqrt{3}}(\gamma_1-\gamma_2) 
 {\bf k}_\rho\cdot{\bf k}_\lambda\right\}\right] 
\nonumber\\ &=& \widetilde{N}'_3\ \widetilde{N}_3\
 \left(\frac{6\pi}{(\mu'+\mu)}\right)^{3/2}\
 \int d^3k_\rho \int d^3k_\lambda\ 
 \exp\left[-\frac{1}{6}\frac{\mu'\mu}{\mu'+\mu} 
 ({\bf k}_\rho^2+{\bf k}_\lambda^2)\right]
 \cdot\nonumber\\ && \times 
 \exp\left[-\left\{\frac{1}{2}(\gamma_1+\gamma_2) {\bf k}_\rho^2
 +\frac{1}{6}(\gamma_1+\gamma_2+4\gamma_3) {\bf k}_\lambda^2 
 +\frac{1}{\sqrt{3}}(\gamma_1-\gamma_2) 
 {\bf k}_\rho\cdot{\bf k}_\lambda\right\}\right] 
\label{app:GDD.5}\end{eqnarray}
\noindent 2.\ With e.g. a component of the ${\bf q}_\rho$-vector 
in the integrand we define the integral
\begin{eqnarray}
 {\bf H}({\bf q}_\rho) &\equiv& 
 \lim_{{\bf d}\rightarrow 0}\widetilde{N}'_3\ \widetilde{N}_3\
 \int d^3q_\rho  d^3q_\lambda \int d^3k_\rho d^3k_\lambda\cdot
 F({\bf k}_\rho,{\bf k}_\lambda)\cdot 
 {\bf q}_\rho\ e^{-{\bf d}\cdot{\bf q}_\rho}
 \cdot\nonumber\\ && \times
 \exp\left[-\frac{1}{6}\left\{(\mu'+\mu){\bf q}_\rho^2
 +(\mu'-\mu){\bf q}_\rho\cdot{\bf k}_\rho\right\} \right] 
 \cdot\nonumber\\ && \times
 \exp\left[-\frac{1}{6}\left\{(\mu'+\mu){\bf q}_\lambda^2
 +(\mu'-\mu){\bf q}_\lambda\cdot{\bf k}_\lambda\right\} \right] 
\label{app:GDD.6}\end{eqnarray}
Here we first make the move ${\bf q}_\rho \rightarrow -\bm{\nabla}_d$ and 
execute the $d^3q_\rho$-integral, which gives
\begin{eqnarray*}
&& \left(\frac{6\pi}{\mu'+\mu}\right)^{3/2}\ \exp\left[\frac{1}{24(\mu'+\mu)}
\left\{(\mu'-\mu) {\bf k}_\rho+6{\bf d}\right\}^2\right].
\end{eqnarray*}
Then, 
\begin{eqnarray*}
&& \lim_{{\bf d} \rightarrow 0} \bm{\nabla}_d \Rightarrow 
 \left(\frac{6\pi}{\mu'+\mu}\right)^{3/2}\ \exp\left[
 \frac{(\mu'-\mu)^2}{24(\mu'+\mu)} {\bf k}_\rho^2\right]\cdot
 \frac{(\mu'-\mu)}{2(\mu'+\mu)}\ {\bf k}_\rho.
\end{eqnarray*}
Performing also the $d^3q_\lambda$-integration we arrive at
\begin{eqnarray}
 {\bf H}({\bf q}_\rho) &=& \widetilde{N}'_3\ \widetilde{N}_3\
 \left(\frac{6\pi}{\mu'+\mu}\right)^{3}\ 
 \int d^3q_\rho  d^3q_\lambda \int d^3k_\rho d^3k_\lambda\cdot
 F({\bf k}_\rho,{\bf k}_\lambda) 
 \cdot\nonumber\\ && \times
 \exp\left[\frac{(\mu'-\mu)^2}{24(\mu'+\mu)} {\bf k}_\rho^2\right]\cdot
 \exp\left[\frac{(\mu'-\mu)^2}{24(\mu'+\mu)} {\bf k}_\lambda^2\right]\cdot
 \frac{(\mu'-\mu)}{2(\mu'+\mu)}\ {\bf k}_\rho, 
\label{app:GDD.7}\end{eqnarray}
and a similar expression for ${\bf H}({\bf q}_\lambda)$.
It is easy to verify that 
${\bf H}({\bf k}_\rho)= {\bf H}({\bf k}_\lambda)=0$.\\
 
\noindent 3.\ With bilinear components of ${\bf k}_\rho$
and ${\bf k}_\lambda$, in the integrand we obtain results similar to
those for the case $\mu'=\mu$. Comparing the basic integral (\ref{app:GDD.4})
with that for $\mu'=\mu$ in Eqn.~\ref{eq:T.21} we see that the change is
\begin{eqnarray*}
&& \frac{1}{12\lambda} \rightarrow 
 \frac{1}{6}\frac{\mu'\mu}{\mu'+\mu}\ {\rm or}\ 
 \lambda \rightarrow \frac{(\mu'+\mu)}{2\mu'\mu}= 
 \frac{1}{2}(\lambda'+\lambda).
\end{eqnarray*}
Then, using again the formula
\begin{eqnarray*}
&& \int d^3 k_\rho d^3 k_\lambda\ e^{-a{\bf k}_\rho^2
 -c {\bf k}_\rho\cdot{\bf k}_\lambda -b {\bf k}_\lambda^2} =
 \left(\frac{4\pi}{4ab-c^2}\right)^{3/2}, 
\end{eqnarray*}
with,               
\begin{subequations}
\label{app:GDD.8}
\begin{eqnarray}
 a &\equiv& \frac{1}{2}\left(A+\alpha+\beta\right),\ 
 A= \frac{\mu'\mu}{3(\mu'+\mu)}
 +(\hat{\gamma}_1 +\hat{\gamma}_2), \\
 b &\equiv& \frac{1}{6}\left(B+\alpha+\beta\right),\ 
 B= \frac{\mu'\mu}{(\mu'+\mu)}
 +(\hat{\gamma}_1 +\hat{\gamma}_2+4\hat{\gamma}_3), \\
 c &\equiv& \frac{1}{\sqrt{3}}\left[C+(\alpha-\beta)\right],\ 
 C=(\hat{\gamma}_1-\hat{\gamma}_2), 
 \end{eqnarray}
\end{subequations}
where again $\gamma_1 = \hat{\gamma}_1+\alpha+\beta$, 
$\gamma_2 = \hat{\gamma}_2+\alpha+\beta$, and $\gamma_3 = \hat{\gamma}_3$.

\noindent With this result we finally obtain, 
\begin{eqnarray}
 G_3 &=& \widetilde{N}'_3\ \widetilde{N}_3\
\left(\frac{6\pi}{(\mu'+\mu)}\right)^{3} 
 \int_0^\infty d\alpha\ \int_0^\infty d\beta\
 e^{-\alpha m_1^2} e^{-\beta m_2^2}\cdot\nonumber\\ && \times 
 \left(\frac{12\pi}{(A+\alpha+\beta)(B+\alpha+\beta)-(C+\alpha-\beta)^2}\right)^{3/2}, 
\label{app:GDD.9}\end{eqnarray}
For the $J_{[\lambda,\lambda]},J_{[\lambda,\rho]}$, and $J_{[\rho,\rho]}$,
similar to the case $\mu'=\mu$ the formulas given in Appendix~\ref{app:T.d}
apply.

\noindent {\bf 4.\ $H_0$ in Cartesian momenta:} 
Recalling the inverse of (\ref{eq:T.17}) 
\begin{eqnarray}
 {\bf k}_\lambda = \sqrt{\frac{3}{2}}({\bf k}_1+{\bf k}_2)\ ,\ 
 {\bf k}_\rho = \sqrt{\frac{1}{2}}({\bf k}_1-{\bf k}_2), 
\label{app:GDD.15}\end{eqnarray}
we write (\ref{app:GDD.4}) into the form
\begin{eqnarray}
 H_0 &=& \widetilde{N}'_3\ \widetilde{N}_3\
 \left(\frac{6\pi}{(\mu'+\mu)}\right)^{3}\
 \int d^3k_\rho \int d^3k_\lambda\ 
 \exp\left[-\frac{1}{6}\frac{\mu'\mu}{\mu'+\mu} 
 ({\bf k}_\rho^2+{\bf k}_\lambda^2)\right]
 \cdot\nonumber\\ && \times 
 \exp\left[-\left\{(\gamma_1+\gamma_3) {\bf k}_1^2
 +(\gamma_1+\gamma_3) {\bf k}_2^2 +2\gamma_3 {\bf k}_1\cdot{\bf k}_2 
\right\}\right] \nonumber\\
 &=& (\sqrt{3})^3\widetilde{N}'_3\ \widetilde{N}_3\
 \left(\frac{6\pi}{(\mu'+\mu)}\right)^{3}\
 \int d^3k_1 \int d^3k_2\ 
 \exp\left[-\frac{1}{6}\frac{\mu'\mu}{\mu'+\mu} 
 ({\bf k}_1^2+{\bf k}_1\cdot{\bf k}_2+{\bf k}_2^2)\right]
 \cdot\nonumber\\ && \times 
 \exp\left[-\left\{(\gamma_1+\gamma_3) {\bf k}_1^2
 +(\gamma_1+\gamma_3) {\bf k}_2^2 +2\gamma_3 {\bf k}_1\cdot{\bf k}_2 
\right\}\right] \nonumber\\
  &=& \widetilde{N}'_3\ \widetilde{N}_3\
 \left(\frac{6\pi}{(\mu'+\mu)}\right)^{3}\
 \left(\frac{12\pi}{4ab-c^2}\right)^{3/2}, 
\label{app:GDD.16}\end{eqnarray}
where 
\begin{subequations}
\begin{eqnarray}
 a &=& \alpha + \frac{1}{6\lambda_{red}}
 +\hat{\gamma}_1+\hat{\gamma}_3 \equiv A_c+\alpha, \\
 b &=& \beta  + \frac{1}{6\lambda_{red}}
 +\hat{\gamma}_2+\hat{\gamma}_3 \equiv B_c+\beta, \\
 c &=& \frac{1}{6\lambda_{red}}+2 \hat{\gamma}_3 \equiv C_c, 
\label{app:GDD.17}\end{eqnarray}
\end{subequations}
with 
 $\lambda_{red} = (\mu^\prime+\mu)/(2\mu^\prime´\mu)$.
Analogous to (\ref{app:GDD.8}),
\begin{eqnarray}
 G_3 &=& \widetilde{N}'_3\ \widetilde{N}_3\
\left(\frac{6\pi}{(\mu'+\mu)}\right)^{3} 
 \int_0^\infty d\alpha\ \int_0^\infty d\beta\
 e^{-\alpha m_1^2} e^{-\beta m_2^2}\cdot\nonumber\\ && \times 
 \left(\frac{12\pi}{4(A_c+\alpha)(B_c+\beta)-C_c^2}\right)^{3/2}, 
\label{app:GDD.18}\end{eqnarray}

\begin{flushleft}
\rule{16cm}{0.5mm}
\end{flushleft}
\section{One-Boson-Exchange Quark-quark Potentials}
\label{app:OBE}
\subsection{Non-strange Meson-exchange}
\label{app:OBE.a}
In this section we treat non-strange meson exchange. The strange meson exchange 
is readily obtained using the prescriptions given in \cite{CW10502} for
the strange meson exchange potentials.\\ 

\noindent {\bf Two-body system:}
In the two-body center of mass system (CM), we denoted the initial- and final-state 
CM-momenta by ${\bf p}_i$ and ${\bf p}_f$.
Using rotational invariance and parity conservation we expand
the $V$-matrix, which is a $4\times 4$-matrix in Pauli-spinor space,
 into a complete set of Pauli-spinor invariants (\cite{MRS89,SNRV71}) 
Introducing the momenta           
\begin{equation}
  {\bf q}=\frac{\small{1}}{\small{2}}({\bf p}_{f}+{\bf p}_{i}),\hspace{2em}
  {\bf k}={\bf p}_{f}-{\bf p}_{i},\hspace{2em}
  {\bf n}={\bf p}_{i}\times {\bf p}_{f}={\bf q}\times {\bf k}
\label{eq:LS.9}\end{equation}
with , of course, ${\bf n}={\bf q}\times {\bf k}$,
we choose for the operators $P_{i}$ in spin-space
\begin{subequations}
\label{eq:LS.10}
\begin{eqnarray}
  P_{1} &=& 1\ , \\ 
  P_{2} &=& \mbox{\boldmath $\sigma$}_1\cdot\mbox{\boldmath $\sigma$}_2\ , \\       
  P_{3} &=& (\mbox{\boldmath $\sigma$}_1\cdot{\bf k})
  (\mbox{\boldmath $\sigma$}_2\cdot{\bf k})
  -\frac{1}{3}(\mbox{\boldmath $\sigma$}_1\cdot\mbox{\boldmath $\sigma$}_2)
  {\bf k}^{2}\ , \\
  P_{4} &=& \frac{i}{2}(\mbox{\boldmath $\sigma$}_1+\mbox{\boldmath $\sigma$}_2)
  \cdot{\bf n}\ , \\[0.1cm]
  P_{5} &=& (\mbox{\boldmath $\sigma$}_1\cdot{\bf n})(\mbox{\boldmath $\sigma$}_2
  \cdot{\bf n})\ , \\
  P_{6} &=& \frac{i}{2} (\mbox{\boldmath $\sigma$}_1-\mbox{\boldmath $\sigma$}_2)
  \cdot{\bf n}\ ,  \\[0.1cm]
  P_{7} &=& (\mbox{\boldmath $\sigma$}_1\cdot{\bf q})(\mbox{\boldmath $\sigma$}_2
  \cdot{\bf k})
   +(\mbox{\boldmath $\sigma$}_1\cdot{\bf k}) (\mbox{\boldmath $\sigma$}_2\cdot{\bf q})\
  \\[0.1cm]
  P_{8} &=& (\mbox{\boldmath $\sigma$}_1\cdot{\bf q})(\mbox{\boldmath $\sigma$}_2
  \cdot{\bf k})
   -(\mbox{\boldmath $\sigma$}_1\cdot{\bf k}) (\mbox{\boldmath $\sigma$}_2\cdot{\bf q})\ .
\end{eqnarray}
\end{subequations}
Here we follow \cite{MRS89,SNRV71}, except that we have chosen
here $P_{3}$ to be a purely `tensor-force' operator.
For the axial-vector mesons there also occurs the invariant 
$P'_{5} = (\bm{\sigma}_1\cdot{\bf q}(\bm{\sigma}_2\cdot{\bf q})
-(\bm{\sigma}_1\cdot\bm{\sigma}_2)\ {\bf q}^2/3$, see \cite{ESC16a} for its treatment.
For the non-strange mesons the mass differences at the vertices are neglected,
we take at the $YYM$- and the $NNM$-vertex the average hyperon and the average
nucleon mass respectively. This implies that we do not include contributions
to the Pauli-invariants $P_7$ and $P_8$.
Then, the potentials are expanded as
\begin{equation}
  V=\sum_{i=1}^{6}V_{i}({\bf k}\,^{2},{\bf q}\,^{2}) P_{i}\ .
\label{eq:LS.11}\end{equation}
For the non-strange quarks also the antisymmetric spin-orbit we will neglect.\\

\noindent {\bf Three-body system:}
The generalization of the Pauli-invariants from the two-body- to a N-body-system, 
in particular to a three-body system is as follows.
In the three-body system it is appropriate to introduce, e.g. for the 12-subsystem
 the momenta
\begin{subequations} \label{eq:tb.1}
\begin{eqnarray}
&& {\bf q}_1 = \frac{1}{2}({\bf p}_1'+{\bf p}_1,\ {\bf k}_1={\bf p}_1'-{\bf p}_1, \\
&& {\bf q}_2 = \frac{1}{2}({\bf p}_2'+{\bf p}_2,\ {\bf k}_2={\bf p}_2'-{\bf p}_2     
\end{eqnarray} \end{subequations}
For the $V_{12;3}$ potential momentum conservation 
${\bf p}_1+{\bf p}_2= {\bf p}_1'+{\bf p}_2'$ gives ${\bf k}_2=-{\bf k}_1$, and
therefore in the expressions below for the $\Omega^{(X)}_i$, where $X=P,V,S,D,A$,
${\bf k} \equiv {\bf k}_1=-{\bf k}_2$. Since for the two-body 12-subsystem 
${\bf q}_1 \neq {\bf q}_2$ for the three-body system we have the generalization
\begin{subequations} \label{eq:tb.2}
\begin{eqnarray}
&& (\bm{\sigma}_1+\bm{\sigma}_2)\cdot{\bf q}\times{\bf k} \rightarrow \frac{1}{2}\bigl[
   \bm{\sigma}_1\cdot{\bf q}_1\times{\bf k}_1 + \bm{\sigma}_2\cdot{\bf q}_2\times{\bf k}_2 \bigr], \\
&& \bm{\sigma}_1\cdot{\bf q}\times{\bf k}\ \bm{\sigma}_2\cdot{\bf q}\times{\bf k} 
 \rightarrow 
 \bm{\sigma}_1\cdot{\bf q}_1\times{\bf k}_1\ \bm{\sigma}_2\cdot{\bf q}_2\times{\bf k}_2 
\end{eqnarray} \end{subequations}
As for the non-local potentials, which are related to the ${\bf q}^2$-terms,
we note that in the three-body system for $V_{12;3}$ we must take
${\bf q}^2=({\bf q}_1^2+{\bf q}_2^2)/2$. Accordingly, the potentials are splitted as
$V_i({\bf k},{\bf q})= V_{i,a}({\bf k}+({\bf q}_1^2+{\bf q}_2^2+{\bf k}^2/2)\ V_{i,b}/2$.  
The appropriate Pauli-invariants for the 12-subsystem in an N-body system are
we choose for the operators $P_{i}$ in spin-space
\begin{subequations}
\label{eq:LS.10b}
\begin{eqnarray}
  P_{1} &=& 1\ ,  
  P_{2}  =  \mbox{\boldmath $\sigma$}_1\cdot\mbox{\boldmath $\sigma$}_2\ , \\       
  P_{3} &=& -(\mbox{\boldmath $\sigma$}_1\cdot{\bf k}_1)
  (\mbox{\boldmath $\sigma$}_2\cdot{\bf k}_2)
  +\frac{1}{3}(\mbox{\boldmath $\sigma$}_1\cdot\mbox{\boldmath $\sigma$}_2)
  ({\bf k}_1\cdot{\bf k}_2)\ , \\
  P_{4} &=& \frac{i}{2}(\mbox{\boldmath $\sigma$}_1\cdot{\bf n}_1
  +\mbox{\boldmath $\sigma$}_2\cdot{\bf n}_2)\ , 
  P_{5}  =  (\mbox{\boldmath $\sigma$}_1\cdot{\bf n}_1)(\mbox{\boldmath $\sigma$}_2
  \cdot{\bf n}_2).      
\end{eqnarray}
\end{subequations}
We skipped here $P_6, P_7, P_8$ since we do not use them in this paper.
Note that these $P_i$ are chosen such that they correspond to the set in (\ref{eq:LS.10})
in the case that ${\bf k}_1 = -{\bf k}_2={\bf k}$ and ${\bf q}_1=-{\bf q}_2={\bf q}$.\\
The potentials for the 12-subsystem are expanded as
\begin{equation}
  V=\sum_{i=1}^{5}V_{i}({\bf k}_1,{\bf k}_2;{\bf q}_1,{\bf q}_2)\ P_{i}.
\label{eq:LS.11b}\end{equation}

\noindent {\bf Listing non-strange meson exchange $\Omega^{(X)}_i\ (X=P,V,S,D,A,B)$:}
\begin{enumerate}
 \item[(a)]   Pseudoscalar-meson exchange:
      \begin{subequations}
      \begin{eqnarray}
       \Omega^{(P)}_{2a} & = & -g^p_{13}g^p_{24}\left( \frac{{\bf k}^{2}}
           {12M_yM_n} \right) \ \ ,\ \ 
       \Omega^{(P)}_{3a}  =  -g^p_{13}g^p_{24}\left( \frac{1}
           {4M_yM_n}  \right), \label{eq1a} \\
       \Omega^{(P)}_{2b} & = & +g^p_{13}g^p_{24}\left( \frac{{\bf k}^{2}}
           {24M_y^2M_n^2} \right) \ \ ,\ \ 
       \Omega^{(P)}_{3b}  =  +g^p_{13}g^p_{24}\left( \frac{1}
           {8M_y^2M_n^2}  \right), \label{eq1b}.    
         \end{eqnarray}
       \end{subequations}
 \item[(b)]   Vector-meson exchange:
     \begin{eqnarray}  
       \Omega^{(V)}_{1a}&=&
   \left\{g^v_{13}g^v_{24}\left( 1-\frac{{\bf k}^{2}}{2M_yM_n}\right)
           -g^v_{13}f^v_{24}\frac{{\bf k}^{2}}{4{\cal M}M_n} 
      -f^v_{13}g^v_{24}\frac{{\bf k}^{2}}{4{\cal M}M_y}
 \vphantom{\frac{A}{A}}\right. \nonumber\\ && \left. \vphantom{\frac{A}{A}}
           +f^v_{13}f^v_{24}\frac{{\bf k}^{4}}
           {16{\cal M}^{2}M_yM_n}\right\},\ \                 
    \Omega^{(V)}_{1b} =  g^v_{13}g^v_{24}\left(
    \frac{3}{2M_yM_n}\right), \nonumber\\
  \Omega^{(V)}_{2a} &=& -\frac{2}{3} {\bf k}^{2}\,\Omega^{(V)}_{3a}, \ \ 
  \Omega^{(V)}_{2b}  =  -\frac{2}{3} {\bf k}^{2}\,\Omega^{(V)}_{3b}, 
 \nonumber\\
    \Omega^{(V)}_{3a}&=& \left\{
           (g^v_{13}+f^v_{13}\frac{M_y}{{\cal M}})
           (g^v_{24}+f^v_{24}\frac{M_n}{{\cal M}}) 
          -f^v_{13}f^v_{24}\frac{{\bf k}^{2}}{8{\cal M}^{2}} \right\}
            /(4M_yM_n), \nonumber\\                 
    \Omega^{(V)}_{3b}&=& -
           (g^v_{13}+f^v_{13}\frac{M_y}{{\cal M}})
           (g^v_{24}+f^v_{24}\frac{M_n}{{\cal M}}) 
            /(8M_y^2M_n^2), \nonumber\\                 
    \Omega^{(V)}_{4}&=&-\left\{12g^v_{13}g^v_{24}+8(g^v_{13}f^v_{24}+f^v_{13}g^v_{24})
           \frac{\sqrt{M_yM_n}}{{\cal M}} 
     - f^v_{13}f^v_{24}\frac{3{\bf k}^{2}}{{\cal M}^{2}}\right\}
            /(8M_yM_n)              \nonumber\\
       \Omega^{(V)}_{5}&=&- \left\{
           g^v_{13}g^v_{24}+4(g^v_{13}f^v_{24}+f^v_{13}g^v_{24})
           \frac{\sqrt{M_yM_n}}{{\cal M}}  
           +8f^v_{13}f^v_{24}\frac{M_yM_n}{{\cal M}^{2}}\right\}
          /(16M_y^{2}M_n^{2})        \nonumber\\
       \Omega^{(V)}_{6}&=&-\left\{(g^v_{13}g^v_{24}
           +f^v_{13}f^v_{24}\frac{{\bf k}^{2}}{4{\cal M}^{2}})
    \frac{(M_n^{2}-M_y^{2})}{4M_y^{2}M_n^{2}} 
      -(g^v_{13}f^v_{24}-f^v_{13}g^v_{24})
      \frac{1}{\sqrt{{\cal M}^{2}M_yM_n}}\right\}.
 \nonumber\\
 \label{eq2}\end{eqnarray}
 \item[(c)]   Scalar-meson exchange:  \hspace{2em}
      \begin{eqnarray} 
      \Omega^{(S)}_{1a} & = & 
      -g^s_{13} g^s_{24} \left( 1+\frac{{\bf k}^{2}}{4M_yM_n}\right)
  \ \ ,\ \ 
      \Omega^{(S)}_{1b}  =  +g^s_{13} g^s_{24} \left(\frac{1}{2M_yM_n}\right)
       \nonumber\\ &&\nonumber\\
      \Omega^{(S)}_{4} &=& -g^s_{13} g^s_{24} \left(\frac{1}{2M_yM_n}\right)
  \hspace{7mm} \ \ ,\ \ 
      \Omega^{(S)}_{5} = g^s_{13} g^s_{24}
        \left(\frac{1}{16M_y^{2}M_n^{2} }\right) 
\nonumber\\ 
      \Omega^{(S)}_{6} &=& -g^s_{13} g^s_{24}
        \frac{(M_n^{2}-M_y^{2})}{4M_y^2M_n^2}.
       \label{Eq:scal} \end{eqnarray}
\item[(d)] Axial-vector-exchange $J^{PC}=1^{++}$:
      \begin{eqnarray} 
      \Omega^{(A)}_{2a} & = & -g^a_{13}g^a_{24}\left[
         1-\frac{2{\bf k}^2}{3M_yM_n}\right]
         +\left[\left(g_{13}^A f_{24}^A\frac{M_n}{{\cal M}}
         +f_{13}^A g_{24}^A \frac{M_y}{{\cal M}}\right)
         -f_{13}^A f_{24}^A \frac{{\bf k}^2}{2{\cal M}^2}\right]\
         \frac{{\bf k}^2}{6M_yM_n}
       \nonumber\\ && \nonumber\\
      \Omega^{(A)}_{2b} &=& 
        -g^a_{13}g^a_{24} \left(\frac{3}{2M_yM_n}\right) 
        \nonumber\\ && \nonumber\\
      \Omega^{(A)}_{3}&=&
        -g^a_{13}g^a_{24} \left[\frac{1}{4M_yM_n}\right]
         +\left[\left(g_{13}^A f_{24}^A\frac{M_n}{{\cal M}}
         +f_{13}^A g_{24}^A \frac{M_y}{{\cal M}}\right)
         -f_{13}^A f_{24}^A \frac{{\bf k}^2}{2{\cal M}^2}\right]\
         \frac{1}{2M_yM_n}
       \nonumber\\ && \nonumber\\
	\Omega^{(A)}_{4}  &=&
     -g^a_{13}g^a_{24}   \left[\frac{1}{2M_yM_n}\right] 
      \ \ ,\ \
      \Omega^{(A)}_{6} = 
     -g^a_{13}g^a_{24} \left[\frac{(M_n^{2}-M_y^{2})}{4M_y^2M_n^2}\right]
       \nonumber\\ && \nonumber\\
      \Omega^{(A)'}_{5} & = &
     -g^a_{13}g^a_{24}   \left[\frac{2}{M_yM_n}\right] 
         \label{eq:axi1} \end{eqnarray}
Here, we used the B-field description with $\alpha_r=1$, 
see \cite{ESC16a} Appendix A.              
 The detailed treatment of the potential proportional to $P_5'$, i.e. 
 with $\Omega_5^{(A)'}$, is given in \cite{ESC16a}, Appendix~B.
\item[(e)] Axial-vector mesons with $J^{PC}=1^{+-}$: 
      \begin{eqnarray} 
       \Omega^{(B)}_{2a} & = & +f^B_{13}f^B_{24}\frac{(M_n+M_y)^2}{m_B^2}
       \left(1-\frac{{\bf k}^2}{4M_yM_n}\right)
       \left( \frac{{\bf k}^{2}}{12M_yM_n} \right), \nonumber\\ 
       \Omega^{(B)}_{2b}  &=&  +f^B_{13}f^B_{24}\frac{(M_n+M_y)^2}{m_B^2}
       \left( \frac{{\bf k}^{2}}{8M_y^2M_n^2} \right)     
     \nonumber\\ 
       \Omega^{(B)}_{3a} & = & +f^B_{13}f^B_{24}\frac{(M_n+M_y)^2}{m_B^2}
       \left(1-\frac{{\bf k}^2}{4M_yM_n}\right)
       \left( \frac{1}{4M_yM_n} \right), \nonumber\\    
       \Omega^{(B)}_{3b}  &=&  +f^B_{13}f^B_{24}\frac{(M_n+M_y)^2}{m_B^2}
       \left( \frac{3}{8M_y^2M_n^2} \right). \nonumber\\     
     \label{eq:bxi1} \end{eqnarray}
 \item[(f)]   Diffractive-exchange (pomeron, $f, f', A_{2}$): \\
      \begin{eqnarray} 
      \Omega^{(D)}_{1a} & = & 
      +g^d_{13} g^d_{24} \left( 1+\frac{{\bf k}^{2}}{4M_yM_n}\right)
  \ \ ,\ \ 
      \Omega^{(D)}_{1b}  =  -g^d_{13} g^d_{24} \left(\frac{1}{2M_yM_n}\right)
       \nonumber\\ &&\nonumber\\
      \Omega^{(D)}_{4} &=& +g^d_{13} g^d_{24} \left(\frac{1}{2M_yM_n}\right)
  \hspace{7mm} \ \ ,\ \ 
      \Omega^{(D)}_{5} = -g^d_{13} g^d_{24}
        \left(\frac{1}{16M_y^{2}M_n^{2} }\right) 
       \nonumber\\ &&\nonumber\\
      \Omega^{(D)}_{6} &=& +g^d_{13} g^d_{24}
        \frac{(M_n^{2}-M_y^{2})}{4M_y^2M_n^2}.
       \label{Eq:diff} \end{eqnarray}
\item[(g)] Odderon-exchange:              
         The $\Omega^{O}_{i}$ are the same as for vector-meson-exchange
         Eq.(ref{eq2}), but with
         $ g_{13}^{V}\rightarrow g_{13}^{O}$, 
         $ f_{13}^{V}\rightarrow f_{13}^{O}$ and similarly for the couplings
         with the 24-subscript.

\end{enumerate}

As in Ref.~\cite{MRS89} in the derivation of the expressions for $\Omega_i^{(X)}$, 
given above, $M_y$ and $M_n$ denote the mean hyperon and nucleon
mass, respectively \begin{math} M_y=(M_{1}+M_{3})/2 \end{math}
and \begin{math} M_n=(M_{2}+M_{4})/2 \end{math},
 and $m$ denotes the mass of the exchanged meson.
Moreover, the approximation                            
        \begin{math}
              1/ M^{2}_{N}+1/ M^{2}_{Y}\approx
              2/ M_nM_y,
        \end{math}
is used, which is rather good since the mass differences
between the baryons are not large.

\subsection{Non-strange Meson Momentum-space Potentials I}
\label{app:OBE.b}
The local potentials are given below.\\

\noindent {\bf (a)} Pseudoscalar-meson exchange:
\begin{eqnarray}
 V^{(P)}_{12;3}({\bf k},{\bf q}) &=& -g^p_{13}g^p_{24}\
 \biggl\{{\bf k}^2 P_2+ 3 P_3\biggr\} 
  \left(\frac{1}{12M_yM_n}\right)\ G_0({\bf k}^2,\Lambda_P^2).
\label{app.obeps}\end{eqnarray}
\noindent {\bf (b)} Vector-meson exchange:
\begin{eqnarray}
&& V^{(V)}_{12;3}({\bf k},{\bf q}) =  g^v_{13} g^v_{24}\ \biggl(
   \biggl\{\left( 1-\frac{{\bf k}^{2}}{2M_yM_n}\right)
           -\left(\kappa^v_{24}\frac{M_y}{\cal M} 
      +\kappa^v_{13}\frac{M_n}{\cal M}\right) \frac{{\bf k}^{2}}{4M_nM_y}
\nonumber\\ && 
  +\kappa^v_{13}\kappa^v_{24}\frac{{\bf k}^{4}}{16{\cal M}^{2}M_yM_n}\biggr\}                     
    -\frac{2}{3}\biggl\{ (1+\kappa^v_{13}\frac{M_y}{{\cal M}})
           (1+\kappa^v_{24}\frac{M_n}{{\cal M}}) 
          -\kappa^v_{13}\kappa^v_{24}\frac{{\bf k}^{2}}{8{\cal M}^{2}} \biggr\}\ 
 \frac{{\bf k}^2}{4M_yM_n}\ P_2
\nonumber\\ && 
    +\biggl\{ (1+\kappa^v_{13}\frac{M_y}{{\cal M}}) (1+\kappa^v_{24}\frac{M_n}{{\cal M}}) 
   -\kappa^v_{13}\kappa^v_{24}\frac{{\bf k}^{2}}{8{\cal M}^{2}} \biggr\}\ P_3 /(4M_yM_n)           
\nonumber\\ && 
    -\biggl\{12+8(\kappa^v_{24}+\kappa^v_{13}) \frac{\sqrt{M_yM_n}}{{\cal M}} 
     - \kappa^v_{13}\kappa^v_{24}\frac{3{\bf k}^{2}}{{\cal M}^{2}}\biggr\} P_4/(8M_yM_n) 
\nonumber\\ && 
       -\biggl\{ 1+4(\kappa^v_{24}+\kappa^v_{13}) \frac{\sqrt{M_yM_n}}{{\cal M}}  
           +8\kappa^v_{13}\kappa^v_{24}\frac{M_yM_n}{{\cal M}^{2}}\biggr\} P_5/(16M_y^{2}M_n^{2})
\nonumber\\ && 
      -\biggl\{(1+\kappa^v_{13}\kappa^v_{24}\frac{{\bf k}^{2}}{4{\cal M}^{2}})
    \frac{(M_n^{2}-M_y^{2})}{4M_y^{2}M_n^{2}} 
      -(\kappa^v_{24}-\kappa^v_{13})/(\sqrt{{\cal M}^{2}M_yM_n})\biggr\}\ P_6
  \biggr) \nonumber\\ && \times G_0({\bf k}^2,\Lambda_V^2).
\label{app.obevc1}\end{eqnarray}
\noindent {\bf (c)} Scalar-meson exchange:
\begin{eqnarray}
 V^{(S)}_{12;3}({\bf k},{\bf q}) &=& -g^s_{13}g^s_{24} \biggl(
      \left( 1+\frac{{\bf k}^{2}}{4M_yM_n}\right)
      +\left[\frac{1}{2M_yM_n}\right]\ P_4
      -\left[\frac{1}{16M_y^{2}M_n^{2} }\right]\ P_5
      \biggr) \nonumber\\ && \times G_0({\bf k}^2,\Lambda_S^2). 
\label{app.obesc1}\end{eqnarray}
\noindent {\bf (d)} Axial-vector-meson exchange $J^{PC}=1^{++}$:
\begin{eqnarray}
 V^{(A)}_{12;3}({\bf k},{\bf q}) &=& -g^a_{13}g^a_{24} \biggl( \bigg\{
       \left[ 1-\frac{2{\bf k}^2}{3M_yM_n}\right]
         -\left[\left(\kappa_{24}^a \frac{M_n}{{\cal M}}
         +\kappa_{13}^a  \frac{M_y}{{\cal M}}\right)
         -\kappa_{13}^a \kappa_{24}^a \frac{{\bf k}^2}{2{\cal M}^2}\right]\
         \frac{{\bf k}^2}{6M_yM_n}\biggr\}\ P_2 
      \nonumber\\ && +\biggl\{
         \left[\frac{1}{4M_yM_n}\right]
         -\left[\left( \kappa_{24}^a\frac{M_n}{{\cal M}}
         +\kappa_{13}^a \frac{M_y}{{\cal M}}\right)
         -\kappa_{13}^a \kappa_{24}^A \frac{{\bf k}^2}{2{\cal M}^2}\right]\
         \frac{1}{2M_yM_n}\biggr\}\ P_3
       \nonumber\\ &&     
     +\left[\frac{1}{2M_yM_n}\right]\ P_4  
     +\left[\frac{(M_n^{2}-M_y^{2})}{4M_y^2M_n^2}\right]\ P_6
     +\left[\frac{2}{M_yM_n}\right]\ P'_5\biggl)\ G_0({\bf k}^2,\Lambda_A^2).
\label{app.obeax1}\end{eqnarray}
\noindent {\bf (e)} Axial-vector-meson exchange $J^{PC}=1^{+-}$:
\begin{eqnarray}
 V^{(B)}_{12;3}({\bf k},{\bf q}) &=& + 
       f^B_{13}f^B_{24}\frac{(M_n+M_y)^2}{m_B^2}
       \left(1-\frac{{\bf k}^2}{4M_yM_n}\right)
       \left( \frac{{\bf k}^{2}}{12M_yM_n}\right)  
       \cdot\nonumber\\ && \times
       \biggl\{P_2 + 3 P_3\biggr\}\ G_0({\bf k}^2,\Lambda_B^2).
\label{app.obeax2}\end{eqnarray}
\noindent {\bf (f)} Diffractive exchange $J^{PC}=0^{++}$:
\begin{eqnarray}
 V^{(D)}_{12;3}({\bf k},{\bf q}) &=& +g^d_{13}g^d_{24} \biggl(
      \left( 1+\frac{{\bf k}^{2}}{4M_yM_n}\right)
      +\left[\frac{1}{2M_yM_n}\right]\ P_4
      -\left[\frac{1}{16M_y^{2}M_n^{2} }\right]\ P_5
      \biggr) \nonumber\\ && \times \exp\biggl[-{\bf k}^2/{\cal M}^2\biggr].
\label{app.diffr1}\end{eqnarray}

\subsection{Non-strange Meson Momentum-space Potentials II}
\label{app:OBE.c}
As for the non-local potentials, which are related to the ${\bf q}^2$-terms,
we note the following. In the three-body system for $V_{12;3}$ we must take
${\bf q}^2=({\bf q}_1^2+{\bf q}_2^2)/2$. Accordingly, the potentials are splitted as
$V_i({\bf k},{\bf q})= V_{i,a}({\bf k}+({\bf q}_1^2+{\bf q}_2^2+{\bf k}^2/2)\ V_{i,b}/2$

\noindent {\bf (a)} Pseudoscalar-meson exchange:
\begin{eqnarray}
&& V^{(P)}_{12;3}({\bf k},{\bf q}) = 
 V^{(P)}_{12;3}({\bf k},{\bf q}) +
\frac{1}{2}({\bf q}_1^2+{\bf q}_2^2+{\bf k}^2/2)\ g^p_{13}g^p_{24}
\cdot\nonumber\\ && \times  
  \biggl\{ {\bf k}^2 P_2 + 3 P_3\biggr\}
  \left[\frac{{\bf k}^2}{24M_yM_n}\right]\ G_0({\bf k}^2,\Lambda_P^2).
\label{app.obeps2}\end{eqnarray}
\noindent {\bf (b)} Vector-meson exchange:
\begin{eqnarray}
 && V^{(V)}_{12;3}({\bf k},{\bf q}) = 
 V^{(V)}_{12;3}({\bf k},{\bf q})  
 -\frac{1}{2}({\bf q}_1^2+{\bf q}_2^2+{\bf k}^2/2) 
 \cdot\nonumber\\ && \times  \biggl(
     (g^v_{13}+f^v_{13}\frac{M_y}{{\cal M}})
     (g^v_{24}+f^v_{24}\frac{M_n}{{\cal M}})
  \biggl\{-\frac{2}{3}{\bf k}^2 P_2+ P_3\bigg\}
            /(8M_y^2M_n^2)\biggr)\ G_0({\bf k}^2,\Lambda_V^2).
\label{app.obevc2}\end{eqnarray}
\noindent {\bf (c)} Scalar-meson exchange:
\begin{eqnarray}
 && V^{(S)}_{12;3}({\bf k},{\bf q}) = V^{(S)}_{12;3}({\bf k},{\bf q})     
 +\frac{1}{2}({\bf q}_1^2+{\bf q}_2^2+{\bf k}^2/2)\
      g^s_{13} g^s_{24} \biggl(\frac{1}{2M_yM_n}\biggr)\ G_0({\bf k}^2,\Lambda_S^2).
\label{app.obesc2}\end{eqnarray}
\noindent {\bf (d)} Axial-vector-meson exchange $J^{PC}=1^{++}$:
\begin{eqnarray}
 && V^{(A)}_{12;3}({\bf k},{\bf q}) = 
 V^{(A)}_{12;3}({\bf k},{\bf q})  
 -\frac{1}{2}({\bf q}_1^2+{\bf q}_2^2+{\bf k}^2/2)\ 
      g^a_{13} g^a_{24}\ \biggl(\frac{3}{2M_yM_n}\biggr)\ P_2\ G_0({\bf k}^2,\Lambda_A^2).
\label{app.obeax3}\end{eqnarray}
\noindent {\bf (e)} Axial-vector-meson exchange $J^{PC}=1^{+-}$:
\begin{eqnarray}
 && V^{(B)}_{12;3}({\bf k},{\bf q}) = 
 V^{(B)}_{12;3}({\bf k},{\bf q})     
 +\frac{1}{2}({\bf q}_1^2+{\bf q}_2^2+{\bf k}^2/2) 
 \cdot\nonumber\\ && \times  \biggl(
       f^B_{13}f^B_{24}\frac{(M_n+M_y)^2}{m_B^2}\biggr)\biggl\{P_2+3 P_3\biggr\}
       \left( \frac{{\bf k}^{2}}{8M_y^2M_n^2}\right)\ G_0({\bf k}^2,\Lambda_B^2).
\label{app.obeax6}\end{eqnarray}
\noindent {\bf (f)} Diffractive exchange $J^{PC}=0^{++}$:
\begin{eqnarray}
 V^{(D)}_{12;3}({\bf k},{\bf q}) &=& V^{(D)}_{12;3}({\bf k},{\bf q})      
 -\frac{1}{2}({\bf q}_1^2+{\bf q}_2^2+{\bf k}^2/2)\
      g^d_{13} g^d_{24} \biggl(\frac{1}{2M_yM_n}\biggr)\ 
 \exp\biggl[-{\bf k}^2/{\cal M}^2\biggr].
\label{app.diffr2}\end{eqnarray}
\begin{flushleft}
\rule{16cm}{0.5mm}
\end{flushleft}
\section{Additional One-Boson-Exchange QQ-Potentials}
\label{app:OBE2}
The extra vertices at the quark-level generate additional OBE-potentials.  
Neglecting the ${\bf k}^4$ etc terms we obtain the following contributions:
\begin{enumerate}
 \item[(a)]   Pseudoscalar-meson exchange: no additional potentials.
 \item[(b)]   Vector-meson exchange:
     \begin{eqnarray}  
       \Delta\Omega^{(V)}_{1a}&=&
   -\bigl\{g^v_{13}f^v_{24}+f^v_{13}g^v_{24}\bigr]\ 
    \frac{{\bf k}^2}{4{\cal M}m_Q},\ \ 
    \Delta\Omega^{(V)}_{1b} =  0, \nonumber\\          
  \Delta\Omega^{(V)}_{2a} &=& -\frac{2}{3} {\bf k}^{2}\,\Delta\Omega^{(V)}_{3a}=0, \ \ 
  \Delta\Omega^{(V)}_{2b}  =  -\frac{2}{3} {\bf k}^{2}\,\Delta\Omega^{(V)}_{3b}=0, 
 \nonumber\\
    \Delta\Omega^{(V)}_{3a}&=& -\left\{
            (g^v_{13}+f^v_{13}\frac{M_y}{{\cal M}})\
           f^v_{24}\left(1+\frac{M_y}{m_Q}\right)
           +(g^v_{24}+f^v_{24}\frac{M_n}{{\cal M}})\
           f^v_{13}\left(1+\frac{M_n}{m_Q}\right) \right\}\
            \frac{{\bf k}^2}{4{\cal M} m_Q}/(4M_yM_n), \nonumber\\                 
    \Delta\Omega^{(V)}_{4}&=& +\biggl\{
     \left(3+2\frac{\sqrt{M_yM_n}}{m_Q}\right)(g^v_{13}f^v_{24}+f^v_{13}g^v_{24})
     +4f^v_{13}f^v_{24}\frac{\sqrt{M_yM_n}}{{\cal M}}\biggr\}
     \left(\frac{{\bf k}^2}{4{\cal M}m_Q}\right)/(2M_yM_n), 
                        \nonumber\\
    \Delta\Omega^{(V)}_{5}&=& +\biggl\{
     \left(1+4\frac{\sqrt{M_yM_n}}{m_Q}\right)(g^v_{13}f^v_{24}+f^v_{13}g^v_{24})
     +8f^v_{13}f^v_{24}\frac{\sqrt{M_yM_n}}{{\cal M}}\biggr\}
     \left(\frac{{\bf k}^2}{4{\cal M}m_Q}\right)/(16M_y^2M_n^2), 
                        \nonumber\\
       \Delta\Omega^{(V)}_{6}&=& 0.                      
 \label{obe2.1}\end{eqnarray}
 \item[(c)]   Scalar-meson exchange:  \hspace{2em}
      \begin{eqnarray} 
      \Delta\Omega^{(S)}_{1a} & = & 
      -g^s_{13} g^s_{24}\ \frac{{\bf k}^2}{2m_Q^2}\ ,\        
      \Delta\Omega^{(S)}_{1b}  = 0,                                    
       \nonumber\\ &&\nonumber\\
      \Delta\Omega^{(S)}_{4} &=& -g^s_{13} g^s_{24}\ 
       \frac{{\bf k}^2}{4m_Q^2}\left[\frac{1}{M_y^2M_n^2}\right]\ ,\
      \Delta\Omega^{(S)}_{5} = g^s_{13} g^s_{24}\
        \frac{{\bf k}^2}{4m_Q^2}\left[\frac{1}{8M_y^2M_n^2}\right], 
       \nonumber\\ &&\nonumber\\
      \Delta\Omega^{(S)}_{6} &=& -g^s_{13} g^s_{24}\ 
        \frac{(M_n^{2}-M_y^{2})}{4M_y^2M_n^2}\ \frac{{\bf k}^2}{2m_Q^2}.
 \label{obe2.2}\end{eqnarray}
 \item[(d)]   Axial-vector-meson exchange: \hspace{2em}
      \begin{eqnarray} 
      \Delta\Omega^{(A)}_{4} & = & 
      +g^a_{13} g^a_{24}\ \left[\frac{4}{M_yM_n}\right].           
 \label{obe2.3}\end{eqnarray}
\end{enumerate}
The transcription to configuration space potentials of these additional 
Pauli-invariants is similar to that in section~\ref{app:OBE} 
and is readily done. 
\begin{flushleft}
\rule{16cm}{0.5mm}
\end{flushleft}

\section{Isospin- and Spin-operators in Three-Quark Space}              
\label{app:iso}  
\noindent 1.\ Baryon octet $J^P=(1/2)^+$ 3 spin-isospin quark wave functions are of the 
symmetric form
\begin{eqnarray}
 \Psi_B &=& \frac{1}{\sqrt{2}}\left(\vphantom{\frac{A}{A}} 
 \phi_{M,S}\otimes\chi_{M,S}+\phi_{M,A}\otimes\chi_{M,A}\right),
\label{eq:iso.1}\end{eqnarray}
where in $\phi_{M,S}$ and $\phi_{M,A}$ the isospin of the 12-subsystems, 
which in the case of the nucleon is 1 and 0 respectively, see e.g. \cite{Close79}. 
In Table~\ref{tab.iso3} the explicit states are given. 
Similarly for the spin wave functions 
$\chi_{M,S}$ and $\chi_{M,A}$.\\
\begin{table}[hbt]
\squeezetable
\begin{tabular}{c|c|c}  \hline\hline 
   & $\phi_{M,S}$  & $ \phi_{M,A}$ \\[2mm] \hline
 "P" & $+\frac{1}{\sqrt{6}}\left[\vphantom{\frac{A}{A}} \left(
 u_1 d_2+d_1u_2\right) u_3 - 2 u_1 u_2 d_3\right]$ &
 $\frac{1}{\sqrt{2}}\left(u_1 d_2-d_1 u_2\right) u_3$ \\[2mm]
 "N" & $-\frac{1}{\sqrt{6}}\left[\vphantom{\frac{A}{A}} \left(
 p_1 d_2+d_1p_2\right) d_3 - 2 d_1 d_2 u_3\right]$ &
 $\frac{1}{\sqrt{2}}\left(u_1 d_2-d_1 u_2\right) d_3$ \\[2mm]
 \hline\hline 
\end{tabular}
\caption{Isospin states for the proton (P) and the neutron (N).}          
\label{tab.iso3}    
\end{table}
The nucleon  consists of three (constituent) quarks, which are 
in the ground state has J=1/2, and T=1/2. The ground-state is symmetric
w.r.t. the $(L,S,I)$ quantum numbers for the permutation of the quarks.   
It is antisymmetric in color.
The total symmetric spin-isospin state we generate by application of the 
symmetrizer ${\cal S}$ to e.g. the state
\begin{equation}
 \Psi_0 = u^{\uparrow} d^{\uparrow} u^{\downarrow}.
\label{eq:iso.2}\end{equation}
Using the S$_3$-projection operator one has 
\begin{equation}
  {\cal S} = \sum_{p_i \in S_3} p_i,
\label{eq:iso.3}\end{equation}
where $\delta_i$ is the signum of the permutation $p_i$. 
The 6 permutations $p_i$ of S$_3$, listed according to the 
conjugation classes, are
\begin{equation}
 S_3 : e; (12), (13); (23), (123), (132). 
\label{eq:iso.4}\end{equation}
Then, the fully symmetrized "P"-state is
\begin{eqnarray}
 \Psi &=& {\cal S} \Psi_0 = 
 \frac{1}{\sqrt{6}}\left\{\vphantom{\frac{A}{A}}
 u^{\uparrow} d^{\uparrow}u^{\downarrow}
 +d^{\uparrow} u^{\uparrow}u^{\downarrow}
 +u^{\downarrow} nd{\uparrow}u^{\uparrow}
 +u^{\uparrow} u^{\downarrow} d^{\uparrow}
 +u^{\downarrow} u^{\uparrow} d^{\uparrow}
 +d^{\uparrow} u^{\downarrow} u^{\uparrow}
 \vphantom{\frac{A}{A}} \right\}.
\label{eq:iso.5}\end{eqnarray}
It is easily verified that (\ref{eq:iso.5}) coincides with (\ref{eq:iso.1}).\\

\noindent 2.\ The matrix elements of the spin-operators
can easily be evaluated explicitly. Using
\begin{eqnarray}
 \bm{\sigma}_i\cdot\bm{\sigma}_j &=& 
 2\left(\sigma_{+,i} \sigma_{-,j}+\sigma_{-,i}\sigma_{+,j}\right) 
+\sigma_{i,z} \sigma_{j,z} 
\label{eq:iso.6}\end{eqnarray}
we derive, working things out for the "P"-state,
\begin{subequations}
\label{eq:iso.7}
\begin{eqnarray}
 \bm{\sigma}_1\cdot\bm{\sigma}_2\ \chi_{M,A} &=& -3 \chi_{M,A}\ ,\ 
 \bm{\sigma}_1\cdot\bm{\sigma}_2\ \chi_{M,S} = \chi_{M,S}, \\ 
 \bm{\sigma}_1\cdot\bm{\sigma}_3\ \chi_{M,A} &=& +\sqrt{3}\ \chi_{M,S}\ ,\ 
 \bm{\sigma}_1\cdot\bm{\sigma}_3\ \chi_{M,S} = \frac{1}{\sqrt{3}}\ \chi_{M,A} 
 +\frac{4}{\sqrt{3}}\ \chi^{\prime}_{M,A}, \\
 \bm{\sigma}_2\cdot\bm{\sigma}_3\ \chi_{M,A} &=& -\sqrt{3}\ \chi_{M,S}\ ,\ 
 \bm{\sigma}_2\cdot\bm{\sigma}_3\ \chi_{M,S} = \frac{1}{\sqrt{3}}\ \chi_{M,A} 
 +\frac{4}{\sqrt{3}}\ \chi^{\prime\prime}_{M,A}, 
\end{eqnarray}
\end{subequations}
where 
\begin{eqnarray}
 \chi_{M,A}^{\prime} &=& \frac{1}{\sqrt{2}}\left(u_1 u_2 d_3- d_1 u_2u_3\right)\ ,\ 
 \chi_{M,A}^{\prime\prime} = 
 \frac{1}{\sqrt{2}} u_1 \left(u_2 d_3- d_2u_3\right),    
\label{eq:iso.8}\end{eqnarray}
with the matrix elements
\begin{subequations}
\label{eq:iso.9}
\begin{eqnarray}
 \chi^\dagger_{M,A}\ \chi^{\prime}_{M,A} &=& +\frac{1}{2}\ ,\ 
 \chi^\dagger_{M,S}\ \chi^{\prime}_{M,A} = -\frac{1}{2}\sqrt{3}, \\
 \chi^\dagger_{M,A}\ \chi^{\prime\prime}_{M,A} &=& -\frac{1}{2}\ ,\ 
 \chi^\dagger_{M,S}\ \chi^{\prime\prime}_{M,A} = -\frac{1}{2}\sqrt{3}.    
\end{eqnarray}
\end{subequations}
The individual matrix elements are
\begin{subequations}
\label{eq:spin.15}
\begin{eqnarray}
 \left(\chi_{M,S}|\mbox{\boldmath $\sigma$}_1\cdot
 \mbox{\boldmath $\sigma$}_2|\chi_{M,S}\right) = +1 &,&
 \left(\chi_{M,A}|\mbox{\boldmath $\sigma$}_1\cdot
 \mbox{\boldmath $\sigma$}_2|\chi_{M,A}\right) = -3, \\
 \left(\chi_{M,S}|\mbox{\boldmath $\sigma$}_1\cdot
 \mbox{\boldmath $\sigma$}_2|\chi_{M,A}\right) =  0 &,&
 \left(\chi_{M,A}|\mbox{\boldmath $\sigma$}_1\cdot
 \mbox{\boldmath $\sigma$}_2|\chi_{M,S}\right) =  0, \\
 \left(\chi_{M,S}|\mbox{\boldmath $\sigma$}_1\cdot
 \mbox{\boldmath $\sigma$}_3|\chi_{M,S}\right) = -2 &,&
   \left(\chi_{M,A}|\mbox{\boldmath $\sigma$}_1\cdot
 \mbox{\boldmath $\sigma$}_3|\chi_{M,A}\right) =  0,\\
 \left(\chi_{M,S}|\mbox{\boldmath $\sigma$}_1\cdot
 \mbox{\boldmath $\sigma$}_3|\chi_{M,A}\right) = +\sqrt{3} &,&
   \left(\chi_{M,A}|\mbox{\boldmath $\sigma$}_1\cdot
 \mbox{\boldmath $\sigma$}_3|\chi_{M,S}\right) = +\sqrt{3},\\
 \left(\chi_{M,S}|\mbox{\boldmath $\sigma$}_2\cdot
 \mbox{\boldmath $\sigma$}_3|\chi_{M,S}\right) = -2 &,&
   \left(\chi_{M,A}|\mbox{\boldmath $\sigma$}_2\cdot
 \mbox{\boldmath $\sigma$}_3|\chi_{M,A}\right) =  0,\\
 \left(\chi_{M,S}|\mbox{\boldmath $\sigma$}_2\cdot
 \mbox{\boldmath $\sigma$}_3|\chi_{M,A}\right) = -\sqrt{3} &,&
   \left(\chi_{M,A}|\mbox{\boldmath $\sigma$}_2\cdot
 \mbox{\boldmath $\sigma$}_3|\chi_{M,S}\right) =-\sqrt{3}. 
\end{eqnarray}
\end{subequations}
These matrix elements apply to all $J^P=(1/2)^+$-baryons.

\noindent 4.\ Baryon octet $J^P=(1/2)^+$ spin-isospin matrix elements:
 From the baryon wave function (\ref{eq:iso.1}) one has
\begin{eqnarray}
 \left(\Psi_B|\ O_I\ O_S\ |\Psi_B\right) &=& \frac{1}{2}\left\{
 \vphantom{\frac{A}{A}} 
 \left(\phi_{M,S}| O_I|\phi_{M,S}\right)
 \left(\chi_{M,S}| O_S|\chi_{M,S}\right)
\right.\nonumber\\ && \left. \hspace{5mm}
 +\left(\phi_{M,S}| O_I|\phi_{M,A}\right)
 \left(\chi_{M,S}| O_S|\chi_{M,A}\right)
\right.\nonumber\\ && \left. \hspace{5mm}
 +\left(\phi_{M,A}| O_I|\phi_{M,S}\right)
 \left(\chi_{M,A}| O_S|\chi_{M,S}\right)
\right.\nonumber\\ && \left. \hspace{5mm}
 +\left(\phi_{M,A}| O_I|\phi_{M,A}\right)
 \left(\chi_{M,A}| O_S|\chi_{M,A}\right)
 \vphantom{\frac{A}{A}} \right\}.
\label{eq:spin.17} \end{eqnarray}
 
\noindent {\bf 5.\ P}: The isospin matrix elements are similar to the spin-operator
matrix element. This leads to the proton matrix elements of the isospin-spin operators:  
\begin{subequations} \label{eq:spin.16}
\begin{eqnarray}
&& \left(\Psi_N| \bm{\tau}_1\cdot\bm{\tau}_2 |\Psi_N\right) =  
   \left(\Psi_N| \bm{\tau}_1\cdot\bm{\tau}_3 |\Psi_N\right) =  
   \left(\Psi_N| \bm{\tau}_2\cdot\bm{\tau}_3 |\Psi_N\right) = -1, \\
&& \left(\Psi_N| \bm{\sigma}_1\cdot\bm{\sigma}_2 |\Psi_N\right) =  
   \left(\Psi_N| \bm{\sigma}_1\cdot\bm{\sigma}_3 |\Psi_N\right) =  
   \left(\Psi_N| \bm{\sigma}_2\cdot\bm{\sigma}_3 |\Psi_N\right) = -1, \\
&& \left(\Psi_N| \bm{\tau}_1\cdot\bm{\tau}_2\ \bm{\sigma}_1\cdot\bm{\sigma}_2|\Psi_N\right) =  
   \left(\Psi_N| \bm{\tau}_1\cdot\bm{\tau}_3\ \bm{\sigma}_1\cdot\bm{\sigma}_3|\Psi_N\right) =  
\nonumber\\ &&
   \left(\Psi_N| \bm{\tau}_2\cdot\bm{\tau}_3\ \bm{\sigma}_2\cdot\bm{\sigma}_3|\Psi_N\right) = +5.  
\end{eqnarray}
\end{subequations}
Of course, these matrix elements are the same for the neutron.\\

\noindent {\bf 6.} $\bm{\Lambda}$: The flavor part of the
wave function is 
\begin{equation}
 \phi_{MS}(\Lambda) = -\frac{1}{2} \left|
 \begin{array}{cccc} \cline{1-3} 
  1 & | & 2 & | \\ \cline{1-3} 
  3 & | &   &   \\ \cline{1-1} \end{array}\right. (uds),
\label{eq:spin.41}\end{equation}
where the Young-operator is $ Y=PQ=[e+(12)][e-(13)]$.
For the explicit derivation of the matrix elements it is
useful to introduce the wave function components
\begin{eqnarray}
&& \phi_1 = \frac{1}{\sqrt{2}}\left(dsu-usd\right)\ ,\ 
 \phi_2 = \frac{1}{\sqrt{2}}\left(sdu-sud\right)\ ,\ 
 \phi_3 = \frac{1}{\sqrt{2}}\left(dus-uds\right).    
\label{eq:spin.21}\end{eqnarray}
These wave functions are orthogonal. The mixed symmetry states for the 
$\Lambda$ are, see \cite{Close79} section 3.3,
\begin{equation}
 \phi_{M,S} = \frac{1}{\sqrt{2}}\left(\phi_1+\phi_2\right)\ ,\ 
 \phi_{M,A} = -\frac{1}{\sqrt{6}}\left(\phi_1-\phi_2+2\phi_3\right).    
\label{eq:spin.52}\end{equation}
The operation of $\bm{\tau}_i\cdot\bm{\tau}_j$ on the components, using (\ref{eq:iso.6}),             
is readily evaluated. The results are
\begin{subequations}
\label{eq:spin.23}
\begin{eqnarray}
&& (\bm{\tau}_1\cdot\bm{\tau}_2)\ \phi_1 = 0\ ,\ (\bm{\tau}_1\cdot\bm{\tau}_2)\ \phi_2 = 0\ ,\
 (\bm{\tau}_1\cdot\bm{\tau}_2)\ \phi_3 = -3 \phi_3, \\
&& (\bm{\tau}_1\cdot\bm{\tau}_3)\ \phi_1 = -3 \phi_1\ ,\ 
 (\bm{\tau}_1\cdot\bm{\tau}_3)\ \phi_2=0\ ,\
 (\bm{\tau}_1\cdot\bm{\tau}_3)\ \phi_3 = 0, \\
&& (\bm{\tau}_2\cdot\bm{\tau}_3)\ \phi_1 = 0\ ,\ 
 (\bm{\tau}_2\cdot\bm{\tau}_3)\ \phi_2=-3\phi_2\ ,\
 (\bm{\tau}_2\cdot\bm{\tau}_3)\ \phi_3 = 0.    
\end{eqnarray}
\end{subequations}
With these we find 
\begin{subequations}
\label{eq:spin.24}
\begin{eqnarray}
&& (\bm{\tau}_1\cdot\bm{\tau}_2)\ \phi_{M,S} = 0\ ,\
   (\bm{\tau}_1\cdot\bm{\tau}_2)\ \phi_{M,A} = +\sqrt{6} \phi_3, \\
&& (\bm{\tau}_1\cdot\bm{\tau}_3)\ \phi_{M,S} = -\frac{3}{\sqrt{2}} \phi_1\ ,\
   (\bm{\tau}_1\cdot\bm{\tau}_3)\ \phi_{M,A} = +\frac{3}{\sqrt{6}} \phi_1, \\
&& (\bm{\tau}_2\cdot\bm{\tau}_3)\ \phi_{M,S} = -\frac{3}{\sqrt{2}} \phi_2\ ,\
   (\bm{\tau}_2\cdot\bm{\tau}_3)\ \phi_{M,A} = -\frac{3}{\sqrt{6}} \phi_2,   
\end{eqnarray}
\end{subequations}
which give the matrix elements
\begin{subequations}
\label{eq:spin.25}
\begin{eqnarray}
&& \left(\phi_{M,S}|\bm{\tau}_1\cdot\bm{\tau}_2|\phi_{M,S}\right) = 0\ ,\
   \left(\phi_{M,A}|\bm{\tau}_1\cdot\bm{\tau}_2|\phi_{M,A}\right) = -2, \\
&& \left(\phi_{M,S}|\bm{\tau}_1\cdot\bm{\tau}_2|\phi_{M,A}\right) = 0\ ,\
   \left(\phi_{M,A}|\bm{\tau}_1\cdot\bm{\tau}_2|\phi_{M,S}\right) = 0, \\
&& \left(\phi_{M,S}|\bm{\tau}_1\cdot\bm{\tau}_3|\phi_{M,S}\right) = -\frac{3}{2}\ ,\
   \left(\phi_{M,A}|\bm{\tau}_1\cdot\bm{\tau}_3|\phi_{M,A}\right) = -\frac{1}{2}, \\
&& \left(\phi_{M,S}|\bm{\tau}_1\cdot\bm{\tau}_3|\phi_{M,A}\right) = +\frac{1}{2}\sqrt{3}\ ,\
   \left(\phi_{M,A}|\bm{\tau}_1\cdot\bm{\tau}_3|\phi_{M,S}\right) = +\frac{1}{2}\sqrt{3}, \\
&& \left(\phi_{M,S}|\bm{\tau}_2\cdot\bm{\tau}_3|\phi_{M,S}\right) = -\frac{3}{2}\ ,\
   \left(\phi_{M,A}|\bm{\tau}_2\cdot\bm{\tau}_3|\phi_{M,A}\right) = -\frac{1}{2}, \\
&& \left(\phi_{M,S}|\bm{\tau}_2\cdot\bm{\tau}_3|\phi_{M,A}\right) = -\frac{1}{2}\sqrt{3}\ ,\
   \left(\phi_{M,A}|\bm{\tau}_2\cdot\bm{\tau}_3|\phi_{M,S}\right) = -\frac{1}{2}\sqrt{3}.    
\end{eqnarray}
\end{subequations}
Similar results hold for the spin-operators. This gives for the $\Lambda$ matrix elements
of the isospin-spin operators:  
\begin{subequations} \label{eq:spin.18}
\begin{eqnarray}
&& \left(\Psi_\Lambda| \bm{\tau}_1\cdot\bm{\tau}_2 |\Psi_\Lambda\right) =
   \left(\Psi_\Lambda| \bm{\tau}_1\cdot\bm{\tau}_3 |\Psi_\Lambda\right) =  
   \left(\Psi_\Lambda| \bm{\tau}_2\cdot\bm{\tau}_3 |\Psi_\Lambda\right) = -1, \\
&& \left(\Psi_\Lambda| \bm{\sigma}_1\cdot\bm{\sigma}_2 |\Psi_\Lambda\right) =
   \left(\Psi_\Lambda| \bm{\sigma}_1\cdot\bm{\sigma}_3 |\Psi_\Lambda\right) =  
   \left(\Psi_\Lambda| \bm{\sigma}_2\cdot\bm{\sigma}_3 |\Psi_\Lambda\right) = -1, \\
&& \left(\Psi_\Lambda| \bm{\tau}_1\cdot\bm{\tau}_2\
    \bm{\sigma}_1\cdot\bm{\sigma}_2|\Psi_\Lambda\right) =
   \left(\Psi_\Lambda| \bm{\tau}_1\cdot\bm{\tau}_3\  
    \bm{\sigma}_1\cdot\bm{\sigma}_3|\Psi_\Lambda\right) =
\nonumber\\ &&
   \left(\Psi_\Lambda| \bm{\tau}_2\cdot\bm{\tau}_3\ 
    \bm{\sigma}_2\cdot\bm{\sigma}_3|\Psi_\Lambda\right) = +2.  
\end{eqnarray}
\end{subequations}
\noindent {\bf 7.} $\bm{\Sigma^+}$: The flavor part of the wave
function is 
\begin{equation}
 \phi_{MS}(\Sigma^+) = -\frac{1}{\sqrt{6}} \left|
 \begin{array}{cccc} \cline{1-3} 
  1 & | & 2 & | \\ \cline{1-3} 
  3 & | &   &   \\ \cline{1-1} \end{array}\right. (uus).
\label{eq:spin.42}\end{equation}
This state is the same as the proton if we make the substitution
$d \rightarrow s$. But the isospin-operator matrix elements are different.
Explicit calculation gives for the $\Sigma^+$ spin-isospin matrix elements  
\begin{subequations} \label{eq:spin.19}
\begin{eqnarray}
&& \left(\Psi_\Sigma| \bm{\tau}_1\cdot\bm{\tau}_2 |\Psi_\Sigma\right) = 
   \left(\Psi_\Sigma| \bm{\tau}_1\cdot\bm{\tau}_3 |\Psi_\Sigma\right) = 
   \left(\Psi_\Sigma| \bm{\tau}_2\cdot\bm{\tau}_3 |\Psi_\Sigma\right) = +\frac{1}{3}, \\
&& \left(\Psi_\Sigma| \bm{\sigma}_1\cdot\bm{\sigma}_2 |\Psi_\Sigma\right) =  
   \left(\Psi_\Sigma| \bm{\sigma}_1\cdot\bm{\sigma}_3 |\Psi_\Sigma\right) =  
   \left(\Psi_\Sigma| \bm{\sigma}_2\cdot\bm{\sigma}_3 |\Psi_\Sigma\right) = -1, \\
&& \left(\Psi_\Sigma| \bm{\tau}_1\cdot\bm{\tau}_2\ 
    \bm{\sigma}_1\cdot\bm{\sigma}_2|\Psi_\Sigma\right) = 
   \left(\Psi_\Sigma| \bm{\tau}_1\cdot\bm{\tau}_3\ 
    \bm{\sigma}_1\cdot\bm{\sigma}_3|\Psi_\Sigma\right) = 
\nonumber\\ &&
   \left(\Psi_\Sigma| \bm{\tau}_2\cdot\bm{\tau}_3\ 
    \bm{\sigma}_2\cdot\bm{\sigma}_3|\Psi_\Sigma\right) = +\frac{1}{3}.
\end{eqnarray}
\end{subequations}
\noindent {\bf 8.} $\bm{\Xi^0}$: The flavor part of the wave
function is 
\begin{equation}
 \phi_{MS}(\Xi^0) = -\frac{1}{\sqrt{6}} \left|
 \begin{array}{cccc} \cline{1-3} 
  1 & | & 2 & | \\ \cline{1-3} 
  3 & | &   &   \\ \cline{1-1} \end{array}\right. (ssu).
\label{eq:spin.43}\end{equation}
This state is the same as the neutron if we make the substitution
$d \rightarrow s$. The matrix elements of the spin-operators
are the same as for the neutron and the proton. The isospin matrix elements are different,
being simply zero due to double the s-quark component. 
The $\Xi^+$ spin-isospin matrix elements are
\begin{subequations} \label{eq:spin.20}
\begin{eqnarray}
&& \left(\Psi_\Xi| \bm{\tau}_1\cdot\bm{\tau}_2 |\Psi_\Xi\right) = 
   \left(\Psi_\Xi| \bm{\tau}_1\cdot\bm{\tau}_3 |\Psi_\Xi\right) = 
   \left(\Psi_\Xi| \bm{\tau}_2\cdot\bm{\tau}_3 |\Psi_\Xi\right) = 0, \\
&& \left(\Psi_\Xi| \bm{\sigma}_1\cdot\bm{\sigma}_2 |\Psi_\Xi\right) =  
   \left(\Psi_\Xi| \bm{\sigma}_1\cdot\bm{\sigma}_3 |\Psi_\Xi\right) =  
   \left(\Psi_\Xi| \bm{\sigma}_2\cdot\bm{\sigma}_3 |\Psi_\Xi\right) = -1, \\
&& \left(\Psi_\Xi| \bm{\tau}_1\cdot\bm{\tau}_2\ 
    \bm{\sigma}_1\cdot\bm{\sigma}_2|\Psi_\Xi\right) = 
   \left(\Psi_\Xi| \bm{\tau}_1\cdot\bm{\tau}_3\ 
    \bm{\sigma}_1\cdot\bm{\sigma}_3|\Psi_\Xi\right) = 
\nonumber\\ &&
   \left(\Psi_\Xi| \bm{\tau}_2\cdot\bm{\tau}_3\ 
    \bm{\sigma}_2\cdot\bm{\sigma}_3|\Psi_\Xi\right) = 0.            
\end{eqnarray}
\end{subequations}
\noindent {\bf 9.} $\bm{\Delta}_{33}^{++}$: 
The flavor part of the wave function is 
\begin{equation}
 \phi_{MS}(\Delta_{33}^{++}) = \frac{1}{6} \left|
 \begin{array}{cccccc} \cline{1-5} 
  1 & | & 2 & | & 3 & |\\ \cline{1-5} 
\end{array}\right. (uuu).
\label{eq:spin.44}\end{equation}
The states are the completely summetric $\phi_S=uuu$ and $\chi_S=+++$.
This gives
\begin{eqnarray}
 (\bm{\tau}_1\cdot\bm{\tau}_2)\ \phi_S = (\bm{\tau}_1\cdot\bm{\tau}_3)\ \phi_S =
 (\bm{\tau}_2\cdot\bm{\tau}_3)\ \phi_S = \phi_S,
\label{eq:spin.51}\end{eqnarray}
and similarly for the spin operators $\bm{\sigma}_1\cdot\bm{\sigma}_2 \chi_S= \chi_S$ etc.
This gives
\begin{subequations} \label{eq:spin.22}
\begin{eqnarray}
&& \left(\Psi_\Delta| \bm{\tau}_1\cdot\bm{\tau}_2 |\Psi_\Delta\right) = 
   \left(\Psi_\Delta| \bm{\tau}_1\cdot\bm{\tau}_3 |\Psi_\Delta\right) = 
   \left(\Psi_\Delta| \bm{\tau}_2\cdot\bm{\tau}_3 |\Psi_\Delta\right) = +1, \\
&& \left(\Psi_\Delta| \bm{\sigma}_1\cdot\bm{\sigma}_2 |\Psi_\Delta\right) =  
   \left(\Psi_\Delta| \bm{\sigma}_1\cdot\bm{\sigma}_3 |\Psi_\Delta\right) =  
   \left(\Psi_\Delta| \bm{\sigma}_2\cdot\bm{\sigma}_3 |\Psi_\Delta\right) = +1, \\
&& \left(\Psi_\Delta| \bm{\tau}_1\cdot\bm{\tau}_2\ 
    \bm{\sigma}_1\cdot\bm{\sigma}_2|\Psi_\Delta\right) = 
   \left(\Psi_\Delta| \bm{\tau}_1\cdot\bm{\tau}_3\ 
    \bm{\sigma}_1\cdot\bm{\sigma}_3|\Psi_\Delta\right) = 
\nonumber\\ &&
   \left(\Psi_\Delta| \bm{\tau}_2\cdot\bm{\tau}_3\ 
    \bm{\sigma}_2\cdot\bm{\sigma}_3|\Psi_\Delta\right) = 1.            
\end{eqnarray}
\end{subequations}

\begin{flushleft}
\rule{16cm}{0.5mm}
\end{flushleft}
 
\section{Momentum-space Wave Functions II}             
\label{app:mom2}     
The wave function as a function of the momenta ${\bf p}_i, i=1,2,3$ in the 
three-particle CM-system is
\begin{eqnarray}
\Psi_3({\bf p}_1,{\bf p}_2,{\bf p}_3) &=& \widetilde{N}_3
 \exp\biggl[-\frac{1}{6\lambda}\left({\bf p}_1^2+{\bf p}_2^2+{\bf p}_3^2\right)\biggr]\
 \delta^3(\sum_{i=1}^3 {\bf p}_i),
\label{app:mom2.1}\end{eqnarray}
where the normalization factor $\widetilde{N}_3$ we evaluate as follows.      
It is convenient to replace $\delta^3(\sum {\bf p}_i)$ by the gaussian form
\cite{gauss-form}
\begin{eqnarray}
 \delta^3(\sum_{i=1}^3 {\bf p}_i) & =& \lim_{m_\epsilon \rightarrow 0} 
 (4\pi m_\epsilon^2)^{-3/2} \exp\left[-(\sum {\bf p}_i^2)/(4m_\epsilon^2\right].
\label{app:mom2.2}\end{eqnarray}
For the norm $\widetilde{N}_3$ we have to evaluate the integral
\begin{eqnarray}
J_3(a,b,c) &=& \Pi_{i=1}^3 \int d^3p_i\ \exp\biggl[-a{\bf p}_1^2-b{\bf p}_2^2-c{\bf p}_3^2\biggr]
 \exp\biggl[-\alpha\left({\bf p}_1+\bf{p}_2+{\bf p}_3\right)^2\biggr].
\label{app:mom2.3}\end{eqnarray}
with $a=b=c=1/3\lambda, \alpha=1/(4m_\epsilon^2)$. In a more explicit form
\begin{eqnarray}
J_3(a,b,c) &=& \Pi_{i=1}^3 d^3p_i\ \exp\biggl[-(a+\alpha){\bf p}_1^2
-(b+\alpha){\bf p}_2^2-(c+\alpha){\bf p}_3^2\biggr]\cdot\nonumber\\ && \times
\biggl[-2\alpha\left( {\bf p}_1\cdot{\bf p}_2 +{\bf p}_1\cdot{\bf p}_3 
+{\bf p}_2\cdot{\bf p}_3\right)\biggr].
\label{app:mom2.4}\end{eqnarray}
Performing successively the integrals gives:\\
\noindent 1.\ ${\bf p}_1$-integration
\begin{eqnarray*}
&\Rightarrow& \int d^3p_1 \exp\biggl[-(a+\alpha){\bf p}_1^2
 -2\alpha({\bf p}_2+{\bf p}_3)\cdot{\bf p}_1\biggr] = 
 \left(\frac{\pi}{a+\alpha}\right)^{3/2} \exp\biggl[+\frac{\alpha^2}{a+\alpha}
 ({\bf p}_2+{\bf p}_3)^2\biggr].
\label{app:mom2.5}\end{eqnarray*}
\noindent 2.\ ${\bf p}_2$-integration
\begin{eqnarray*}
&\Rightarrow& \int d^3p_2 \exp\biggl[-(b+\alpha){\bf p}_2^2 
-2\alpha ({\bf p}_2\cdot{\bf p}_3)
 +\frac{\alpha^2}{a+\alpha} ({\bf p}_2+{\bf p}_3)^2\biggr]
 =\left(\frac{\pi(a+\alpha)}{ab+\alpha(a+b)}\right)^{3/2}
\cdot\nonumber\\ && \times
\exp\left( \frac{\alpha^2a^2}{(a+\alpha)(ab+\alpha(a+b))}\ {\bf p}_3^2\right).
\label{app:mom2.6}\end{eqnarray*}
\noindent 3.\ ${\bf p}_3$-integration
\begin{eqnarray*}
&\Rightarrow& \int d^3p_3 \exp\biggl[-(c+\alpha){\bf p}_3^2 +\frac{\alpha^2}{a+\alpha}
{\bf p}_3^2 +
\frac{\alpha^2a^2}{(a+\alpha)(ab+\alpha(a+b))}\ {\bf p}_3^2\biggr] =
\nonumber\\ && 
\left(\pi\frac{ab+\alpha(a+b)}{abc+(ab+ac+bc)\alpha}\right)^{+3/2}.
\label{app:mom2.7}\end{eqnarray*}
Collecting factors we obtain
\begin{eqnarray}
 J_3(a,b,c) &=& \pi^{9/2}\biggl\{abc+\alpha(ab+ac+bc)\biggr\}^{-3/2} 
 = \pi^{9/2} \biggl\{\gamma^3+3\alpha\gamma^2\biggr\}^{-3/2},
\label{app:mom2.8}\end{eqnarray}
with the notation $a=b=c=\gamma \equiv 1/3\lambda$.                           
From $\widetilde{N}_3^2\ (4\pi m_\epsilon^2)^{-3} J_3(a,b,c)=1$ follows
\begin{eqnarray}
 \widetilde{N}_3^2 &=& 
(4\pi m_\epsilon^2)^{3}\ J_3^{-1}(a,b,c)\biggl|_{a=b=c=\gamma} =
 \pi^{-9/2}\ (4\pi m_\epsilon^2)^{3}\ 
 \biggl\{\gamma^3+3\alpha\gamma^2\biggr\}^{+3/2},
\label{app:mom2.9}\end{eqnarray}
The expectation of the kinetic-energy operator becomes
\begin{eqnarray}
\big\langle T \bigr\rangle &=& (2m_Q)^{-1} (4\pi m_\epsilon^2)^{-3}
\widetilde{N}_3^2 \int \Pi_{i=1}^3 d^3p_i\ \left(\sum_{i=1}^3 {\bf p}_i^2\right)\
\exp\biggl[-\frac{1}{3\lambda} \left(\sum_{i=1}^3 {\bf p}_i^2\right)\biggr] 
\cdot\nonumber\\ && \times 
\exp\biggl[\left(\sum_{i=1}^3 {\bf p}_i\right)^2/(4m_\epsilon^2)\biggr]
 \equiv (2m_Q)^{-1} (4\pi m_\epsilon^2)^{-3}\ \widetilde{N}_3 I_3 
\nonumber\\ &=&
 -(2m_Q)^{-1} (4\pi m_\epsilon^2)^{-3}\ \widetilde{N}_3  
\left(\frac{d}{da}+\frac{d}{db}+\frac{d}{dc}\right)\ J_3(a,b,c), 
\label{app:mom2.10}\end{eqnarray}
\begin{eqnarray}
 I_3(a,b,c) &=& \frac{3}{2}\pi^{9/2}\ \bigl[ab+ac+bc+2\alpha(a+b+c)\bigr]
\biggl\{abc+\alpha(ab+ac+bc)\biggr\}^{-5/2},
\label{app:mom2.11}\end{eqnarray}
which gives
\begin{eqnarray}
\big\langle T \bigr\rangle &=& (2m_Q)^{-1}\cdot \frac{3}{2}
\bigl[ 3\gamma^2+6\alpha \gamma\bigr]\ \biggl[\gamma^3+3\alpha\gamma^2\biggr]^{-1} =
 \frac{9}{2} \gamma^{-1}\left(1+2\frac{\alpha}{\gamma}\right)\
\left(1+3\frac{\alpha}{\gamma}\right)^{-1}/(2m_Q) \nonumber\\ &=&
 \frac{27}{2}\lambda\ \left(1+\frac{9}{2m_\epsilon^2 R_N^2}\right)
 \left(1+\frac{27}{4m_\epsilon^2 R_N^2}\right)^{-1}/(2m_Q) \rightarrow
 (27/2) (m_Q R_N)^{-2}\ m_Q.
\label{app:mom2.12}\end{eqnarray}
This gives for $R_N = 1 $ fm and $m_Q=321.75$ MeV approximately $3m_Q/2 = 469$ MeV, 
giving the same answer as in (\ref{eq:KIN2}). 

\begin{flushleft}
\rule{16cm}{0.5mm}
\end{flushleft}
\section{Relativistic expansion factors}                          
\label{app:relfacts}
In the Pauli-spinor expansion of the Dirac-spinors occur the $(E+M)^{-1}$
factors, which show up as $(4M'M)^{-1}$ coefficients in the spin-spin,
tensor, and spin-orbit potentials. In the quadratic-spin-orbit as $(4M'M)^{-2}$
coefficients. Comparing these coefficients for the nucleon-nucleon and the
quark-quark potentials there is a difference of 9 and 81, making these 
potentials much stronger in the quark-quark case. This seems artificial   
in realizing that the quarks are moving relativistically inside a nucleon.
A way to include these $(E+M)^{-1}$-factors in an exact way within the 
context of the harmonic-oscillator quark-model of the baryons is described in 
this Appendix.\\
\noindent Starting from the integral presentation
\begin{eqnarray}
 \frac{1}{E({\bf p})+M} &=& \frac{2}{\pi}\int_0^\infty d\lambda\
 \frac{\lambda^2}{(\lambda^2+M^2)} \frac{1}{(E^2({\bf p})+\lambda^2)}
\nonumber\\ &=& \frac{2}{\pi}\int_0^\infty d\alpha\ e^{-\alpha({\bf p}^2+M^2)}\
 \int_0^\infty \lambda^2d\lambda\ \frac{e^{-\alpha \lambda^2}}{\lambda^2+M^2}
\label{app:facts1}\end{eqnarray}
The $\lambda$-integral is
\begin{equation}
 \int_0^\infty \lambda^2d\lambda\ \frac{e^{-\alpha \lambda^2}}{\lambda^2+M^2}
 = \frac{1}{2}\sqrt{\frac{\pi}{\alpha}}\biggl(1 - \sqrt{\pi\alpha M^2}\
 e^{\alpha M^2} {\it Erfc}\left(\sqrt{\alpha} M\right)\biggr).
\label{app:facts2} \end{equation}
This leads to the exact expression
\begin{equation}
 \frac{1}{E({\bf p})+M} = \frac{1}{\sqrt{\pi}}\int_0^\infty 
 \frac{d\alpha}{\sqrt{\alpha}}\ e^{-\alpha M^2}\ \biggl(1-\sqrt{\pi\alpha M^2}\
e^{\alpha M^2}\ {\it Erfc}(\sqrt{\alpha M^2})\biggr)\cdot 
 exp\bigl[-\alpha {\bf p}^2\bigr].
\label{app:facts3} \end{equation}
After making the transformation $ \alpha = y^2$ and subsequently $y=x/M$ one
obtains                    
\begin{eqnarray}
 \frac{1}{E({\bf p})+M} &=& \frac{1}{2M}\frac{4}{\sqrt{\pi}}\int_0^\infty dx\
e^{-x^2}\left[1-\sqrt{\pi}\ x e^{x^2}{\it Erfc}(x)\right]\cdot
 \exp\left[-\frac{x^2}{M^2} {\bf p}^2\right] 
 \nonumber\\ &\equiv& (2M)^{-1}\ f({\bf p}^2,M^2),
\label{app:facts4} \end{eqnarray}
and the non-relativistic approximation means $f({\bf p}^2,M^2)=1$.\\
Note, that again the momentum behavior is Gaussian, and can be incorporated 
in the calculations of the matrix elements of the $V_2$ potentials.
Of course, for $[(E(p_1)+M)(E(p_2)+M)]^{-1}$,
 this at the cost of two-extra numerical integrals. 

\section{SU(3) NJL-form Instanton Lagrangian}                     
\label{app:instnlag}
The 't Hooft instanton-determinant generated quark-quark interaction
 \cite{GtH76,Bo10} in the (u,d,s)-sector
\begin{eqnarray}
 {\cal L}_{det} &=& 8G_2\ e^{i\theta}\ \det(\bar{\psi}_R\psi_L) + h.c. 
 \nonumber\\ &=& G_2\biggl[ (\bar{\psi}\lambda_0 \psi)^2
+(\bar{\psi}\lambda_0 \gamma_5 \psi)^2 
-(\bar{\psi}\lambda_i \psi)^2
-(\bar{\psi}\lambda_i \gamma_5 \psi)^2 
\biggr]. 
\label{app:inst2.1}\end{eqnarray}
Here, we have taken in the last expression $\theta=0$. 
The convention used for the right- and left-hand quarks is
\begin{equation}
 q_R = \frac{1}{2}(1+\gamma_5)\ q,\ q_L= \frac{1}{2}(1-\gamma_5)\ q.
\label{app:inst2.2}\end{equation}
where q is the generic for u,d, and s. 
In \cite{SHUR84} the (u,d,s)-sector Lagrangian reads 
\begin{eqnarray}
 {\cal L}_4 &=& \lambda_{ud}(\bar{u}_Ru_L)(\bar{d}_Rd_L)
 +\lambda_{su}(\bar{s}_Rs_L)(\bar{u}_Ru_L)
 +\lambda_{sd}(\bar{s}_Rs_L)(\bar{d}_Rd_L) + (R \leftrightarrow L), 
\nonumber\\ \lambda_{ud}&=& 2n_+/(\langle\bar{\psi}\psi\rangle)^2,\
\lambda_{su}=\lambda_{sd}=\lambda_{ud} \langle \bar{u}u\rangle/
\biggl[\langle \bar{s}s\rangle-3 m_s/(2\pi^2\rho_c)\biggr],
\label{app:inst2.3}\end{eqnarray}
which implies for the (u,d)-sector the Lagrangian 
\begin{eqnarray}
 {\cal L}_{ud} &=& \lambda_{ud}\biggl[(\bar{u}_Ru_L)(\bar{d}_Rd_L)
                   +(\bar{u}_Lu_R)(\bar{d}_Ld_R)\biggr] \nonumber\\
 &=& \frac{1}{2}\lambda_{ud}\biggl[
 (\bar{u}u)(\bar{d}d)+(\bar{u}\gamma_5u)(\bar{d}\gamma_5 d)\biggr].
\label{app:inst2.4}\end{eqnarray}
In the (ud)-sector the Lagrangian (\ref{app:inst2.1}) is 
\begin{eqnarray}
 {\cal L}_{det}(ud) &\Rightarrow& 
G_2\biggl[ (\bar{\psi}\psi)^2 +(\bar{\psi}\gamma_5 \psi)^2 
-(\bar{\psi}\bm{\tau} \psi)^2 -(\bar{\psi}\bm{\tau}\gamma_5 \psi)^2 
\biggr]. 
\label{app:inst2.5}\end{eqnarray}
Working out the Lagrangian (\ref{app:inst2.1}) for the (u,d)-sector 
 one obtains 
\begin{subequations}
\label{app:inst2.6}
\begin{eqnarray}
&& (\bar{\psi}\psi)^2 = (\bar{u}u)(\bar{u}u) +2(\bar{u}u)(\bar{d}d)
+ (\bar{d}d)(\bar{d}d), \\
&& (\bar{\psi}\gamma_5\psi)^2 = (\bar{u}\gamma_5u)(\bar{u}\gamma_5u) 
 +2(\bar{u}\gamma_5u)(\bar{d}\gamma_5d) + (\bar{d}\gamma_5d)(\bar{d}\gamma_5d), \\
&& (\bar{\psi}\bm{\tau}\psi)^2 = 
(\bar{u}u)(\bar{u}u)+(\bar{d}d)(\bar{d}d)
-2(\bar{u}u)(\bar{d}d) +4(\bar{u}d)(\bar{d}u), \\
&& (\bar{\psi}\bm{\tau}\gamma_5\psi)^2 = 
(\bar{u}\gamma_5u)(\bar{u}\gamma_5u)+(\bar{d}\gamma_5d)(\bar{d}\gamma_5d)
-2(\bar{u}\gamma_5u)(\bar{d}\gamma_5d) +4(\bar{u}\gamma_5d)(\bar{d}\gamma_5u).   
\end{eqnarray}
\end{subequations}
Now,  
\begin{eqnarray*}
&& (\bar{u}d)(\bar{d}u)+(\bar{u}\gamma_5 d)(\bar{d}\gamma_5 u) \sim 
 \frac{1}{2}\bigl[ (\bar{u}u)(\bar{d}d)+(\bar{u}\gamma_5u)(\bar{d}\gamma_5 d)\bigr],
\end{eqnarray*}
where the Fierz-identities, see Appendix in \cite{Ok82}, have been used.
This also generates an tensor-type of term which as is usual neglected,
see {\it e.g.} \cite{Glozman00b}.
Then, from (\ref{app:inst2.5}) and (\ref{app:inst2.6}) we obtain
\begin{eqnarray}
 {\cal L}_{det}(ud) &\approx & 
2G_2\biggl[ (\bar{u}u)(\bar{d}d)+(\bar{u}\gamma_5u)(\bar{d}\gamma_5d)\biggr].
\label{app:inst2.7}\end{eqnarray}
This corresponds with Eq.~(\ref{app:inst2.4}), and 
implies the relation $4 G_2=\lambda_{ud}$.

\noindent The complete instanton Lagrangian reads, see \cite{SHUR84} Eqn.~(6.9),
\begin{eqnarray}
 {\cal L}_{uds} &=& 
\lambda_{ud} (\bar{u}_Ru_L)(\bar{d}_Rd_L) +
\lambda_{su} (\bar{s}_Rs_L)(\bar{u}_Ru_L) +
\lambda_{sd} (\bar{s}_Rs_L)(\bar{d}_Rd_L) + (R \leftrightarrow L)
\nonumber\\ &=& {\cal L}_{det}(ud) + {\cal L}_{det}(su) + {\cal L}_{det}(sd),
\label{app:inst2.8}\end{eqnarray}
where 
\begin{equation}
 \lambda_{ud} \approx 2n_+/\langle \bar{\Psi}\Psi\rangle^2;\ \ \lambda_{su}=\lambda_{sd}=
\lambda_{ud} \langle\bar{u}u\rangle/[\langle\bar{s}s\rangle-3m_s/2\pi^2\rho_c].
\label{app:inst2.9}\end{equation}
Here, $\langle \bar{\Psi}\Psi\rangle$ etc the vacuum is the chiral spontaneously
broken vacuum. (Note that the vacuum $|0\rangle$ in the CQM is
"trivial", i.e. $\langle0| \bar{q}q|0\rangle =0$.)
\noindent We now work out the SU(3)-symmetric Lagrangian in 
Eq.~(\ref{app:inst2.1}), and take $(\lambda_0)_{i,j} = (2/\sqrt{3})\ \delta_{i,j}$.
For the scalar current terms we get
\begin{eqnarray}
 {\cal L}_{det}(S) &=& G_2\biggl[ (\bar{\psi}\lambda_0 \psi)^2
-(\bar{\psi}\lambda_i \psi)^2 \biggr] \nonumber\\ &=& 8G_2\biggl[
\biggl( \bar{u}u\cdot\bar{d}d+ \bar{u}u\cdot\bar{s}s+ \bar{d}d\cdot\bar{s}s\biggr)
\nonumber\\ && -
\biggl( \bar{u}d\cdot\bar{d}u+ \bar{u}s\cdot\bar{s}u+ \bar{d}s\cdot\bar{s}d\biggr)
\nonumber\\ & \Rightarrow& G_2\biggl[
3\biggl( \bar{u}u\cdot\bar{d}d+ \bar{u}u\cdot\bar{s}s+ \bar{d}d\cdot\bar{s}s\biggr)
\nonumber\\ && -
\biggl( \bar{u}\gamma_5u\cdot\bar{d}\gamma_5d+ 
 \bar{u}\gamma_5u\cdot\bar{s}\gamma_5s+ \bar{d}\gamma_5d\cdot\bar{s}\gamma_5s\biggr)
 + \ldots \biggr]
\label{app:inst2.10}\end{eqnarray}
Here, for arriving at the last expression we used the Fierz-transformation.
Similarly, for the pseudoscalar current terms 
\begin{eqnarray}
 {\cal L}_{det}(P) &\Rightarrow& G_2\biggl[ 
3\biggl( \bar{u}\gamma_5u\cdot\bar{d}\gamma_5d+ \bar{u}\gamma_5u\cdot\bar{s}\gamma_5s
 + \bar{d}\gamma_5d\cdot\bar{s}\gamma_5s\biggr)
\nonumber\\ && -
\biggl( \bar{u}u\cdot\bar{d}d+ \bar{u}u\cdot\bar{s}s+ \bar{d}d\cdot\bar{s}s\biggr)
 - \ldots \biggr]
\label{app:inst2.11}\end{eqnarray}
For ${\cal L}_{det}={\cal L}_{det}(S)+{\cal L}_{det}(P)$ the dotted terms cancel, except
for the tensor terms. Then, the result for ${\cal L}_{det}$ is   
\begin{equation}
{\cal L}_{det} \approx \biggl[ {\cal L}_{ud}+{\cal L}_{us}+{\cal L}_{ds}\biggr],
\label{app:inst2.12}\end{equation}
where ${\cal L}_{det}(us)$ and ${\cal L}_{det}(ds)$ are defined similarly as
${\cal L}_{det}(ud)$. 

\noindent {\bf Naive considerations:} 
Assuming that $\lambda_{ud}=\lambda_{su}=\lambda_{sd}\equiv \lambda_I$ the instanton couples
as follows: $P,N \sim uud, ddu \rightarrow 2\lambda_{I}$, $\Lambda,\Sigma^0 \rightarrow 3\lambda_I$,
$\Delta_{33} \sim 0$, and $\Xi^0 \sim uss \rightarrow 2 \lambda_I$. 
Also, $\langle \Delta_{33}|{\cal L}_{det}|\Delta_{33}\rangle =0$.
Furthermore, $\Sigma^+ \sim uus \rightarrow 2\lambda_{us}$, and 
$\Sigma^- \sim dds \rightarrow 2 \lambda_{ds}$. Then we expect 
$\Lambda, \Sigma^0 \sim uds \rightarrow \lambda_{us}+\lambda_{ds}$ ({\it calculation?}).
Therefore, instantons break SU(3)-symmetry when {\it e.g.}  
$\lambda_{ud} \neq \lambda_{us}=\lambda_{ds}$.\\
\noindent {\it In the next Appendix we give the results from an explicit calculation
of the matrix elements, which is clearing up the questions raised here!!}

\section{Baryon SU(3)-flavor- and Spin-operators}              
\label{app:uni}  
\noindent {\bf 1.} For the evaluation of the instanton two-body  
quark-quark interaction (\ref{app:inst2.1}), see e.g. \cite{Bub05,Wei75},
\begin{eqnarray}
 {\cal L}_{det} &=& G_2\biggl[ 
 (\bar{\psi}\lambda_0 \psi)^2 -(\bar{\psi}\bm{\lambda} \psi)^2
+(\bar{\psi}\lambda_0 \gamma_5 \psi)^2 -(\bar{\psi}\bm{\lambda} \gamma_5 \psi)^2 
\biggr]. 
\label{app:uni.1}\end{eqnarray}
where $\lambda_0 =\sqrt{2/3} {\bf 1}$, and $\lambda_a,\ a=1,8$ are the Gell-Mann matrices.\\
For the baryons the matrix elements of the SU(3)-flavor operators $\bm{\lambda}_i\cdot\bm{\lambda}_j$
and the spin operators $\bm{\sigma}_i\cdot\bm{\sigma}_j$ are given in this Appendix for
(i,j)=(1,2),(1,3), and (2,3).\\
\noindent {\bf 2.} The baryon octet $J^P=(1/2)^+$ 3 spin-isospin quark wave 
functions are  symmetric in spin-flavor space, see (\ref{eq:iso.1}),
\begin{eqnarray}
 \Psi_B &=& \frac{1}{\sqrt{2}}\left(\vphantom{\frac{A}{A}} 
 \phi_{M,S}\otimes\chi_{M,S}+\phi_{M,A}\otimes\chi_{M,A}\right)\ ,\
 \Psi_{\Delta_{33}} = \Phi_S\ \chi_S, 
\label{eq:uni.2}\end{eqnarray}
where in $\phi_{M,S}$ and $\phi_{M,A}$ the isospin of the 12-subsystems, 
which in the case of the nucleon is 1 and 0 respectively, see e.g. \cite{Close79}. 

\noindent {\bf 3.}\ Baryon octet $J^P=(1/2)^+$ spin-isospin matrix elements:
 From the baryon wave function (\ref{eq:uni.2}) one has
\begin{eqnarray}
 \left(\Psi_B|\ O_I\ O_S\ |\Psi_B\right) &=& \frac{1}{2}\left\{
 \vphantom{\frac{A}{A}} 
 \left(\phi_{M,S}| O_I|\phi_{M,S}\right)
 \left(\chi_{M,S}| O_S|\chi_{M,S}\right)
\right.\nonumber\\ && \left. \hspace{5mm}
 +\left(\phi_{M,S}| O_I|\phi_{M,A}\right)
 \left(\chi_{M,S}| O_S|\chi_{M,A}\right)
\right.\nonumber\\ && \left. \hspace{5mm}
 +\left(\phi_{M,A}| O_I|\phi_{M,S}\right)
 \left(\chi_{M,A}| O_S|\chi_{M,S}\right)
\right.\nonumber\\ && \left. \hspace{5mm}
 +\left(\phi_{M,A}| O_I|\phi_{M,A}\right)
 \left(\chi_{M,A}| O_S|\chi_{M,A}\right)
 \vphantom{\frac{A}{A}} \right\}.
\label{eq:uni.3} \end{eqnarray}
 
\noindent {\bf 4.\ N}: The unitary matrix elements are similar to the spin-operator
matrix element. The nucleon (P,N) proton matrix elements of the unitary-spin and spin 
two-body-operators are:
\begin{subequations} \label{eq:uni.4}
\begin{eqnarray}
&& \left(\Psi_N| \bm{\lambda}_1\cdot\bm{\lambda}_2 |\Psi_N\right) =  
   \left(\Psi_N| \bm{\lambda}_1\cdot\bm{\lambda}_3 |\Psi_N\right) =  
   \left(\Psi_N| \bm{\lambda}_2\cdot\bm{\lambda}_3 |\Psi_N\right) = -2/3, \\
&& \left(\Psi_N| \bm{\sigma}_1\cdot\bm{\sigma}_2 |\Psi_N\right) =  
   \left(\Psi_N| \bm{\sigma}_1\cdot\bm{\sigma}_3 |\Psi_N\right) =  
   \left(\Psi_N| \bm{\sigma}_2\cdot\bm{\sigma}_3 |\Psi_N\right) = -1, \\
&& \left(\Psi_N| \bm{\lambda}_1\cdot\bm{\lambda}_2\ \bm{\sigma}_1\cdot\bm{\sigma}_2|\Psi_N\right) =  
   \left(\Psi_N| \bm{\lambda}_1\cdot\bm{\lambda}_3\ \bm{\sigma}_1\cdot\bm{\sigma}_3|\Psi_N\right) =  
\nonumber\\ &&
   \left(\Psi_N| \bm{\lambda}_2\cdot\bm{\lambda}_3\ \bm{\sigma}_2\cdot\bm{\sigma}_3|\Psi_N\right) = +14/3.
\end{eqnarray}
\end{subequations}
Explicit calculation shows that these matrix elements are the same for $\Lambda, \Sigma$, and $\Xi$,
which is not surprising in view of the complete spin-flavor symmetry of the baryon states.
 
\noindent {\bf 5.\ $\bm{\Delta}_{33}$}: 
The matrix elements of the unitary-spin and spin two-body-operators are:
\begin{subequations} \label{eq:uni.5}
\begin{eqnarray}
&& \left(\Psi_\Delta| \bm{\lambda}_1\cdot\bm{\lambda}_2 |\Psi_\Delta\right) =  
   \left(\Psi_\Delta| \bm{\lambda}_1\cdot\bm{\lambda}_3 |\Psi_\Delta\right) =  
   \left(\Psi_\Delta| \bm{\lambda}_2\cdot\bm{\lambda}_3 |\Psi_\Delta\right) = +4/3, \\
&& \left(\Psi_\Delta| \bm{\sigma}_1\cdot\bm{\sigma}_2 |\Psi_\Delta\right) =  
   \left(\Psi_\Delta| \bm{\sigma}_1\cdot\bm{\sigma}_3 |\Psi_\Delta\right) =  
   \left(\Psi_\Delta| \bm{\sigma}_2\cdot\bm{\sigma}_3 |\Psi_\Delta\right) = +1, \\
&& \left(\Psi_\Delta| \bm{\lambda}_1\cdot\bm{\lambda}_2\ \bm{\sigma}_1\cdot\bm{\sigma}_2|\Psi_\Delta\right) =  
   \left(\Psi_\Delta| \bm{\lambda}_1\cdot\bm{\lambda}_3\ \bm{\sigma}_1\cdot\bm{\sigma}_3|\Psi_\Delta\right) =  
\nonumber\\ &&
   \left(\Psi_\Delta| \bm{\lambda}_2\cdot\bm{\lambda}_3\ \bm{\sigma}_2\cdot\bm{\sigma}_3|\Psi_\Delta\right) = +4/3.
\end{eqnarray}
\end{subequations}
 


\subsection{ Miscellaneous Material}                                          
\label{sec:15f} 
In Table~\ref{table11} the ESC16 energies are displayed. To arrive 
at the values shown in Table~\ref{table6} for T=0 these values have to be
multiplied with the expectation values of the operators
$ 1, (\bm{\sigma}_i\cdot\bm{\sigma}_j)$ for $E_C$ and 
($E_\sigma,E_T,E_{Q_{12}}$ respectively for each baryon. Similarly for T=1 
$E_C$ and ($E_\sigma, E_T, E_{Q_{12}}$ are multiplied by the
values of the operators $(\bm{\tau}_1\cdot\bm{\tau}_2)$, and 
$(\bm{\sigma}_1\cdot\bm{\sigma}_2), (\bm{\tau}_1\cdot\bm{\tau}_2)$
respectively.\\
In Table~\ref{table6} the contributions to the baryon mass of the 
 ESC16 central $(C_0,C_1)$, spin-spin $(\sigma_0,\sigma_1)$are shown. 
The latter get contributions from $V_{\sigma\sigma}, V_T$ and $V_{Q_{12}}$.
Also the contributions from the confinement and OGE are tabulated.
The constant $C_0= 760$ MeV in the confinement potential is taken 
from Novikov et al \cite{Nov78} in their work on Charmonium.\\
In Table~\ref{table7} the baryon masses are tabulated coming from the 
ESC16 QQ-potentials, OGE-potentials, the confinement potential,
the quark kinetic energies, the CM-energy subtraction, and the quark masses.
The subtracted by the CM-energy is $231$ MeV. 

\begin{table}
\caption{Coefficients of the ESC16 contributions to the potential
energies in the expansion 
$E_{ESC} = \bigl[E_{C,0}+ E_{\sigma,0}(\bm{\sigma}_1\cdot\bm{\sigma}_2))\biggr]
+ \bigl[E_{C,1}+ E_{\sigma,1}\bm{\sigma}_1\cdot\bm{\sigma}_2)\biggr]\
 (\bm{\tau}_1\cdot\bm{\tau}_2)$.
The quark masses are $m_N=312.75$ and $m_S=500$ in MeV.
Quark-radii are $R= 1.0$ fm for P, $\Delta_{33}$, $\Lambda$, $\Sigma^+$. }       
\label{table11}
\begin{center}
\begin{ruledtabular}
\begin{tabular}{cc|cccc} 
   QQ & T & $E_C$ & $E_\sigma$ & $E_T$ & $E_{Q_{12}}$ \\
\hline
   NN & 0 &+72.9 & -5.16 & +1.42 & +72.9 \\                       
   SN & 0 &+67.7 & -8.18 & +1.75 & +67.7 \\                       
   SS & 0 &+42.2 & -6.65 & +1.64 & +0.05 \\                       
\hline
   NN & 1 &+3.10 & -2.00 & +0.79 & -0.00 \\                       
   SN & 1 &+0.00 & -0.00 & +0.00 & +0.00 \\                       
   SS & 1 &+0.00 & +0.00 & +0.00 & +0.00 \\                       
\hline
\end{tabular}
\end{ruledtabular}
\end{center}
\end{table}
\begin{table}
\caption{Contributions Baryon masses using ESC16-parameters.
$C_0, \sigma_0$ denote the isospin 0 
contributions for the operators $1, (\bm{\sigma}_i\cdot\bm{\sigma}_j)$, and
$C_1, \sigma_1$ denote the isospin 1 
contributions for the operators $\bm{\tau}_i\cdot\bm{\tau}_j, 
 (\bm{\tau}_i\cdot\bm{\tau}_j)(\bm{\sigma}_i\cdot\bm{\sigma}_j)$. Note that
the spin-operator gets contributions from the spin-spin, tensor, and 
quadratic spin-orbit potentials.
The quark masses are $m_N=312.75$ and $m_S=500$ in MeV.
Quark-radii are $R= 1.0$ fm for P, $\Delta_{33}$, $\Lambda$, $\Sigma^+$.                         
}
\label{table6}
\begin{center}
\begin{ruledtabular}
\begin{tabular}{c|cc|cc|cc|c} \hline\hline
baryon & $C_0$ & $\sigma_0$ & $C_1$ & $\sigma_1$ & 
         $CNF_c$ & $CNF_\sigma$ & OGE \\                
\hline
 $P(939)$            &+72.9 & -69.1 &-3.1 & -1.2 & -659.0 & -161.0 &+5.9 \\
 $\Delta_{33}(1236)$ &+69.1 & +69.9 &+3.1 & +0.3 & -659.0 & +161.0 &-9.9 \\
 $\Lambda$(1115)     &+67.7 & -61.2 &+0.0 & +0.0 & -659.0 & -161.0 &+5.9 \\
 $\Sigma^+$(1189)    &+67.7 & -61.2 &+0.0 & +0.0 & -659.0 & -158.0 &+5.9 \\
 $\Xi^0$(1321)       &+42.2 & -61.2 &+0.0 & +0.0 & -659.0 & -53.8  &+0.6 \\
\hline
\end{tabular}
\end{ruledtabular}
\end{center}
\end{table}
\begin{table}
\caption{Contributions Baryon masses from the ESC QQ-potential (V$_{ESC}$),
the confinement central potential and the "magnetic" spin-spin interaction (V$_{conf}$),
the one-gluon-exchange interactions (OGE), the kinetic energy (E$_{kin}$), and 
constituent quark masses. Quark-radii are $R= 1.0$ fm for 
P, $\Delta_{33}$, $\Lambda$, $\Sigma^+$.                         
The quark masses are $m_N=312.75$ and $m_S=500$ in MeV.
}
\label{table7}
\begin{center}
\begin{ruledtabular}
\begin{tabular}{c|ccc|ccc|c} \hline\hline
 baryon & $ V_{ESC}$ & $V_{conf}$ & OGE & $V_{tot}$ & $E_{kin}$ & 
 $\sum_{i=1}^3 m_i$ & Mass \\
\hline
 $P(939)$            & -0.50& -820  & +5.90& -831 & +827 & 938.26 &  935 \\    
 $\Delta_{33}(1236)$ & +432& -498  & -9.90& -76  & +624 & 938.26 & 1486 \\    
 $\Lambda(1115)$     &+6.50 & -820  & +5.90& -808 & +833 &1125.50 & 1155 \\   
 $\Sigma(1189)$      &+6.50 & -820  & +5.90& -808 & +925 &1125.50 & 1253 \\   
 $\Xi(1321)$         &-19.0 & -820  & +0.06& -839 & +944 &1312.75 & 1381 \\   
\hline
\end{tabular}
\end{ruledtabular}
\end{center}
\end{table}



\end{document}